\documentclass{jfm_arxiv}
\usepackage{graphicx}
\usepackage{epstopdf,epsfig}
\usepackage{newtxtext}
\usepackage{newtxmath}
\usepackage{natbib}
\usepackage{float}
\usepackage{caption,subcaption}
\usepackage{titlesec}
\usepackage{amsmath, amssymb}
\usepackage{booktabs}
\usepackage{placeins}
\usepackage{xcolor}
\numberwithin{equation}{section}
\usepackage{hyperref}

\hypersetup{
    colorlinks = true,
    urlcolor   = blue,
    citecolor  = blue,
}

\usepackage{tikz}
\usetikzlibrary{arrows.meta}
\usepackage{pgfplots}
\pgfplotsset{compat=1.16}

\newcommand{\RomanNumeralCaps}[1]
\linenumbers

\title{Catheter-modulated transient solute dispersion in micropolar annular flow with a retentive and absorptive arterial wall}

\author{
Sohel Ahmed\aff{1},
Nanda Poddar\aff{2}\corresp{\email{nandapoddarcr7@gmail.com;nandap@srmist.edu.in}},
Kajal Kumar Mondal\aff{1}\corresp{\email{kkmondol@yahoo.co.in}}
\and
Subham Dhar\aff{3},
}

\affiliation{
\aff{1}Department of Mathematics, Cooch Behar Panchanan Barma University,
Cooch Behar 736101, India
\aff{2}Department of Mathematics, SRM Institute of Science and Technology,
Kattankulathur, Chennai 603203, India
\aff{3}Department of Civil Engineering, National Taiwan University,
Taipei City 10617, Taiwan
}

\begin{document}
\maketitle

\begin{abstract}
We study transient solute dispersion in pressure-driven micropolar flow through a concentric annulus representing a catheterized artery. The catheter is impermeable to solute, whereas the arterial wall combines irreversible removal with reversible surface retention. The exact steady micropolar velocity field is coupled to the Gill--Sankarasubramanian generalized-dispersion framework \citep{gill1970royal}; the resulting bulk--surface hierarchy is advanced with Rannacher-damped Crank--Nicolson time stepping to determine the exchange, convection and dispersion coefficients, \(K_0(t)\), \(K_1(t)\) and \(K_2(t)\). The exchange coefficient is independent of the micropolar parameters and satisfies \(-K_0(0^+)=2(\beta+\theta Da)/(1-\lambda^2)\), whereas the initial growth of shear-induced dispersion is \(K_2-Pe^{-2}\sim \sigma_v^2 t\), with \(\sigma_v^2\) the cross-sectional velocity variance; hence the leading short-time dispersion is hydrodynamic and independent of wall kinetics. In the weak-reaction regime, micropolarity modifies convection approximately linearly and shear dispersion quadratically through the velocity-amplitude factor. Most notably, a narrow-gap analysis with \(\varepsilon=1-\lambda\) gives \(\bar v\sim (2-N_c)\varepsilon^2/6\) and \(K_2-Pe^{-2}\sim (2-N_c)^2\varepsilon^6/7560\) for a fixed pressure gradient, revealing a sixth-power suppression of shear dispersion as the catheter approaches the arterial wall. Consistently, increasing \(\lambda\) from \(0.01\) to \(0.30\) reduces effective convection by about a factor of \(2.45\), while the chemically passive shear-induced dispersion falls by a factor of \(24\). The coupled bulk--surface formulation also yields the exact partition \(\Phi_m+\Phi_s+\Phi_a=1\), separating mobile, reversibly retained and irreversibly absorbed solute. The reconstructed field further quantifies reaction-modulated transverse non-uniformity, while the overall results distinguish hydrodynamic effects of confinement and microrotation from kinetic wall effects.
\end{abstract}


\section{Introduction}

Fluid mechanics sits at the heart of many physiological processes. Blood travels
through a branched network of vessels. Air travels through the airways. In both
systems, the fate of a dissolved substance depends on the flow that carries it.
This dependence draws researchers from applied mathematics, chemical engineering
and biomedical engineering towards a common set of questions. Arterial
pharmacokinetics offers a clear example. A drug injected into an artery must reach
a downstream site at an appropriate concentration. An insufficient dose may be
ineffective, whereas excessive exposure can produce local or systemic toxicity. Related
problems arise in chromatographic separation, in mechanical ventilation of the
lung and in pollutant transport through natural streams. The common
feature is the interplay of shear, molecular diffusion and chemical reaction at
the boundary. None of the three acts alone, and the outcome cannot be predicted
from any one of them in isolation.

\citet{taylor1953royal} first explained how a solute cloud spreads in laminar
tube flow. He showed that axial convection and radial molecular diffusion act
together, and that the cloud eventually travels with the mean speed of the fluid.
His result applies only after the solute has mixed across the cross-section.
\cite{aris1956royal} recovered the same dispersion coefficient through the method
of moments and removed the restriction on the initial distribution. He later extended the treatment to pulsatile flow \citep{aris1960royal}. Both analyses still
describe the large-time limit. \cite{gill1970royal} closed
this gap with the generalized dispersion model, in which the concentration is
expanded in a series of axial derivatives of the sectional mean. The model yields
time-dependent transport coefficients from the instant of release.
\cite{barton1983jfm} reformulated the moment method and obtained an all-time
solution. \cite{chatwin1970jfm} traced the slow approach of the axial
distribution towards a Gaussian form, and \cite{mukherjee1988acta} examined
the role of a time-dependent pressure gradient. Blood, however, does not behave
as a Newtonian fluid in small vessels. \cite{sharp1993abe} tested Taylor's
theory against the Casson, power-law and Bingham models. \cite{dash2000abe} and
\cite{nagarani2004abe} applied the generalized dispersion model to Casson fluid
flow, and \cite{rana2016jfm, rana2016pof} treated the Casson and Herschel--Bulkley models.

The wall of an artery is not inert. Part of the solute attaches to the endothelium
and to the tissue layer behind it. Some of that solute never returns to the blood;
the rest returns after a delay. The first process is an irreversible reaction, or
wall absorption. The second is a reversible reaction, also described as retention
or phase exchange. \cite{sankara1973royal}
first placed a first-order irreversible absorption inside the generalized
dispersion model, and they defined the exchange, convection and dispersion
coefficients adopted here. \cite{mazumder1992jfm} studied the same reaction in
pulsatile tube flow. \cite{phillips1995jfm} described time-dependent transport
with exchange between two phases. Boundary retention was incorporated into classical
dispersion theory by \cite{purnama1988jfm}, while
\cite{zhang2017jfm} resolved the transient transition produced by kinetic
sorption in Poiseuille flow. \cite{wang2022prf} subsequently examined
pre-asymptotic dispersion with wall adsorption and emphasized that reactive-wall
effects can remain important before the long-time regime is reached.
\cite{ng2006royal} treated both kinetics together
in steady and oscillatory tube flow. \cite{ng2008pof} solved
the problem for a wall that both retains and absorbs, and they followed the
transport coefficients through the entire transient stage. Their results show why
both reactions matter. The irreversible reaction depletes the mobile phase and
steepens the radial concentration gradient. The reversible reaction returns solute
to the flow behind the main cloud, produces a long trailing tail, and can even
reverse the sign of the convection coefficient at intermediate times \citep{ng2008pof}.
Later studies confirmed these features in other flows and other rheologies
\citep{mazumder2012hmt, debnath2019jem, roy2020ichmt, jiang2022jfm, das2022royal}. A model that
retains only one of the two kinetics therefore, cannot represent their combined
transient effects on arterial solute transport.

A catheter is a routine instrument in clinical practice. Physicians use it for
pressure measurement, for angiography, for balloon angioplasty and for the
delivery of a drug to a specific site. Its insertion changes the geometry
of the vessel. The circular lumen becomes an annulus, bounded by the catheter
surface on the inside and by the arterial wall on the outside. The change is not
cosmetic. The annular gap raises the frictional resistance, alters the wall shear
stress and modifies the whole velocity profile \citep{mazumder2005qjmam}. \cite{jayaraman1995mbec} reported a rise in axial wall shear stress in a curved catheterized
artery. \cite{dash1996jbm} analysed steady and unsteady Casson flow in the same
geometry. \cite{sarkar2004acta} examined wall absorption
in oscillatory annular flow, and \cite{mazumder2005qjmam}
obtained the moments and the mean axial concentration for a pulsatile annular flow
with a reactive outer wall. \cite{nagarani2006acta} carried out the corresponding
exact analysis for Casson fluid in an annulus.
\cite{poddar2021karj} subsequently used layer-adapted meshes to resolve steady
solute transport through a catheterized annulus with outer-wall absorption. A consistent message emerges from
this body of work. The aspect ratio $\lambda = b/a$, the ratio of the catheter
radius to the arterial radius, controls the shear available to the solute, and a
larger catheter suppresses the spread of the tracer. A reliable estimate of this
catheter-induced change is essential for the correct interpretation of a pressure
reading and for a dependable dosage at a downstream site.

Recent computational work has also broadened the available tools for wall-reactive
transport: physics-informed neural networks have been used to reconstruct two-dimensional
reactive solute fields and wall-resolved uptake in canonical shear flows
\citep{poddar2026pinn}. The closest strands of the literature nevertheless remain separated. Annular studies with simultaneous reversible and irreversible wall reactions have primarily used Newtonian or Casson descriptions of the carrier fluid \citep{debnath2019jem, roy2020ichmt}. Micropolar reactive transport has been examined in tubes with absorptive walls \citep{shah2020ichmt}, and a micropolar concentric-annulus model has also been used for catheter-related dispersion with a single absorptive wall and a long-time treatment \citep{ramarao2023cma}. More recent annular blood-flow models with other rheologies likewise retain irreversible uptake as the wall chemistry \citep{ajith2026ejmbf}. Thus, the unresolved issue is not annular dispersion, micropolarity, or wall reaction separately, but their full transient coupling when a micropolar annular flow is bounded by an inert catheter and an arterial wall that both retain and irreversibly absorb solute.

The present study addresses this specific gap. Blood is represented phenomenologically as a micropolar fluid in the sense of \cite{eringen1966jmm}, consistent with established micropolar models of arterial blood flow \citep{mekheimer2008acta,jaiswal2019pof}. Microstructural rotational effects then enter through the coupling number \(N_c\) and the micropolar parameter \(m_p\). A steady, fully developed, axisymmetric micropolar flow through the annulus supplies the velocity field, while the outer wall combines irreversible uptake, governed by \(\beta\), with reversible surface retention, governed by \(\theta\) and the Damk\"ohler number \(Da\). The generalized dispersion model \citep{gill1970royal, sankara1973royal} is used to obtain the full transient coefficients \(K_0\), \(K_1\), and \(K_2\), together with the sectional and reconstructed local concentrations. The coupled formulation also gives an exact partition of the injected mass into mobile, reversibly retained, and irreversibly absorbed fractions. The reconstructed field is further used to quantify the transient transverse non-uniformity of the mobile cloud. In addition to recovering established tube limits, the analysis identifies short-time laws for exchange and dispersion and a narrow-gap asymptotic law that quantifies how catheter confinement suppresses shear dispersion under a fixed pressure gradient.

The paper is organised as follows. Section~\ref{sec:formulation} formulates the micropolar flow, the dual wall-reaction transport problem, and the generalized-dispersion hierarchy. Section~\ref{sec:numerics} gives the numerical discretization. Section~\ref{sec:results} presents validation, asymptotic analysis, and the parametric results, and Section~\ref{sec:conclusion} summarizes the main conclusions.

\section{Mathematical formulation}
\label{sec:formulation}

\begin{figure}
    \centering
    \includegraphics[width=\linewidth]{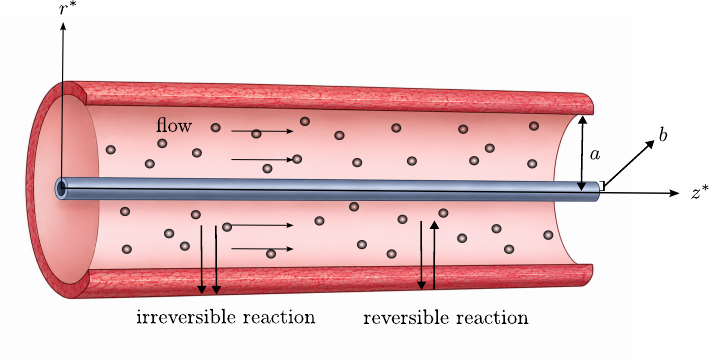}
    \caption{Schematic of the concentric catheter--artery configuration. The micropolar fluid occupies \(b\le r^*\le a\); the catheter surface \(r^*=b\) is impermeable to solute, while the arterial wall \(r^*=a\) supports irreversible uptake and reversible surface retention.}
    \label{fig:1}
\end{figure}

\subsection{Micropolar flow in the catheterized annulus}
\label{subsec:flowmodel}

We consider a long, straight, concentric annulus formed by an inner catheter of radius \(b\) and an outer arterial wall of radius \(a\), with \(0<b<a\), as shown in Figure~\ref{fig:1}. The fluid is assumed incompressible, steady, laminar, axisymmetric, and fully developed. The imposed axial pressure gradient is constant. The axial velocity is denoted by \(v^*(r^*)\), and \(w^*(r^*)\) denotes the azimuthal microrotation. The micropolar description is used here as a phenomenological representation of rotational microstructural effects in the suspension rather than as a literal rigid-particle model of individual erythrocytes.

Micropolar flow between concentric cylindrical surfaces has a classical analytical foundation \citep{ariman1967pof}. Under the present steady pressure-driven assumptions, the reduced linear- and angular-momentum equations of Eringen's micropolar theory \citep{eringen1966jmm} are
\begin{equation}\label{eq:dimensional_linear_momentum}
-\frac{\partial p}{\partial z^*}
+(\mu+\kappa)\frac{1}{r^*}\frac{d}{dr^*}
\left(r^*\frac{dv^*}{dr^*}\right)
+\kappa\frac{1}{r^*}\frac{d}{dr^*}(r^*w^*)=0,
\end{equation}
and
\begin{equation}\label{eq:dimensional_angular_momentum}
\gamma\frac{d}{dr^*}\left[
\frac{1}{r^*}\frac{d}{dr^*}(r^*w^*)\right]
-\kappa\left(\frac{dv^*}{dr^*}+2w^*\right)=0.
\end{equation}
Here \(\mu\) is the Newtonian shear viscosity, \(\kappa\) is the vortex viscosity, and \(\gamma\) is the spin-gradient viscosity. No slip and no spin are imposed at both solid surfaces,
\begin{subequations}\label{eq:dimensional_boundary_conditions}
\begin{align}
v^*(b)=w^*(b)&=0,\\
v^*(a)=w^*(a)&=0.
\end{align}
\end{subequations}
The no-spin condition is adopted here as a strong-anchoring condition in which the microrotation is arrested at the catheter and arterial surfaces.

We introduce
\begin{equation}\label{eq:non_dimensional_parameters_velocity}
r=\frac{r^*}{a},\quad
v=\frac{v^*}{v_0},\quad
w=\frac{aw^*}{v_0},\quad
N_c=\frac{\kappa}{\kappa+\mu},\quad
m_p=a\left(\frac{2\mu+\kappa}{\mu+\kappa}\frac{\kappa}{\gamma}\right)^{1/2},
\end{equation}
where
\begin{equation}\label{eq:reference_velocity}
v_0=-\frac{a^2}{2(2\mu+\kappa)}\frac{\partial p}{\partial z^*}
\end{equation}
is a pressure-gradient velocity scale. It coincides with the centreline velocity only in the Newtonian full-pipe limit; in an annulus, it should not be interpreted as the maximum velocity. With \(\lambda=b/a\), the dimensionless equations are
\begin{equation}\label{eq:dimensionless_linear_momentum}
2(2-N_c)+\frac{1}{r}\frac{d}{dr}\left(r\frac{dv}{dr}\right)
+N_c\frac{1}{r}\frac{d}{dr}(rw)=0,
\end{equation}
\begin{equation}\label{eq:dimensionless_angular_momentum}
\frac{d}{dr}\left[\frac{1}{r}\frac{d}{dr}(rw)\right]
-\frac{m_p^2}{2-N_c}\left(\frac{dv}{dr}+2w\right)=0,
\end{equation}
with
\begin{equation}\label{eq:dimensionless_boundary_conditions}
v(\lambda)=w(\lambda)=v(1)=w(1)=0.
\end{equation}

Solving Eqs.~\eqref{eq:dimensionless_linear_momentum}--\eqref{eq:dimensionless_boundary_conditions} gives
\begin{equation}\label{eq:microrotational_velocity}
w(r)=r+\frac{A_3}{r}-\frac{A_2}{m_p}K_1(m_pr)+\frac{A_1}{m_p}I_1(m_pr),
\end{equation}
and
\begin{equation}\label{eq:axial_velocity}
v(r)=A_4+\frac{2A_1}{m_p^2}-r^2-2A_3\log r
-\frac{N_c}{m_p^2}\left[A_1I_0(m_pr)+A_2K_0(m_pr)\right],
\end{equation}
where \(I_j\) and \(K_j\) are modified Bessel functions.

The coefficients \( A_1 \), \( A_2 \), \( A_3 \), and \( A_4 \) are given by:

\begin{subequations}\label{eq:coefficients}
\begin{equation}
    A_1 = \frac{a_{12} + a_{22} - a_{32} - a_{42} \lambda - a_{12} \lambda^2 - a_{22} \lambda^2 + a_{32} \lambda^2 + a_{42} \lambda^3 - 2 a_{42} \lambda \log(\lambda) + 2 a_{22} \lambda^2 \log(\lambda)}{F_1}
\end{equation}

\begin{equation}
    A_2 = -\frac{a_{11} + a_{21} - a_{31} - a_{41} \lambda - a_{11} \lambda^2 - a_{21} \lambda^2 + a_{31} \lambda^2 + a_{41} \lambda^3 - 2 a_{41} \lambda \log(\lambda) + 2 a_{21} \lambda^2 \log(\lambda)}{F_1}
\end{equation}
\begin{equation}
\begin{aligned}
A_3 = (a_{11} a_{42} \lambda - a_{12} a_{41} \lambda + a_{21} a_{42} \lambda - a_{22} a_{41} \lambda - a_{31} a_{42} \lambda + a_{32} a_{41} \lambda - a_{11} a_{22} \lambda^2 \\+ a_{12} a_{21} \lambda^2 - a_{21} a_{32} \lambda^2 + a_{22} a_{31} \lambda^2- a_{21} a_{42} \lambda^3 + a_{22} a_{41} \lambda^3)/F_1
\end{aligned}
\end{equation}

\begin{equation}
\begin{aligned}
A_4 = (a_{11} a_{32} - a_{12} a_{31} + a_{21} a_{32} - a_{22} a_{31} + a_{31} a_{42} \lambda - a_{32} a_{41} \lambda + a_{11} a_{22} \lambda^2 - a_{12} a_{21} \lambda^2 - a_{11} a_{32} \lambda^2\\ + a_{12} a_{31} \lambda^2 - a_{11} a_{42} \lambda^3 + a_{12} a_{41} \lambda^3 + 2 a_{11} a_{42} \lambda \log(\lambda) - 2 a_{12} a_{41} \lambda \log(\lambda) + 2 a_{21} a_{42} \lambda \log(\lambda) \\- 2 a_{22} a_{41} \lambda \log(\lambda) - 2 a_{11} a_{22} \lambda^2 \log(\lambda) + 2 a_{12} a_{21} \lambda^2 \log(\lambda))/F_1
\end{aligned}
\end{equation}

\begin{equation}
\begin{aligned}
    F_1 = a_{11} a_{22} - a_{12} a_{21} + a_{21} a_{32} - a_{22} a_{31} - a_{11} a_{42} \lambda + a_{12} a_{41} \lambda \\
    + a_{31} a_{42} \lambda - a_{32} a_{41} \lambda + 2 a_{21} a_{42} \lambda \log(\lambda) - 2 a_{22} a_{41} \lambda \log(\lambda)
\end{aligned} 
\end{equation}

\begin{equation}
\begin{aligned}
    a_{11} = \frac{2}{m_p^2} -  \frac{N_c}{m_p^2} I_0(m_p),
    a_{12} = - \frac{N_c}{m_p^2} K_0(m_p), 
    a_{21} = \frac{I_1(m_p)}{m_p}, 
    a_{22} = -\frac{K_1(m_p)}{m_p}, \\
    a_{31} = \frac{2}{m_p^2} -  \frac{N_c}{m_p^2}  I_0(m_p \lambda), 
    a_{32} = - \frac{N_c}{m_p^2}  K_0(m_p \lambda), 
    a_{41} = \frac{I_1(m_p  \lambda)}{m_p}, 
    a_{42} = -\frac{K_1(m_p  \lambda)}{m_p}.
\end{aligned}
\end{equation}
\end{subequations}

The parameters \(N_c\) and \(m_p\) measure, respectively, the coupling between linear and angular motion and the ratio of the arterial length scale to the characteristic micropolar length scale \citep{jaiswal2019pof}. Throughout the parameter studies, the pressure gradient, and hence the reference scale \(v_0\), is held fixed; this point is important when interpreting the effect of changing \(\lambda\).

\subsection{Reactive solute transport}
\label{subsec:transportmodel}

Let \(C^*(t^*,z^*,r^*)\) denote the mobile-phase solute concentration in the annular fluid and \(C_s^*(t^*,z^*)\) the amount retained per unit area of the outer wall. For a dilute solute of molecular diffusivity \(D\),
\begin{equation}\label{eq:dimensional_convection_diffusion}
\frac{\partial C^*}{\partial t^*}
+v^*(r^*)\frac{\partial C^*}{\partial z^*}
=D\left[
\frac{1}{r^*}\frac{\partial}{\partial r^*}
\left(r^*\frac{\partial C^*}{\partial r^*}\right)
+\frac{\partial^2 C^*}{\partial {z^*}^2}
\right].
\end{equation}

A cross-sectionally uniform point injection of total mobile mass \(M\) is prescribed at \(t^*=0\):
\begin{subequations}\label{eq:dimensional_initial_conditions}
\begin{align}
C^*(0,z^*,r^*)&=C_0\,a\delta(z^*),\\
C_s^*(0,z^*)&=0,
\end{align}
\end{subequations}
where
\begin{equation}
C_0=\frac{M}{\pi a^3(1-\lambda^2)}.
\label{eq:C0}
\end{equation}
The factor \(1-\lambda^2\) ensures that integrating the initial concentration over the annular cross-section and over \(z^*\) gives exactly \(M\).

The catheter is taken to be inert and impermeable to solute,
\begin{equation}\label{eq:inner_solute_bc_dim}
\frac{\partial C^*}{\partial r^*}(t^*,z^*,b)=0.
\end{equation}
At the arterial wall, irreversible uptake and reversible retention act in parallel, following the retentive--absorptive wall framework used in reactive shear-dispersion studies \citep{ng2006royal,ng2008pof}. Conservation of solute at \(r^*=a\) gives
\begin{equation}\label{eq:outer_flux_dim}
-D\frac{\partial C^*}{\partial r^*}
=\beta^* C^*+\frac{\partial C_s^*}{\partial t^*},
\qquad r^*=a,
\end{equation}
while the reversible surface kinetics are
\begin{equation}\label{eq:surface_kinetics_dim}
\frac{\partial C_s^*}{\partial t^*}
=k_f C^*-k_b C_s^*
=k_b\left(\theta^* C^*-C_s^*\right),
\qquad
\theta^*=\frac{k_f}{k_b}.
\end{equation}
Here \(\beta^*\) and \(k_f\) have dimensions of velocity, \(k_b\) has dimensions of inverse time, and \(\theta^*\) has dimensions of length. Thus \(C_s^*\) has dimensions of bulk concentration multiplied by length, as required by the flux balance. The axial far-field conditions are
\begin{equation}\label{eq:farfield_dim}
C^*,\ \frac{\partial C^*}{\partial z^*}\longrightarrow0
\qquad\text{as}\qquad |z^*|\to\infty .
\end{equation}

We use
\begin{equation}\label{eq:non_dimensional_parameters_dispersion}
C=\frac{C^*}{C_0},\quad
C_s=\frac{C_s^*}{aC_0},\quad
r=\frac{r^*}{a},\quad
z=\frac{Dz^*}{a^2v_0},\quad
t=\frac{Dt^*}{a^2},
\end{equation}
together with
\begin{equation}
Pe=\frac{av_0}{D},\quad
\beta=\frac{\beta^*a}{D},\quad
\theta=\frac{\theta^*}{a},\quad
Da=\frac{k_ba^2}{D}.
\label{eq:dimensionless_reaction_parameters}
\end{equation}
The transport problem becomes
\begin{equation}\label{eq:dimensionless_convection_diffusion}
\frac{\partial C}{\partial t}
+v(r)\frac{\partial C}{\partial z}
=
\frac{1}{r}\frac{\partial}{\partial r}
\left(r\frac{\partial C}{\partial r}\right)
+\frac{1}{Pe^2}\frac{\partial^2C}{\partial z^2},
\qquad \lambda<r<1,
\end{equation}
with
\begin{subequations}\label{eq:dimensionless_ic_bc_conditions}
\begin{align}
C(0,z,r)&=\frac{\delta(z)}{Pe}, \quad
C_s(0,z)=0,\\
\frac{\partial C}{\partial r}(t,z,\lambda)&=0,\\
-\frac{\partial C}{\partial r}-\beta C
&=\frac{\partial C_s}{\partial t}
=Da(\theta C-C_s),
\quad \mbox{at} \quad r=1,\\
C,\ \frac{\partial C}{\partial z}&\to0,
\quad \mbox{as} \quad |z|\to\infty.
\end{align}
\end{subequations}

\subsection{Generalized dispersion formulation}
\label{subsec:gdm}

Following \citet{gill1970royal} and \citet{sankara1973royal}, the mobile and retained concentrations are expanded in axial derivatives of the sectional mean:
\begin{subequations}\label{eq:series_expansions}
\begin{align}
C(t,z,r)
&=\sum_{n=0}^{\infty}f_n(t,r)\,
\frac{\partial^n C_m}{\partial z^n},\\
C_s(t,z)
&=\sum_{n=0}^{\infty}g_n(t)\,
\frac{\partial^n C_m}{\partial z^n}.
\end{align}
\end{subequations}
The annular sectional mean is
\begin{equation}\label{eq:mean_concentration}
C_m(t,z)=\frac{2}{1-\lambda^2}\int_{\lambda}^{1}C(t,z,r)\,r\,dr.
\end{equation}
For compactness, define
\begin{equation}
\langle q\rangle
=\frac{2}{1-\lambda^2}\int_{\lambda}^{1}q(r)\,r\,dr .
\label{eq:annular_average}
\end{equation}

Averaging Eq.~\eqref{eq:dimensionless_convection_diffusion} over the annulus and using the inner no-flux condition gives
\begin{equation}\label{eq:mean_balance}
\frac{\partial C_m}{\partial t}
=
\frac{1}{Pe^2}\frac{\partial^2C_m}{\partial z^2}
+\frac{2}{1-\lambda^2}\frac{\partial C}{\partial r}(t,z,1)
-\frac{\partial}{\partial z}\langle vC\rangle .
\end{equation}
Substitution of Eq.~\eqref{eq:series_expansions} yields
\begin{equation}\label{eq:mean_gdm}
\frac{\partial C_m}{\partial t}
=\sum_{n=0}^{\infty}K_n(t)\frac{\partial^nC_m}{\partial z^n},
\end{equation}
where
\begin{equation}\label{eq:Kgeneral}
K_n(t)=
\frac{\delta_{n2}}{Pe^2}
+\frac{2}{1-\lambda^2}\frac{\partial f_n}{\partial r}(t,1)
-\frac{2}{1-\lambda^2}\int_{\lambda}^{1}r\,v(r)f_{n-1}(t,r)\,dr,
\end{equation}
with \(f_{-1}\equiv0\). In particular,
\begin{subequations}\label{eq:transport_coefficients}
\begin{align}
K_0(t)
&=\frac{2}{1-\lambda^2}f_{0,r}(t,1),\\
K_1(t)
&=\frac{2}{1-\lambda^2}f_{1,r}(t,1)
-\frac{2}{1-\lambda^2}\int_{\lambda}^{1}rvf_0\,dr,\\
K_2(t)
&=\frac{1}{Pe^2}
+\frac{2}{1-\lambda^2}f_{2,r}(t,1)
-\frac{2}{1-\lambda^2}\int_{\lambda}^{1}rvf_1\,dr .
\end{align}
\end{subequations}
With the present sign convention, \(K_0<0\) represents net mobile-phase depletion and \(K_1<0\) corresponds to downstream advection; therefore the positive quantities \(-K_0\) and \(-K_1\) are plotted below. The shear-induced part of the dispersion is \(K_2-Pe^{-2}\).

Equating equal axial derivatives after substitution into the local transport equation gives, for \(n=0,1,2\),
\begin{equation}\label{eq:fn_hierarchy}
\frac{\partial f_n}{\partial t}
=
\frac{1}{r}\frac{\partial}{\partial r}
\left(r\frac{\partial f_n}{\partial r}\right)
-vf_{n-1}
-\sum_{j=0}^{n}K_jf_{n-j}
+\frac{f_{n-2}}{Pe^2},
\end{equation}
where \(f_{-1}=f_{-2}=0\). The retained-phase expansion is not an independent algebraic boundary value: differentiating the second series in Eq.~\eqref{eq:series_expansions} with respect to time and using Eq.~\eqref{eq:mean_gdm} gives
\begin{equation}\label{eq:gn_hierarchy}
\frac{dg_n}{dt}
+\sum_{j=0}^{n}K_jg_{n-j}
=
Da\left[\theta f_n(t,1)-g_n(t)\right],
\qquad n=0,1,2.
\end{equation}
Accordingly, the coefficient-wise wall conditions are
\begin{subequations}\label{eq:fn_boundary_conditions}
\begin{align}
f_{n,r}(t,\lambda)&=0,\\
-f_{n,r}(t,1)-\beta f_n(t,1)
&=
\frac{dg_n}{dt}
+\sum_{j=0}^{n}K_jg_{n-j}
=
Da\left[\theta f_n(t,1)-g_n(t)\right].
\end{align}
\end{subequations}
The initial and normalization conditions are
\begin{subequations}\label{eq:fn_initial_normalization}
\begin{align}
f_0(0,r)&=1, \quad
f_n(0,r)=0\quad(n=1,2),\\
g_n(0)&=0\quad(n=0,1,2),\\
\frac{2}{1-\lambda^2}\int_{\lambda}^{1}f_n(t,r)\,r\,dr
&=\delta_{n0}.
\end{align}
\end{subequations}
The last relation follows directly from the definition of \(C_m\) and provides an important numerical consistency check.

Retaining the hierarchy through \(K_2\), the sectional mean satisfies
\begin{equation}\label{eq:mean_second_order}
\frac{\partial C_m}{\partial t}
=K_0C_m+K_1\frac{\partial C_m}{\partial z}
+K_2\frac{\partial^2C_m}{\partial z^2}.
\end{equation}
For the point-source initial condition, its Green-function form is
\begin{equation}\label{eq:Cm_green}
C_m(t,z)
=
\frac{\mathcal M(t)}
{\sqrt{4\pi\mathcal D(t)}}
\exp\left[-\frac{(z-z_g(t))^2}{4\mathcal D(t)}\right],
\end{equation}
where
\begin{equation}\label{eq:mean_moments}
\mathcal M(t)=\frac{1}{Pe}\exp\left(\int_0^tK_0(\tau)\,d\tau\right),
\quad
z_g(t)=-\int_0^tK_1(\tau)\,d\tau,
\quad
\mathcal D(t)=\int_0^tK_2(\tau)\,d\tau.
\end{equation}
The local field shown later is reconstructed consistently to second order as
\begin{equation}\label{eq:local_reconstruction}
C(t,z,r)\approx
f_0C_m+f_1\frac{\partial C_m}{\partial z}
+f_2\frac{\partial^2C_m}{\partial z^2}.
\end{equation}

The dual wall kinetics also permit an exact partition of the initially
injected solute. Let \(\Phi_m\), \(\Phi_s\), and \(\Phi_a\) denote,
respectively, the fractions of the initial mass that remain mobile, are
temporarily retained at the arterial wall, and have been irreversibly
absorbed. From Eqs.~\eqref{eq:series_expansions} and
\eqref{eq:mean_moments},
\begin{subequations}\label{eq:mass_fractions}
\begin{align}
\Phi_m(t)
&=
Pe\int_{-\infty}^{\infty}C_m(t,z)\,dz
=
\exp\left(\int_0^tK_0(\tau)\,d\tau\right),\\
\Phi_s(t)
&=
\frac{2Pe}{1-\lambda^2}
\int_{-\infty}^{\infty}C_s(t,z)\,dz
=
\frac{2g_0(t)}{1-\lambda^2}\,\Phi_m(t),\\
\Phi_a(t)
&=
\frac{2\beta Pe}{1-\lambda^2}
\int_0^t\int_{-\infty}^{\infty}
C(\tau,z,1)\,dz\,d\tau \\
&=
\frac{2\beta}{1-\lambda^2}
\int_0^t f_0(\tau,1)\Phi_m(\tau)\,d\tau .
\end{align}
\end{subequations}
All axial-derivative terms integrate to zero under the far-field conditions,
so only \(g_0\) contributes to the retained mass. Integrating the mobile
balance and using the surface balance gives
\begin{equation}\label{eq:mass_conservation}
\Phi_m(t)+\Phi_s(t)+\Phi_a(t)=1.
\end{equation}
Thus \(\Phi_a\) is monotone non-decreasing, whereas \(\Phi_s\) may increase
during surface loading and decrease when desorption returns stored solute to
the mobile phase. Equation~\eqref{eq:mass_conservation} also provides an
independent global check on the coupled bulk--surface computation.

Within the present framework, the Gill--Sankarasubramanian reduction recasts
the original advection--diffusion problem into a coupled bulk--surface
hierarchy for \(f_n\), \(g_n\), and the transient transport coefficients
\(K_n\). This reduced hierarchy is then advanced numerically, enabling the
annular micropolar flow, reversible surface retention, irreversible wall
uptake, and their transient coupling, to be treated within a single
generalized-dispersion formulation.

\section{Numerical scheme}
\label{sec:numerics}

The coupled hierarchy \eqref{eq:fn_hierarchy}--\eqref{eq:fn_initial_normalization} is solved on a uniform radial grid using a second-order finite-difference discretization and implicit time stepping. Crank--Nicolson time centering \citep{crank1947cn} is used after a short damped start-up. This start-up is important here because the initially uniform coefficient \(f_0(0,r)=1\) is, for a reactive wall, not compatible with the wall-flux condition at \(t=0^+\); an initial diffusive boundary layer is therefore generated immediately. To suppress the high-frequency oscillations that Crank--Nicolson can produce for such nonsmooth start-up data, a Rannacher-type damping procedure \citep{rannacher1984} is used. In the computations reported here, four backward-Euler half-steps are first taken on a fine start-up step \(\Delta t_f=\Delta t/20\), Crank--Nicolson is then continued with \(\Delta t_f\) up to \(t=0.01\), and the calculation subsequently proceeds with the production step \(\Delta t\).

For \(N_r\) radial nodes,
\[
r_j=\lambda+j\Delta r,\qquad
j=0,\ldots,N_r-1,\qquad
\Delta r=\frac{1-\lambda}{N_r-1}.
\]
At an interior node,
\begin{equation}\label{eq:radial_operator}
(L_h f)_j
=
\frac{f_{j+1}-2f_j+f_{j-1}}{\Delta r^2}
+\frac{1}{r_j}\frac{f_{j+1}-f_{j-1}}{2\Delta r},
\end{equation}
which is the second-order central approximation to \(r^{-1}(rf_r)_r\). The wall conditions are also imposed with centred second-order formulas by eliminating ghost nodes. At the impermeable catheter,
\begin{equation}\label{eq:inner_discrete}
f_{n,-1}=f_{n,1},
\end{equation}
whereas at the arterial wall, we define
\begin{equation}\label{eq:outer_discrete}
q_n
\equiv f_{n,r}(1)
=-(\beta+\theta Da)f_n(1)+Da\,g_n,
\qquad
f_{n,N_r}=f_{n,N_r-2}+2\Delta r\,q_n .
\end{equation}
Consequently, the discrete radial operator at the two physical boundary nodes becomes
\[
(L_hf_n)_0=\frac{2(f_{n,1}-f_{n,0})}{\Delta r^2},
\qquad
(L_hf_n)_{N_r-1}
=
\frac{2(f_{n,N_r-2}-f_{n,N_r-1})}{\Delta r^2}
+\left(\frac{2}{\Delta r}+1\right)q_n .
\]
This ghost-node elimination preserves a strictly tridiagonal system at every time level.

For compactness, write
\begin{equation}
\mathcal R_n
=
L_hf_n-vf_{n-1}
-\sum_{q=0}^{n}K_qf_{n-q}
+\frac{f_{n-2}}{Pe^2},
\label{eq:source_n}
\end{equation}
with \(f_{-1}=f_{-2}=0\), and let \(h_m=t^{m+1}-t^m\). The bulk update is written in the common weighted form
\begin{equation}\label{eq:CN}
\frac{f_n^{m+1}-f_n^{m}}{h_m}
=
(1-\omega_m)\mathcal R_n^{m}
+\omega_m\mathcal R_n^{m+1},
\end{equation}
where \(\omega_m=1\) during the backward-Euler damping steps and \(\omega_m=\tfrac12\) for the subsequent Crank--Nicolson steps. The retained-phase coefficients are advanced with exactly the same time weighting,
\begin{align}
\frac{g_n^{m+1}-g_n^m}{h_m}
&+(1-\omega_m)\sum_{q=0}^{n}K_q^{m}g_{n-q}^{m}
+\omega_m\sum_{q=0}^{n}K_q^{m+1}g_{n-q}^{m+1}
\nonumber\\
&=
Da\Big[
\theta\big((1-\omega_m)f_n^{m}(1)+\omega_m f_n^{m+1}(1)\big)
-\big((1-\omega_m)g_n^{m}+\omega_m g_n^{m+1}\big)
\Big].
\label{eq:surface_CN}
\end{align}
Thus the complete coupling term \(\sum_{q=0}^{n}K_qg_{n-q}\) in Eq.~\eqref{eq:gn_hierarchy} is retained at both time levels. Because the generalized-dispersion hierarchy is triangular, the equations are advanced sequentially for \(n=0,1,2\). For a fixed Picard iterate, elimination of the ghost nodes and of the new-time surface coefficient leaves a tridiagonal linear system for each \(f_n^{m+1}\), which is solved by the Thomas algorithm. The corresponding \(g_n^{m+1}\) and \(K_n^{m+1}\) are then updated from Eqs.~\eqref{eq:surface_CN} and \eqref{eq:Kgeneral}, and the process is repeated until
\[
\max_{n=0,1,2}
\frac{|K_n^{(k+1)}-K_n^{(k)}|}
{\max(|K_n^{(k+1)}|,10^{-12})}
<10^{-10},
\]
with at most 200 Picard iterations per time step.

All annular radial integrals, including those in Eqs.~\eqref{eq:Kgeneral} and \eqref{eq:fn_initial_normalization}, are evaluated by composite Simpson quadrature; \(N_r\) is therefore chosen odd. The initial generalized-dispersion state is imposed as
\[
f_0(0,r)=1,\qquad f_1(0,r)=f_2(0,r)=0,\qquad
g_0(0)=g_1(0)=g_2(0)=0,
\]
and the initial transport coefficients are evaluated directly from the exact wall and averaging relations,
\[
K_0(0^+)=-\frac{2(\beta+\theta Da)}{1-\lambda^2},
\qquad
K_1(0)=-\bar v,
\qquad
K_2(0)=Pe^{-2}.
\]

After each bulk solve, the exact normalization implied by the definition of \(C_m\) is imposed by projection. If \(\widetilde f_n\) denotes the unconstrained discrete solution at the new time level, then
\begin{equation}\label{eq:moment_projection}
f_0=
\frac{\widetilde f_0}{\langle\widetilde f_0\rangle},
\qquad
f_n=\widetilde f_n-\langle\widetilde f_n\rangle,
\quad n=1,2.
\end{equation}
This projection does not introduce an additional physical assumption; it enforces the exact identities \(\langle f_0\rangle=1\) and \(\langle f_n\rangle=0\) that follow from Eq.~\eqref{eq:mean_concentration}. Its use is numerically important in strongly depleted cases. Once the lower-order normalization constraints are satisfied, a discrete mean error \(e_n=\langle f_n\rangle-\delta_{n0}\) contains an amplification factor proportional to
\(\exp[-\int_0^tK_0(\tau)\,d\tau]=\Phi_m^{-1}\). Hence, an otherwise small quadrature or time-stepping error can contaminate \(K_1\) and \(K_2\) when the mobile fraction becomes small. The projection removes only this spurious mean mode, while the resolved radial structure continues to be determined by the finite-difference equations. Accordingly, the normalization residual is zero to numerical roundoff by construction and is not used as an independent convergence measure.

The time integrals entering \(\Phi_m\), \(\Phi_s\), \(\Phi_a\), \(z_g\), and \(\mathcal D\) are accumulated with the same temporal weights used by the evolution scheme: backward-Euler weighting during the damped start-up and trapezoidal weighting during Crank--Nicolson evolution. The independent global conservation diagnostic is therefore
\begin{equation}\label{eq:convergence_errors}
E_{\mathrm{mass}}(t)
=
\left|\Phi_m(t)+\Phi_s(t)+\Phi_a(t)-1\right|.
\end{equation}

\begin{table}
  \centering
  \caption{Grid- and time-step refinement at $t=1$ for the representative case
  $N_c=0.5$, $m_p=1$, $\lambda=0.01$, $\beta=1$, $\theta=0.5$, $Da=1$, and $Pe=1000$.
  The last column reports the maximum global mass-conservation residual over
  $0\leqslant t\leqslant 1$.}
  \label{tab:convergence}
  \begin{tabular}{rccccc}
    \toprule
    $N_r$ & $\Delta t$ & $-K_0$ & $-K_1$ & $K_2-Pe^{-2}$ & $E_{\mathrm{mass}}$ \\
    \midrule
     51 & $2\times10^{-4}$   & 1.36050 & 0.204261 & $1.39105\times10^{-2}$ & $1.130\times10^{-8}$  \\
    101 & $1\times10^{-4}$   & 1.36053 & 0.204246 & $1.39112\times10^{-2}$ & $2.825\times10^{-9}$  \\
    201 & $5\times10^{-5}$   & 1.36053 & 0.204243 & $1.39114\times10^{-2}$ & $7.060\times10^{-10}$ \\
    401 & $2.5\times10^{-5}$ & 1.36054 & 0.204242 & $1.39114\times10^{-2}$ & $1.762\times10^{-10}$ \\
    \bottomrule
  \end{tabular}
\end{table}

A grid- and time-step refinement study is carried out for the representative case
\(N_c=0.5\), \(m_p=1\), \(\lambda=0.01\), \(\beta=1\), \(\theta=0.5\),
\(Da=1\), and \(Pe=1000\). Here \(\Delta t\) denotes the production
Crank--Nicolson step; the start-up steps are refined proportionally as described
above. The transport coefficients and the independent global mass residual at
\(t=1\) are reported in Table~\ref{tab:convergence}.

The transport coefficients are essentially grid--time independent over the
reported sequence. From the coarsest to the finest calculation, the relative
changes in \(-K_0\), \(-K_1\), and \(K_2-Pe^{-2}\) are
\(2.94\times10^{-5}\), \(9.30\times10^{-5}\), and
\(6.47\times10^{-5}\), respectively. Between the two finest calculations,
the changes in \(-K_0\) and \(-K_1\) fall to \(7.35\times10^{-6}\) and
\(4.90\times10^{-6}\), while \(K_2-Pe^{-2}\) is unchanged to the digits shown.
The maximum global mass residual decreases by approximately a factor of four
at each refinement and reaches \(1.762\times10^{-10}\) on the finest grid.
Together, the coefficient stability and the independently decreasing mass
residual establish adequate numerical resolution for the parameter studies
reported below.

\section{Results and discussion}
\label{sec:results}
\begin{table}
\centering
\caption{Dimensionless parameter values and ranges used in this study.}
\label{tab:parameters}
\begin{tabular}{llll}
\toprule
Parameter & Reference range & Present study & Source \\
\midrule
$N_c$     & $[0,1)$             & $0.1,\,0.5,\,0.9$       & \citet{eringen1966jmm} \\
$m_p$     & $(0,\infty)$        & sweep; $1,\,3,\,5$ in Fig.~\ref{fig:4} & \citet{mekheimer2008acta} \\
$\lambda$ & $0.01$--$0.5$       & $0.01$--$0.5$           & \citet{mazumder2005qjmam} \\
$\beta$   & $0.01$--$100$       & $0.01,\,1,\,10$        & \citet{sankara1973royal} \\
$Da$      & $0.1$--$10$         & $0.1,\,0.5,\,1,\,1.5$  & \citet{das2022royal} \\
$\theta$  & $0$--$1.5$          & $0.1,\,0.5,\,1$         & \citet{ng2008pof} \\
$Pe$      & $10^{2}$--$10^{4}$  & $1000$                  & \citet{mazumder1992jfm} \\
\bottomrule
\end{tabular}
\end{table}

Table~\ref{tab:parameters} summarizes the principal parameter values and ranges used in the main reactive calculations and places them alongside representative ranges from the cited literature. In each parameter sweep, the quantity named in the corresponding figure is varied while the remaining parameters are held fixed at the values stated in the caption; this isolates geometric, rheological, and kinetic effects without changing the pressure-gradient scaling.

\subsection{Validation}
\label{subsec:validation}
\begin{figure}
\centering
\begin{subfigure}{0.48\linewidth}
\centering
    \includegraphics[width=\linewidth]{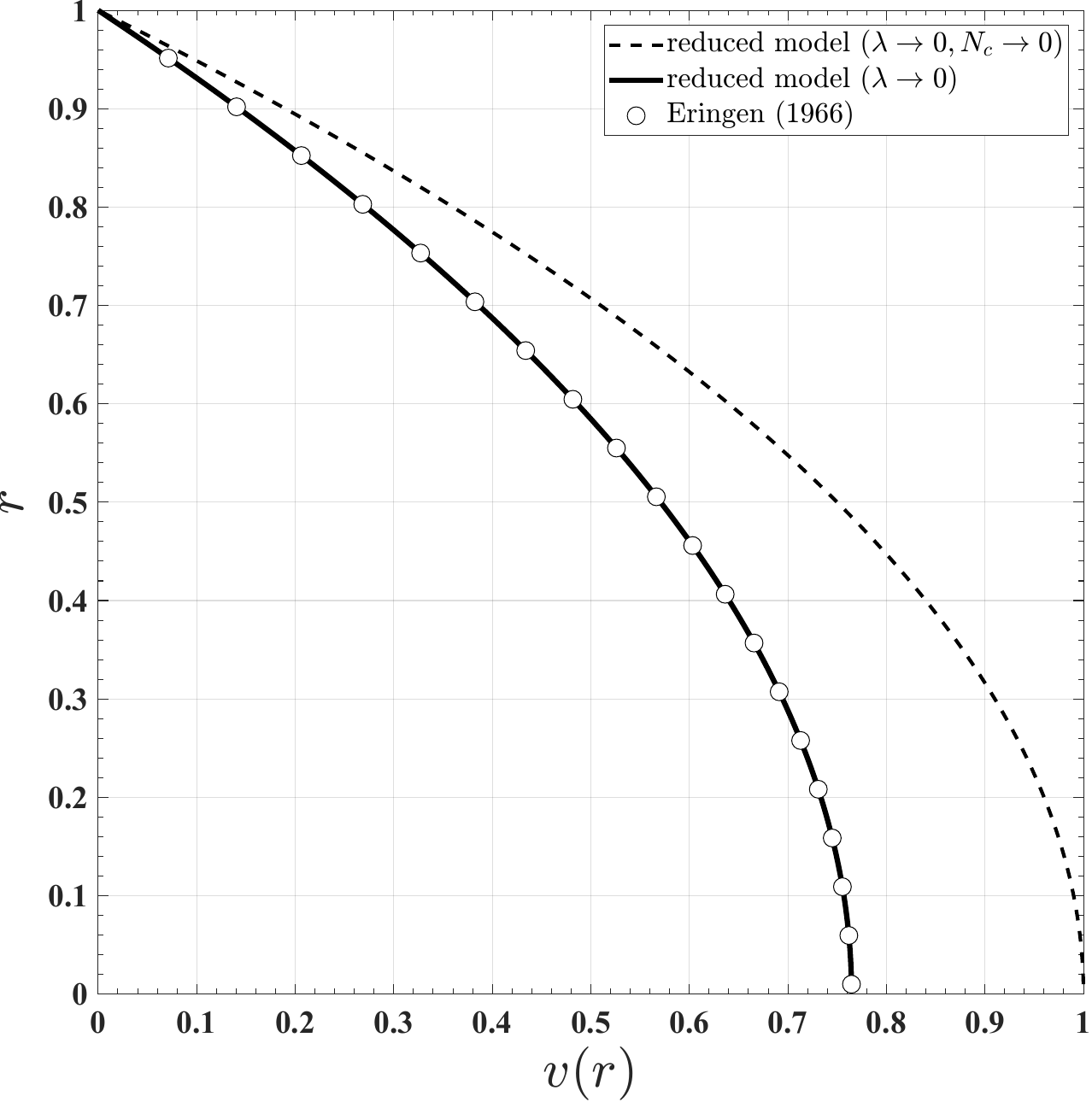}
    \caption{}
    \label{fig:2a}
\end{subfigure}
\begin{subfigure}{0.48\linewidth}
\centering
    \includegraphics[width=\linewidth]{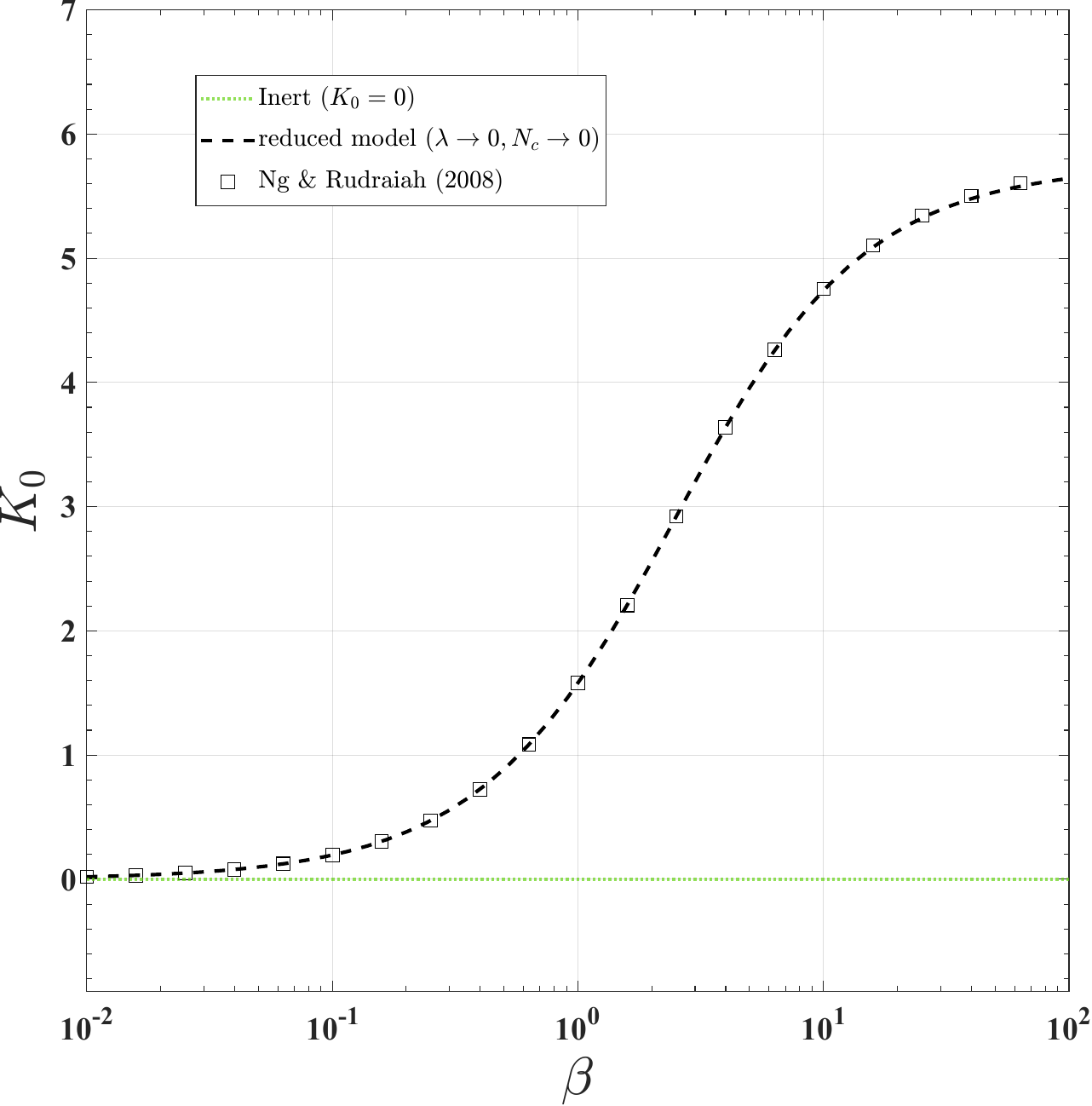}
    \caption{}
    \label{fig:2b}
\end{subfigure}
\begin{subfigure}{0.48\linewidth}
\centering
    \includegraphics[width=\linewidth]{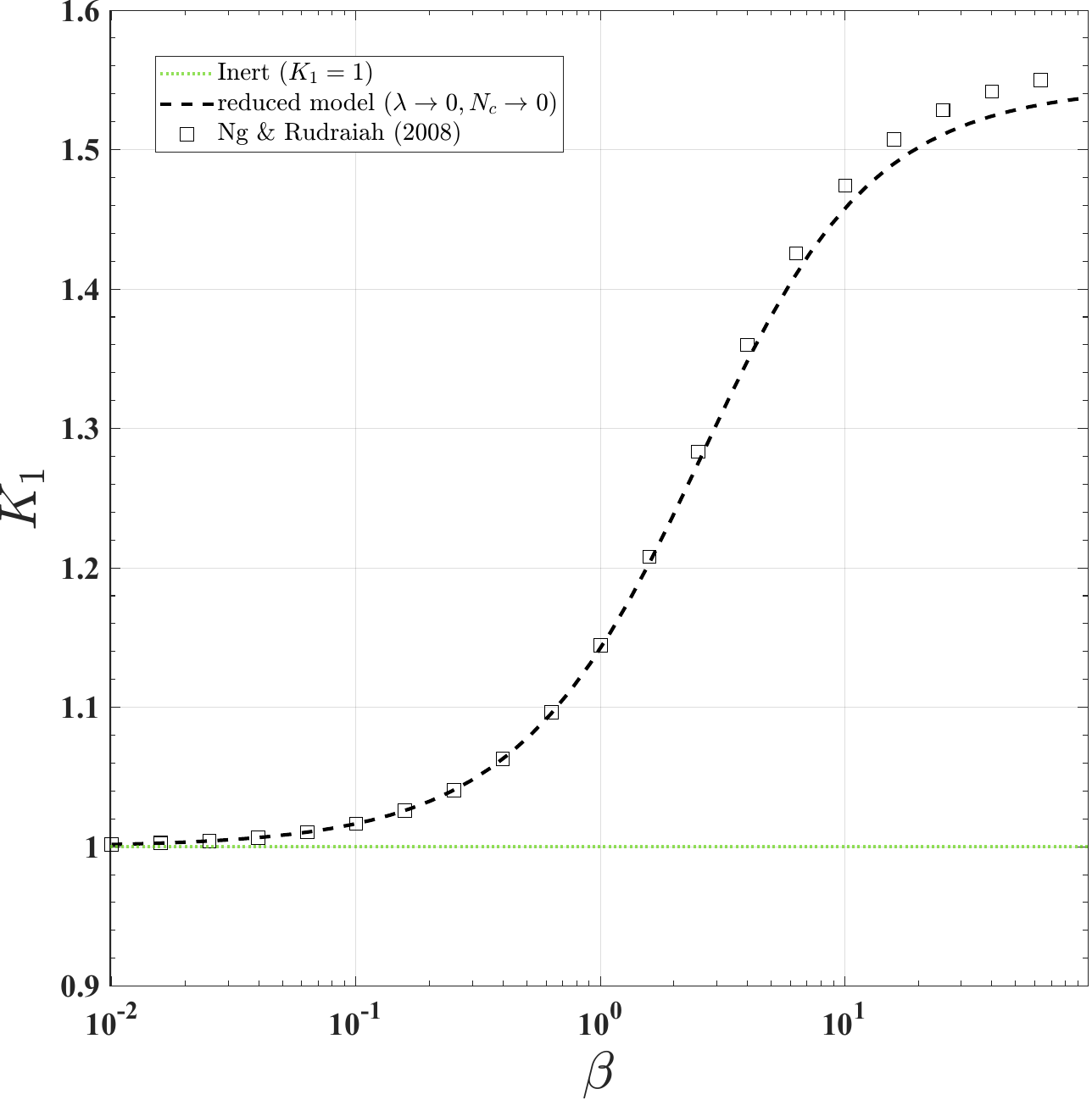}
    \caption{}
    \label{fig:2c}
\end{subfigure}
\begin{subfigure}{0.48\linewidth}
\centering
    \includegraphics[width=\linewidth]{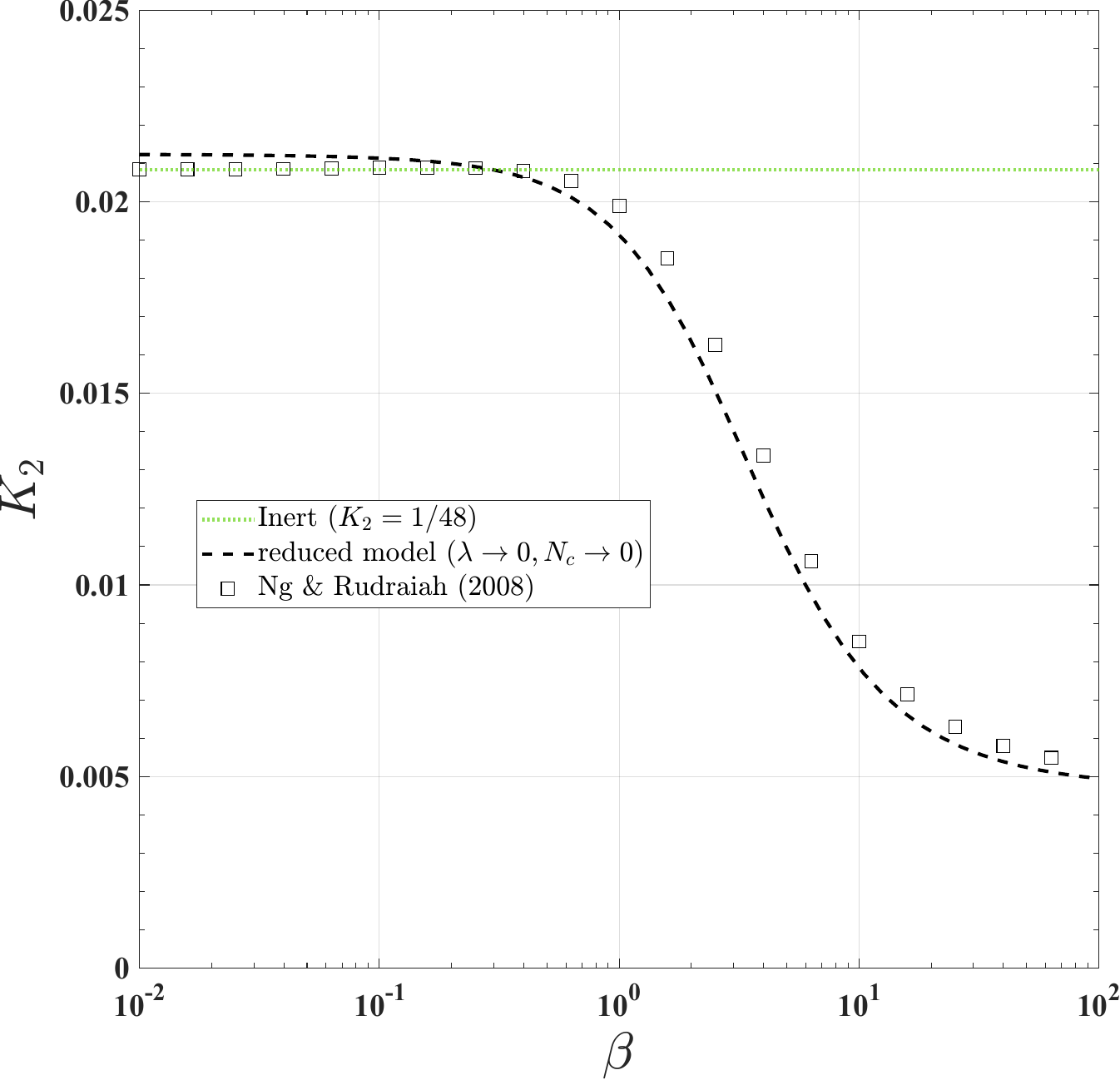}
    \caption{}
    \label{fig:2d}
\end{subfigure}
\caption{For the reduced model (Poiseuille flow, irreversible reaction only, $\lambda \to 0, N_c \to 0)$; (a) the velocity profile, (b) the exchange coefficient $(K_0)$, (c) the convection coefficient $(K_1)$, (d) the dispersion coefficient $(K_2)$.}
    \label{fig:2}
\end{figure}

To establish the credibility of the present formulation, we examine two limiting cases in which the governing equations collapse to configurations that have been independently studied in the literature. The axial velocity given by equation \eqref{eq:axial_velocity} is governed by two geometric and rheological parameters: the aspect ratio $\lambda = b/a$, which measures the relative size of the inner catheter, and the coupling number $N_c = \kappa/(\kappa+\mu)$, which quantifies the strength of the microrotational coupling. First, letting \(\lambda\to0\) removes the catheter and reduces the annular velocity field to micropolar flow in an unobstructed circular tube. The analytical solution in Eq.~\eqref{eq:axial_velocity} agrees with the corresponding pipe-flow result of \citet{eringen1966jmm}, as shown in Figure~\ref{fig:2a}. This comparison tests the Bessel-function representation of the velocity and microrotation fields together with the no-slip/no-spin boundary closure. A further reduction \(N_c\to0\) removes the microrotational coupling and recovers the Newtonian Hagen--Poiseuille profile \(v(r)=1-r^2\). Thus, Figure~\ref{fig:2a} checks the two successive limits in which the present annular micropolar model must collapse to established solutions.

A second and independent validation concerns the reactive transport hierarchy. Taking \(\lambda\to0\), \(N_c\to0\), and suppressing reversible retention reduces the problem to steady Newtonian Poiseuille flow in a circular tube with irreversible wall absorption, the configuration analysed by \citet{ng2008pof}. In this limit the present pressure-gradient scale \(v_0\) is the Newtonian centreline speed and is twice the cross-sectional mean used in their normalization. After accounting for this velocity scaling and the sign convention for the convection coefficient,
\[
K_0^{\rm here}=-K_0^{\rm Ng},\qquad
K_1^{\rm here}=-\frac{1}{2}K_1^{\rm Ng},\qquad
K_2^{\rm here}=\frac{1}{4}K_2^{\rm Ng},
\]
when axial molecular diffusion is omitted for the direct comparison. Figures~\ref{fig:2b}--\ref{fig:2d} are therefore displayed in the benchmark convention. The reduced calculations agree closely with the published data over \(10^{-2}\le\beta\le10^2\), and the inert-wall limits approach \(K_0=0\), \(K_1=1\), and \(K_2=1/48\).

The three benchmark coefficients also provide distinct physical checks on the implementation. Increasing irreversible uptake increases the exchange magnitude because the mobile solute is removed more rapidly at the wall. It also increases the effective advective transport of the surviving solute, since preferential removal of slow near-wall material shifts the mobile concentration toward the faster interior. At the same time, the benchmark dispersivity decreases because the surviving mobile phase samples a reduced range of axial velocities. Agreement in all three quantities therefore tests, separately, the wall-flux coupling, the concentration-weighted advection, and the shear-dispersion term in the generalized-dispersion hierarchy. Together with the independent grid--time refinement in Table~\ref{tab:convergence}, these limiting comparisons establish the numerical and analytical consistency of the formulation used below.

\subsection{Effect of the flow parameters}
\begin{figure}
    \centering
\begin{subfigure}{0.49\linewidth}
    \includegraphics[width=\linewidth]{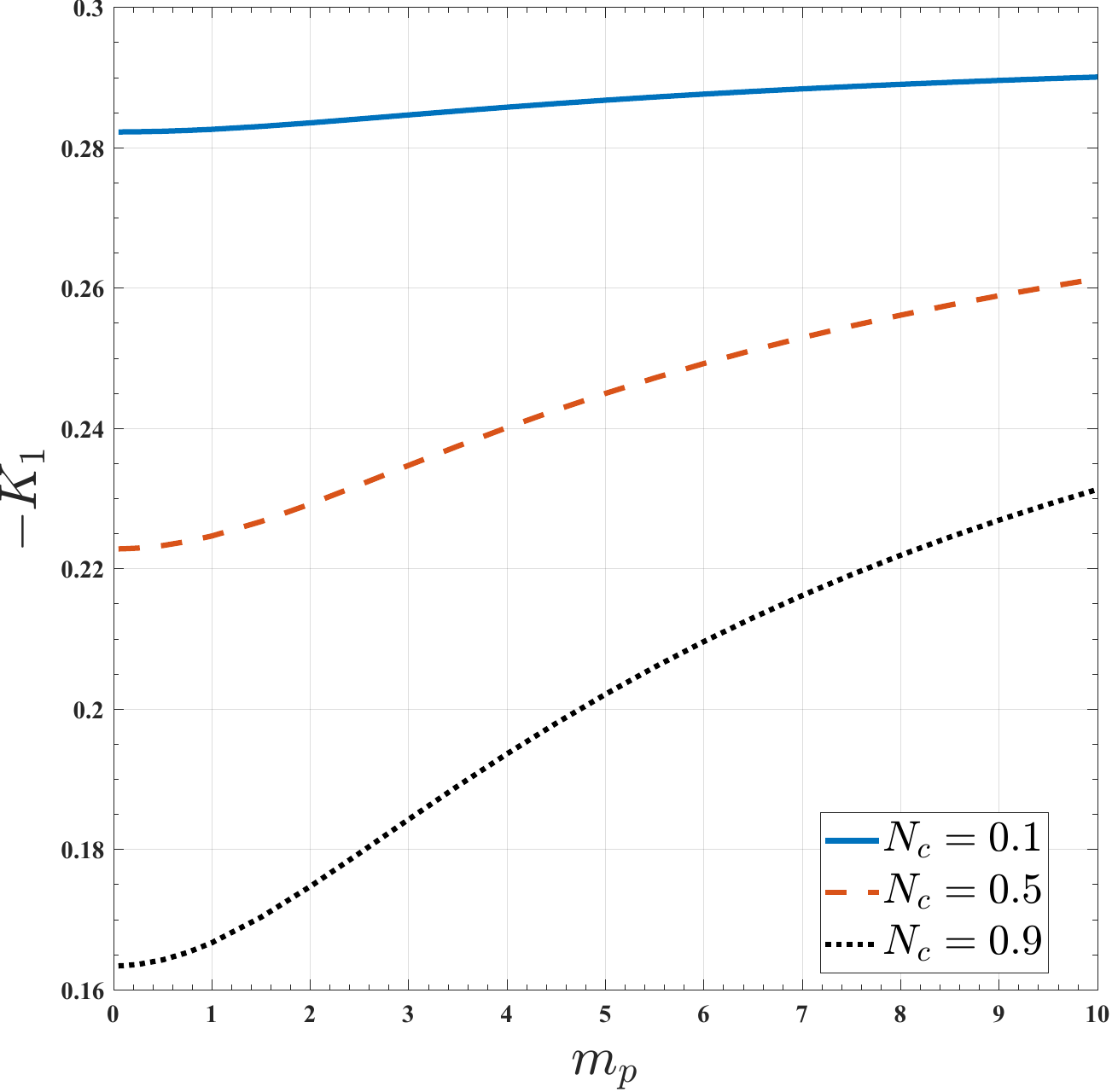}
    \caption{}
    \label{fig:3a}
\end{subfigure}
\begin{subfigure}{0.49\linewidth}
    \centering
    \includegraphics[width=\linewidth]{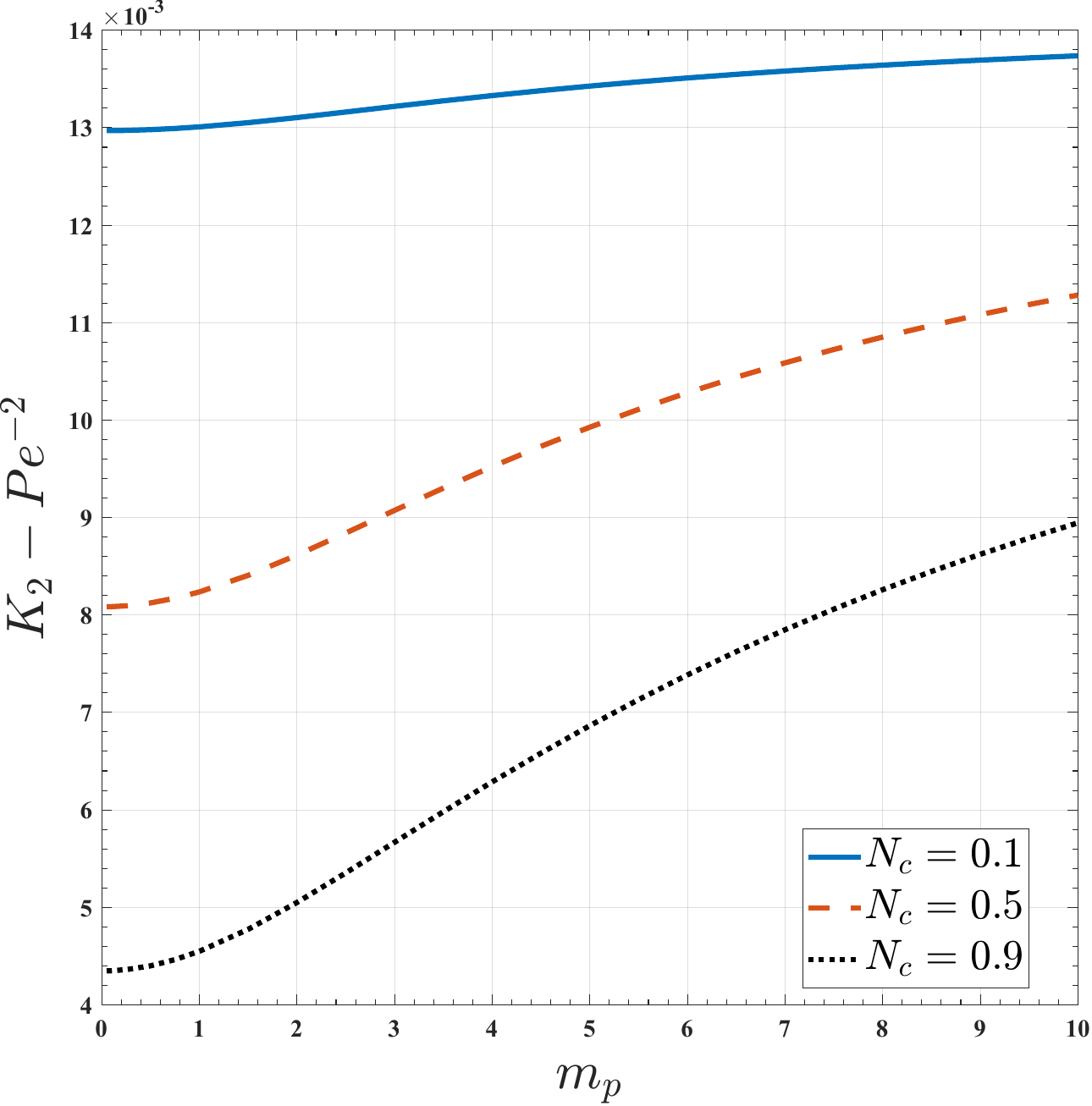}
    \caption{}
    \label{fig:3b}
\end{subfigure}
    \caption{Dependence of (a) the convection magnitude $-K_1$ and (b) the shear-induced dispersion $K_2-Pe^{-2}$ on $m_p$ for $N_c=0.1$, $0.5$, and $0.9$, with $\beta=0.01$, $\theta=0.5$, $Da=1$, and $\lambda=0.01$.}
    \label{fig:3}
\end{figure}

\begin{figure}
    \centering
\begin{subfigure}{0.49\linewidth}
    \centering
    \includegraphics[width=\linewidth]{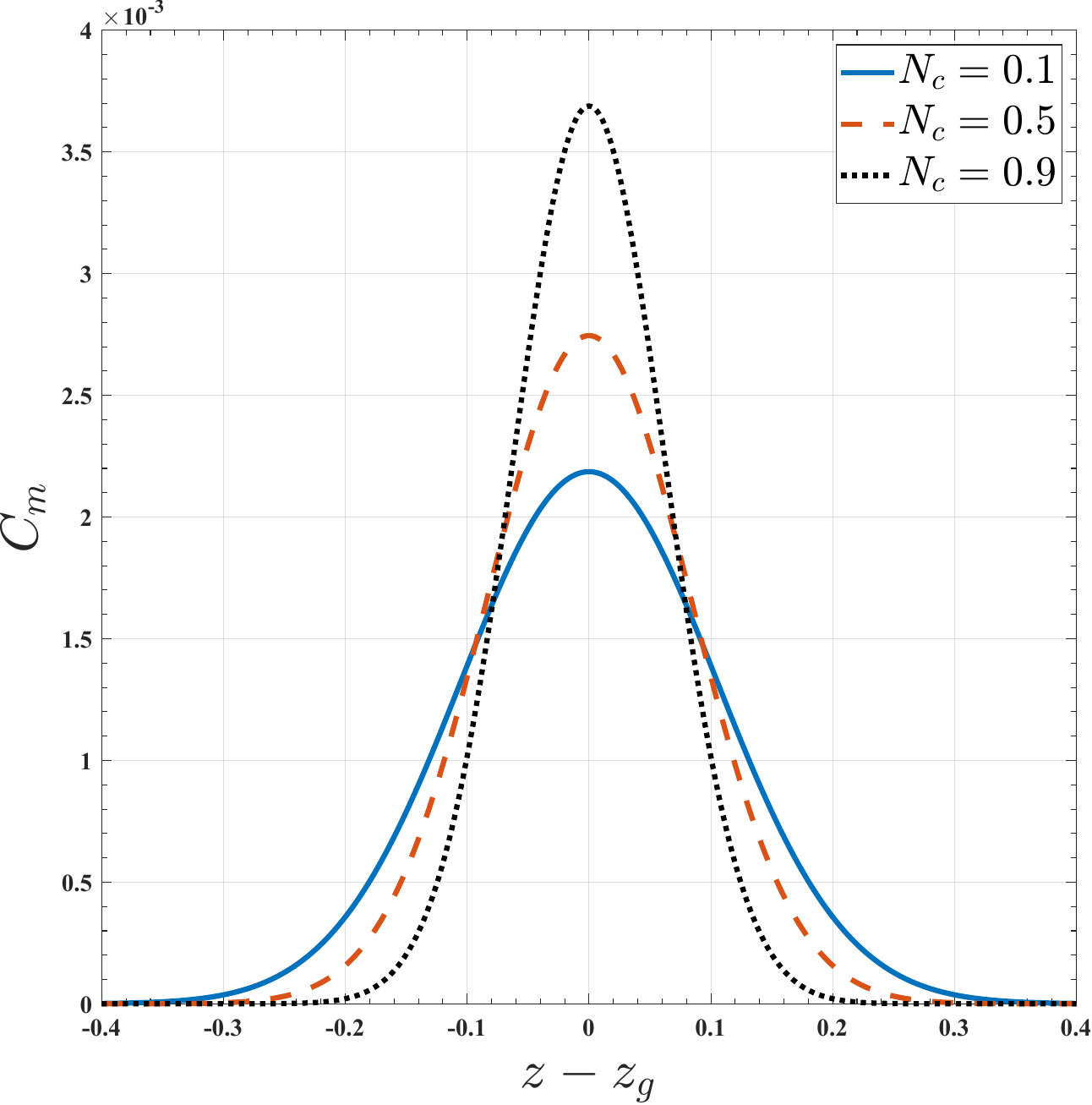}
    \caption{}
    \label{fig:4a}
\end{subfigure}
\begin{subfigure}{0.49\linewidth}
    \centering
    \includegraphics[width=\linewidth]{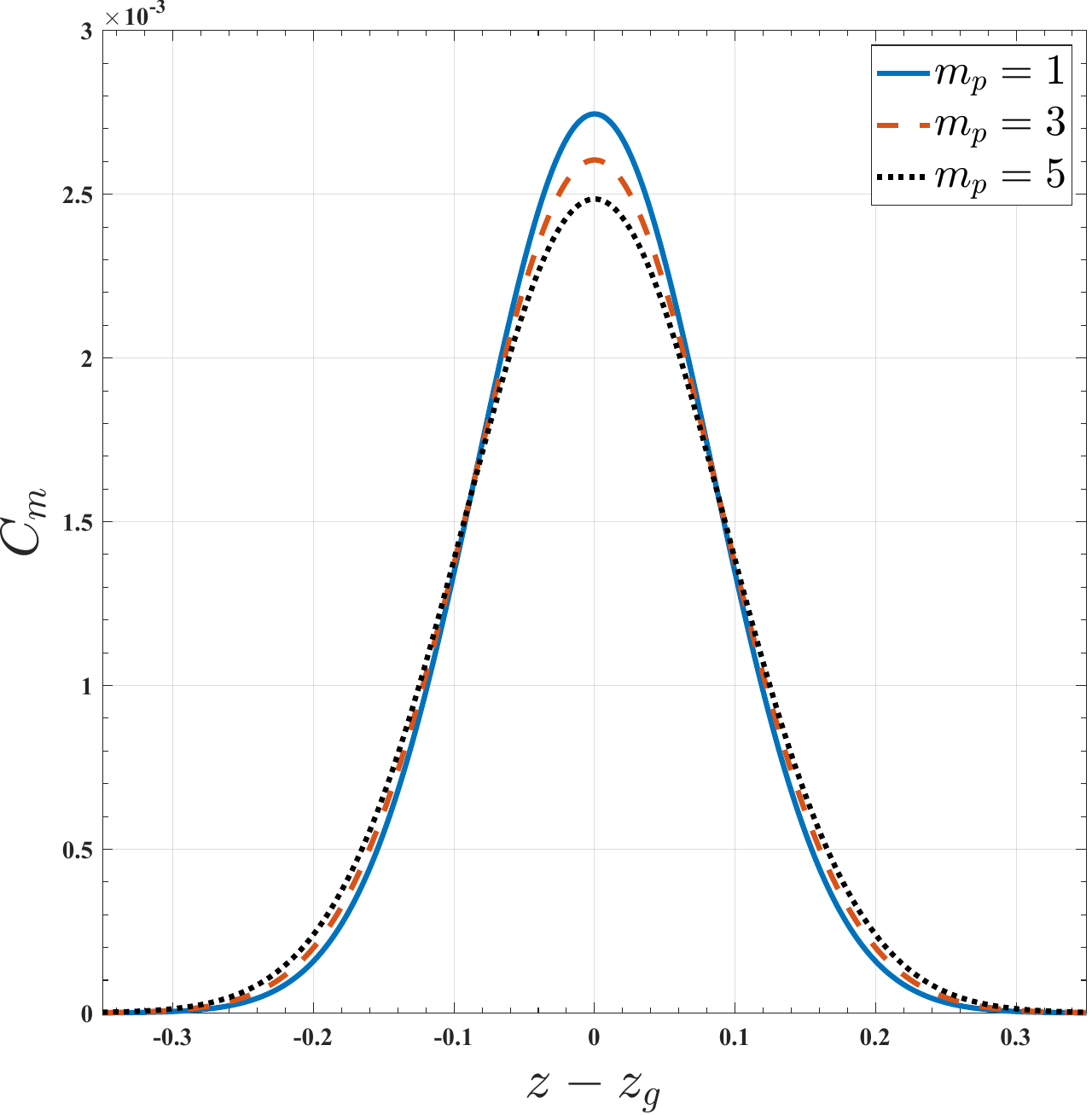}
    \caption{}
    \label{fig:4b}
\end{subfigure}
    \caption{Sectional mean concentration $C_m$ at $t=1$ for variations in (a) $N_c$ at fixed $m_p=1$ and (b) $m_p$ at fixed $N_c=0.5$. Other parameters are $\beta=0.01$, $\theta=0.5$, $Da=1$, $\lambda=0.01$, and $Pe=1000$.}
    \label{fig:4}
\end{figure}

A useful structural property follows directly from the \(n=0\) member of Eq.~\eqref{eq:fn_hierarchy}:
\begin{equation}
\frac{\partial f_0}{\partial t}
=
\frac{1}{r}\frac{\partial}{\partial r}
\left(r\frac{\partial f_0}{\partial r}\right)
-K_0(t)f_0 .
\label{eq:f0}
\end{equation}
Neither the velocity nor the microrotation field appears in this equation or in the corresponding surface equation for \(g_0\). Consequently, for fixed \(\lambda\), \(\beta\), \(\theta\), and \(Da\), the exchange coefficient \(K_0\) is independent of \(N_c\) and \(m_p\). Micropolarity therefore modifies the transport through \(K_1\) and \(K_2\), not through the total mobile-phase depletion. This separation is important when interpreting Figures~\ref{fig:3} and \ref{fig:4}. Changes in cloud translation and width can be attributed directly to the velocity field, while the mobile mass is unchanged within each rheological sweep.

The limiting velocity fields make this dependence explicit. When \(m_p\to0\) (equivalently \(\gamma\to\infty\)), the no-spin conditions force \(w\to0\) and Eq.~\eqref{eq:dimensionless_linear_momentum} reduces to
\begin{equation}
v(r)\longrightarrow \frac{2-N_c}{2}\,v_N(r),
\qquad
v_N(r)=1-r^2+\frac{1-\lambda^2}{\ln(1/\lambda)}\ln r ,
\label{eq:smallmp}
\end{equation}
where \(v_N\) is the Newtonian annular Poiseuille profile under the same pressure-gradient scaling. In the opposite limit \(m_p\to\infty\), microrotation follows the local vorticity away from \(O(m_p^{-1})\) wall layers, \(w\to-\tfrac12v_r\), and the leading axial profile approaches \(v_N\), independently of \(N_c\). Thus, the micropolar annulus continuously connects two velocity-amplitude limits, which may be summarized by
\begin{equation}
A(N_c,m_p)=\frac{\max_r v(r)}{\max_r v_N(r)},
\qquad
A\to1-\frac{N_c}{2}\quad(m_p\to0),
\qquad
A\to1\quad(m_p\to\infty).
\label{eq:amplitude}
\end{equation}

For the weak-reaction case in Figure~\ref{fig:3}, the normalized shape of the velocity profile varies much less than its amplitude, and the computed transport coefficients accordingly satisfy, to a good approximation,
\[
-K_1\propto A,
\qquad
K_2-Pe^{-2}\propto A^2.
\]
Indeed, the ratio \((K_2-Pe^{-2})/K_1^2\) varies by less than 3\% over the whole range of the figure.
Figure~\ref{fig:3a} shows that \(-K_1\) increases monotonically with \(m_p\) for every \(N_c\). As \(m_p\) grows, the spin field adjusts more readily to the local vorticity, and the effective resistance decreases toward the large-\(m_p\) limit; the pressure-driven flow therefore accelerates. At fixed \(m_p\), increasing \(N_c\) has the opposite effect and lowers the axial velocity amplitude. At small \(m_p\), the curves for \(N_c=0.1,0.5,0.9\) are separated approximately in the ratio \(1:0.79:0.58\), matching the factor \(1-N_c/2\) predicted by Eq.~\eqref{eq:smallmp} to within 0.1\% at \(m_p=0.05\). Their separation decreases as \(m_p\) increases, consistent with the \(N_c\)-independent large-\(m_p\) limit.

The shear-induced dispersion in Figure~\ref{fig:3b} displays the same parameter ordering but a substantially stronger sensitivity. Removing the constant molecular contribution \(Pe^{-2}\) isolates the part generated by advection, which here has two sources. Differential advection first creates a transverse concentration contrast, and radial diffusion repeatedly transfers solute between faster and slower streamlines. This is Taylor dispersion. In addition, the retentive wall holds solute stationary while the mobile cloud moves on and releases it later behind the centroid. At \(t=1\) the second mechanism dominates: for \(N_c=0.5\) and \(m_p=1\), \(K_2-Pe^{-2}=8.24\times10^{-3}\), about 6.5 times the inert Taylor value \(1.26\times10^{-3}\). Both mechanisms produce an axial variance that scales with the square of the advective velocity. So at small \(m_p\) the three ordinates are separated in the ratio \(1:0.62:0.34\), in agreement with \((1-N_c/2)^2\). A larger \(N_c\) therefore suppresses axial spreading more strongly than it suppresses the mean advective speed.

Figure~\ref{fig:4} translates these coefficient trends into the observable sectional concentration. The profiles are plotted against \(z-z_g\), so the centroid translation associated with \(K_1\) has been removed. Since \(K_0\) is unchanged when only \(N_c\) or \(m_p\) is varied, all curves within a panel contain the same mobile mass, and their differences are purely redistributive. From Eqs.~\eqref{eq:Cm_green}--\eqref{eq:mean_moments},
\[
C_m(t,z_g)=\frac{\mathcal M(t)}{\sqrt{4\pi\mathcal D(t)}},
\qquad
\mathcal D(t)=\int_0^t K_2(\tau)\,d\tau .
\]
Because \(\mathcal D\) scales approximately as \(A^2\) (the molecular part \(Pe^{-2}t\) is negligible here), the peak scales as \(C_m(t,z_g)\propto A^{-1}\). Hence, the smaller accumulated dispersion produced by larger \(N_c\) gives a taller and narrower cloud: the peak rises by a factor of 1.69 from \(N_c=0.1\) to \(0.9\), matching the inverse ratio 1.70 of \(-K_1\) in Figure~\ref{fig:3a}. Increasing \(m_p\) broadens the cloud as the flow approaches the Newtonian-annulus limit, lowering the peak by 9\% from \(m_p=1\) to \(5\). The stronger variation across \(N_c\) than across the displayed \(m_p\) range confirms that the coupling number is the dominant rheological control in these calculations, while \(m_p\) governs the rate of approach between the two limiting velocity states.

\subsection{Effect of the annular aspect ratio}
\begin{figure}
    \centering
\begin{subfigure}{0.49\linewidth}
    \centering
    \includegraphics[width=\linewidth]{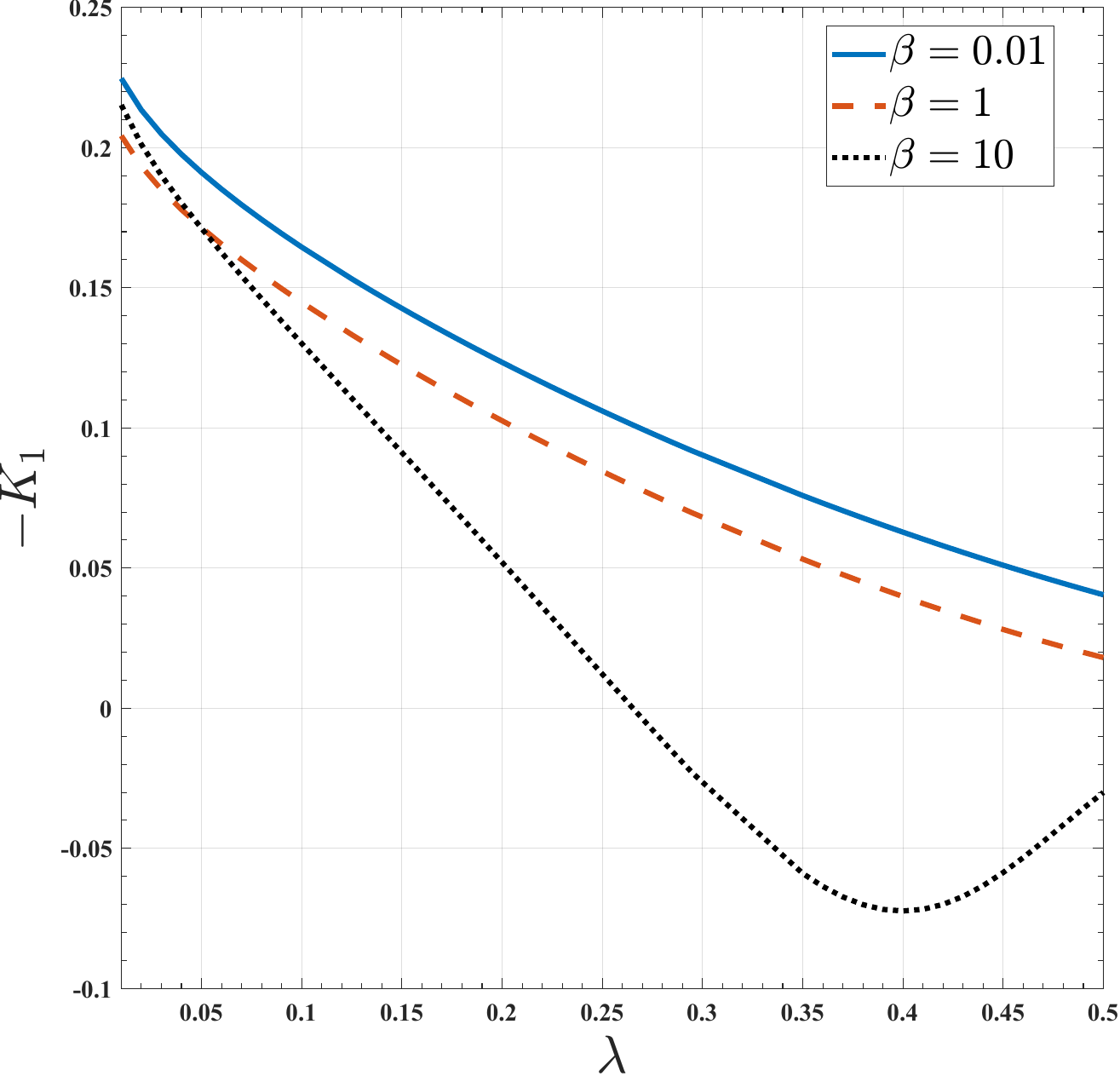}
    \caption{}
    \label{fig:5a}
\end{subfigure}
\begin{subfigure}{0.49\linewidth}
    \centering
    \includegraphics[width=\linewidth]{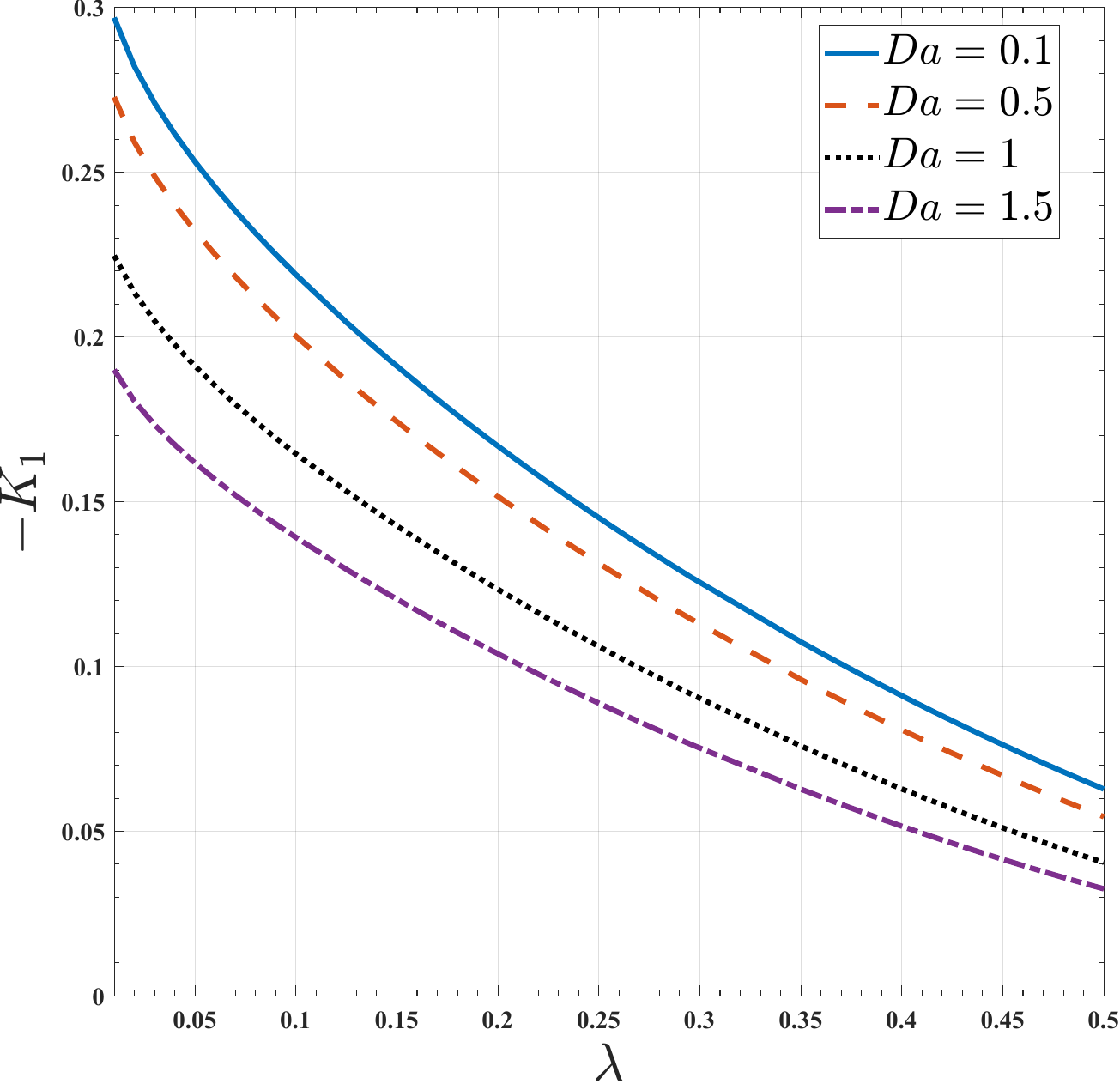}
    \caption{}
    \label{fig:5b}
\end{subfigure}
\begin{subfigure}{0.49\linewidth}
    \centering
    \includegraphics[width=\linewidth]{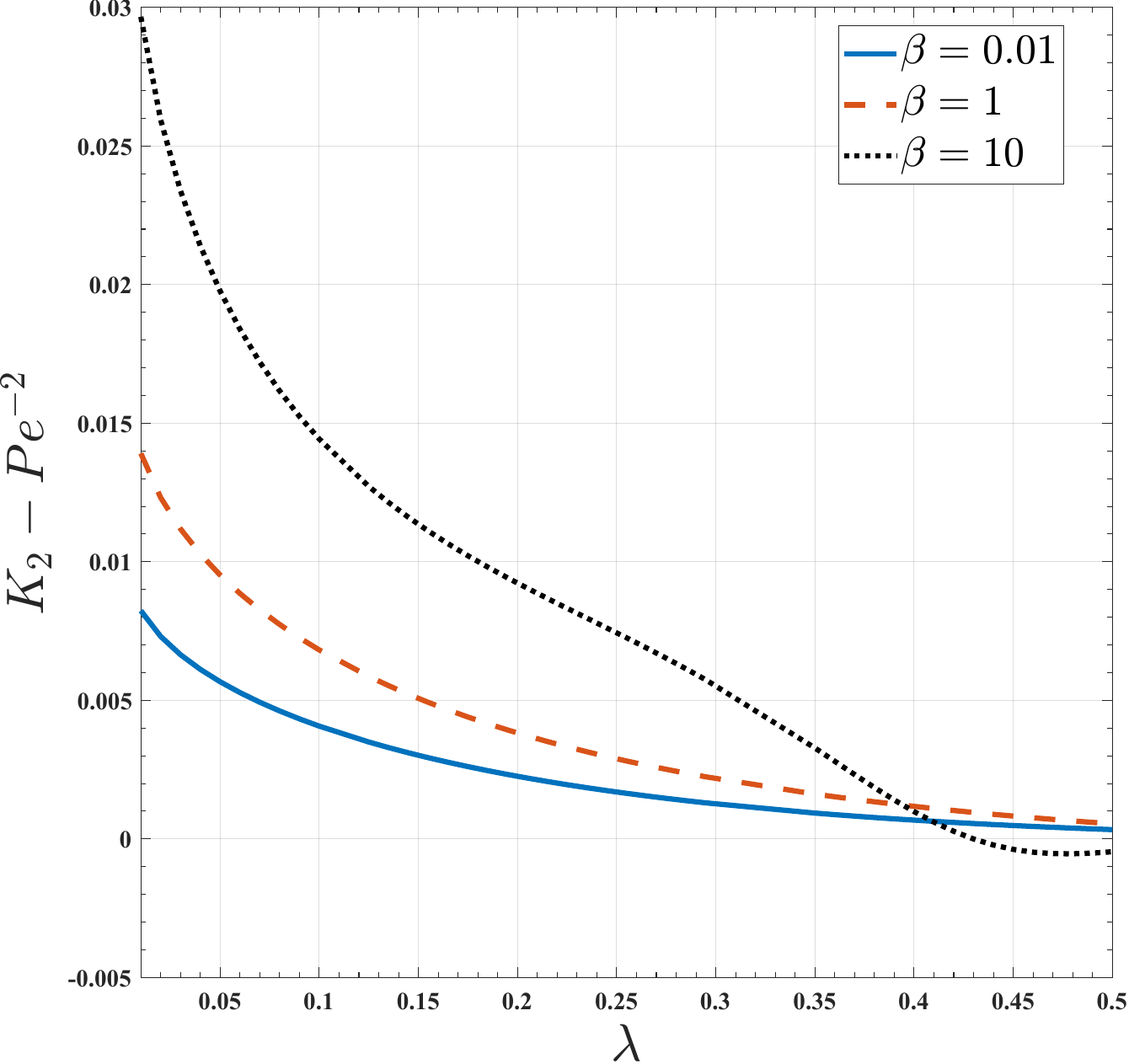}
    \caption{}
    \label{fig:5c}
\end{subfigure}
\begin{subfigure}{0.49\linewidth}
    \centering
    \includegraphics[width=\linewidth]{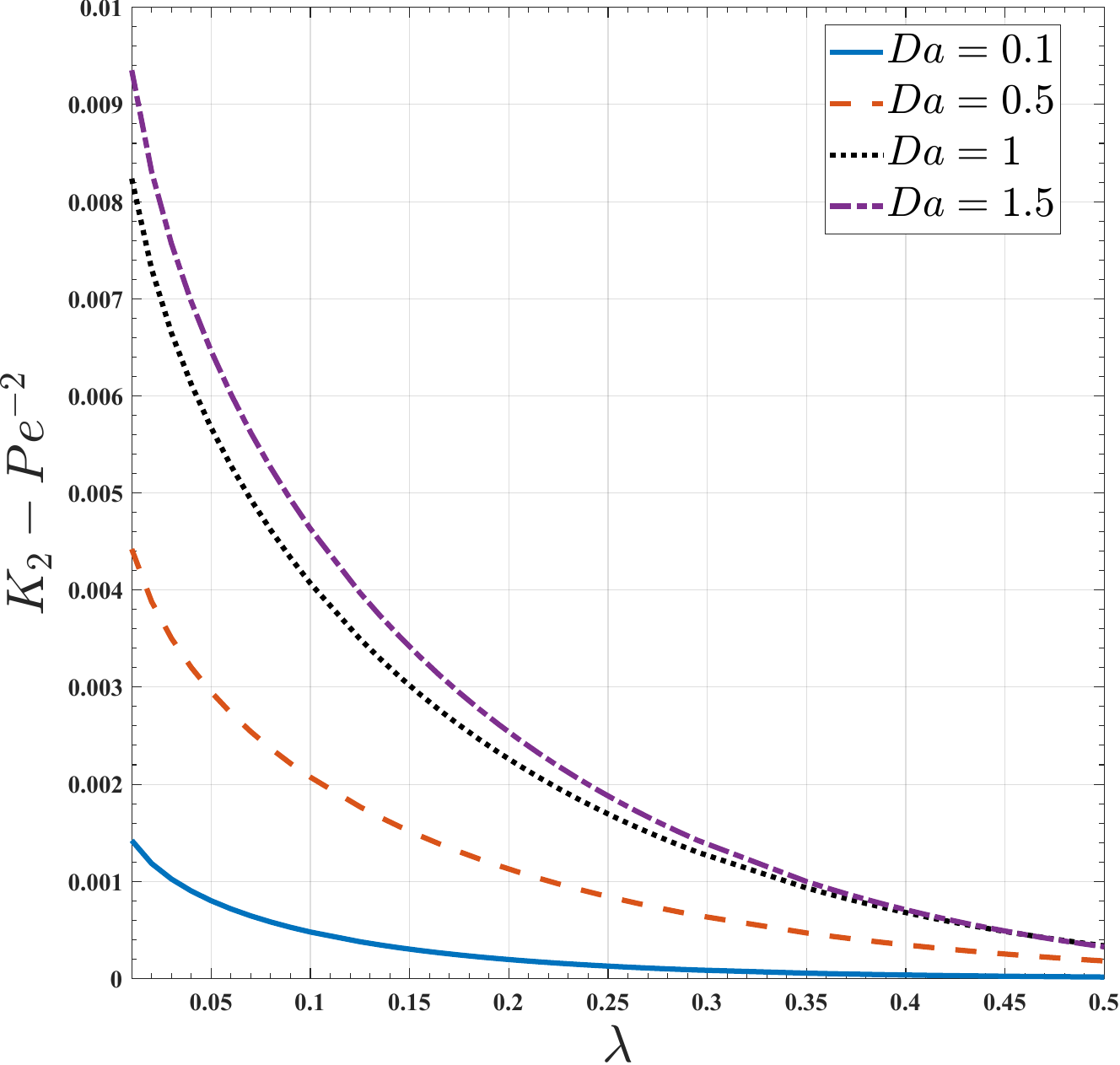}
    \caption{}
    \label{fig:5d}
\end{subfigure}    
    \caption{Effect of catheter ratio $\lambda$ on (a,b) the convection coefficient $-K_1$ and (c,d) the shear-induced dispersion $K_2-Pe^{-2}$ at $t=1$: (a,c) variation with $\beta$ at fixed $Da=1$; (b,d) variation with $Da$ at fixed $\beta=0.01$. Other parameters are $m_p=1$, $N_c=0.5$, and $\theta=0.5$.}
    \label{fig:5}
\end{figure}

The aspect ratio \(\lambda=b/a\) controls the width of the fluid-filled annular gap and is therefore the direct geometric measure of catheter confinement. All comparisons in Figure~\ref{fig:5} are made at a fixed imposed pressure gradient, so increasing \(\lambda\) changes both the transverse length scale and the pressure-driven velocity field rather than merely removing cross-sectional area. For a chemically passive wall,
\begin{equation}
K_1=-\bar v,
\qquad
\bar v=\frac{2}{1-\lambda^2}\int_{\lambda}^{1}r\,v(r)\,dr ,
\label{eq:K1inert}
\end{equation}
which identifies \(-K_1\) directly with the sectional-mean fluid speed. For \(\beta\le1\), Figure~\ref{fig:5a} accordingly shows a strong monotone reduction of advection as the catheter is enlarged. For the baseline case, \(-K_1\) decreases by a factor of about \(2.45\) between \(\lambda=0.01\) and \(0.3\), and by a factor of \(5.5\) by \(\lambda=0.5\). This is faster than the decline of \(\bar v\) itself (factors \(2.32\) and \(4.69\)), because the retentive wall holds part of the solute stationary and therefore retards the cloud: \(-K_1/\bar v\) falls from \(0.76\) at \(\lambda=0.01\) to \(0.64\) at \(\lambda=0.5\). The wall area per unit fluid volume, \(2/(1-\lambda^2)\), grows with \(\lambda\), so a larger share of the solute is held on the wall in a narrower gap. The effect is not proportional simply to the area occupied by the catheter, because the new inner no-slip boundary reorganizes the entire annular velocity profile.

Irreversible uptake acting alone would raise \(-K_1\) at fixed \(\lambda\) by preferentially removing mobile solute from the slow-moving region adjacent to the reactive outer wall; the surviving solute molecules would then be weighted toward faster streamlines. For \(\theta=Da=0\) at \(\lambda=0.01\), \(-K_1\) indeed rises from \(0.296\) to \(0.410\) as \(\beta\) increases from \(0.01\) to \(10\). With retention present, a second effect competes: faster depletion of the mobile phase leaves the stationary retained store relatively larger, and solute released from this store re-enters the flow behind the moving cloud. At \(\lambda=0.01\) the two effects nearly balance, and \(-K_1\) varies non-monotonically with \(\beta\) (\(0.225\), \(0.204\), and \(0.215\) for \(\beta=0.01\), \(1\), and \(10\)). As \(\lambda\) increases, the retention effect prevails. The \(\beta\)-dependent curves separate further, and for \(\beta=10\) the convection coefficient changes sign near \(\lambda\approx0.27\), reaching a minimum of \(-0.072\) at \(\lambda=0.4\). A negative \(-K_1\) means that the centroid of the mobile phase moves upstream: by \(t=1\) less than \(0.3\%\) of the solute is still mobile in these cases, and this remnant is supplied mainly by desorption from the wall store left behind the cloud. This amplification has a clear geometric origin: the transverse diffusion time scales as \((1-\lambda)^2\), while the reactive surface per unit fluid volume scales as \(2/(1-\lambda^2)\). A narrow annular gap is therefore homogenized rapidly and placed in contact with a proportionally larger reactive and retentive surface. Catheter confinement, therefore, suppresses the mean flow but amplifies, rather than suppresses, the influence of wall chemistry on the mobile solute.

The effect on dispersion is considerably stronger than the effect on convection. Between \(\lambda=0.01\) and \(0.3\), \(K_2-Pe^{-2}\) falls by a factor of about \(6\) for the baseline case, and by \(\lambda\simeq0.4\) the shear contribution is already small on the scale of Figure~\ref{fig:5c}. The purely shear-driven part is suppressed much more strongly: for a chemically passive wall it falls by a factor of \(24\) over the same range, from \(1.26\times10^{-3}\) to \(5.29\times10^{-5}\). Taylor dispersion requires both a transverse difference in axial velocity and a finite transverse distance over which diffusion transfers solute between streamlines; narrowing the annulus weakens both ingredients simultaneously. For a chemically passive wall, this strong geometric suppression can be quantified asymptotically. Let
\begin{equation}
\varepsilon=1-\lambda\ll1,
\qquad
y=\frac{r-\lambda}{\varepsilon}\in[0,1].
\label{eq:narrow_coordinates}
\end{equation}
For fixed \(m_p\) and fixed imposed pressure gradient, the dominant balances in Eqs.~\eqref{eq:dimensionless_linear_momentum} and \eqref{eq:dimensionless_angular_momentum} require
\[
v=\varepsilon^2V_0(y)+\cdots,
\qquad
w=\varepsilon^3W_0(y)+\cdots .
\]
At leading order,
\begin{equation}
V_0''+2(2-N_c)=0,
\qquad
V_0(0)=V_0(1)=0,
\end{equation}
and hence
\begin{equation}
V_0(y)=(2-N_c)y(1-y),
\qquad
\bar v\sim\frac{2-N_c}{6}\varepsilon^2 .
\label{eq:narrow_velocity}
\end{equation}
Thus, the mean speed itself vanishes quadratically as the available gap closes.

For any fixed \(t>0\), the transverse diffusion time is \(O(\varepsilon^2)\), so the thin gap is in its leading Taylor state as \(\varepsilon\to0\). Writing \(f_1=\varepsilon^4F(y)+\cdots\), the leading cell problem is
\begin{equation}
F''=V_0-\bar V_0,\qquad
F'(0)=F'(1)=0,\qquad
\int_0^1F\,dy=0,
\end{equation}
with \(\bar V_0=(2-N_c)/6\). Its solution is
\begin{equation}
F(y)=\frac{2-N_c}{360}
\left[1-30y^2(1-y)^2\right].
\end{equation}
Substitution into Eq.~\eqref{eq:transport_coefficients} gives
\begin{equation}
K_2-Pe^{-2}
\sim
\frac{(2-N_c)^2}{7560}(1-\lambda)^6
=
\frac{(1-N_c/2)^2}{1890}(1-\lambda)^6,
\qquad(\lambda\to1).
\label{eq:narrowgap6}
\end{equation}
Equivalently,
\begin{equation}
K_2-Pe^{-2}
\sim
\frac{\bar v^2(1-\lambda)^2}{210},
\label{eq:narrowgap210}
\end{equation}
which is the classical plane-Poiseuille Taylor--Aris coefficient when the full gap width is used as the transverse length scale \citep{dattaghosal2009}. The classical factor \(1/210\) is therefore recovered, while the new result for the present fixed-pressure-gradient catheter limit is the sixth-power collapse: \(\bar v^2=O((1-\lambda)^4)\) and the diminishing transverse mixing length contributes a further \(O((1-\lambda)^2)\). This law describes the shear-driven dispersion alone. With reversible retention, the narrow-gap dispersion is controlled instead by the exchange kinetics, whose capacity per unit fluid volume \(2\theta/(1-\lambda^2)\) grows with \(\lambda\): at \(\lambda=0.5\) the baseline value of \(K_2-Pe^{-2}\) in Figure~\ref{fig:5c}, \(3.36\times10^{-4}\), is about \(60\) times the passive value \(5.50\times10^{-6}\).

\begin{figure}
    \centering
    \begin{subfigure}{0.49\linewidth}
        \centering
        \IfFileExists{eps/narrow_gap_K2_scaling.pdf}
        {\includegraphics[width=\linewidth]{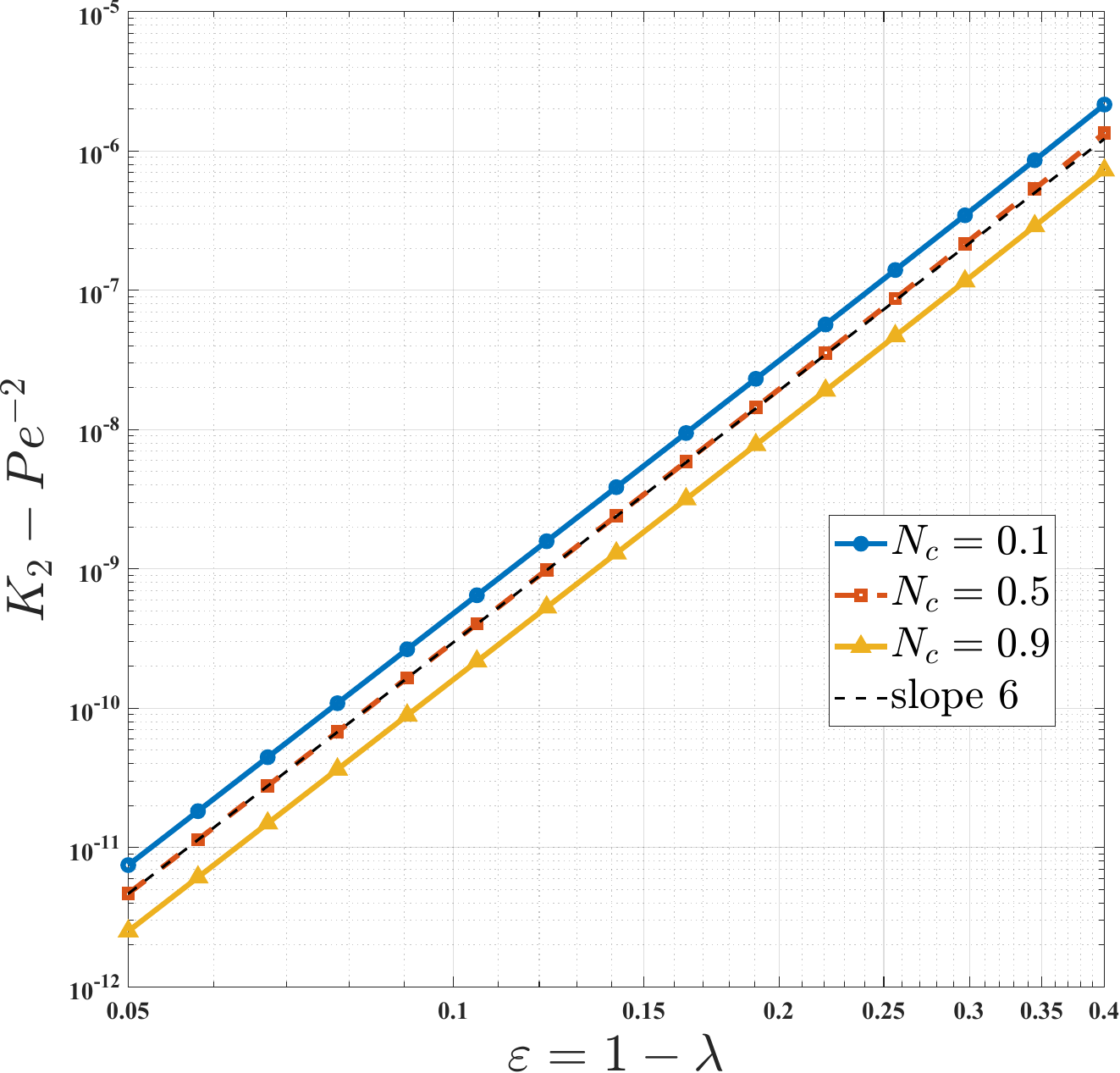}}
        {\fbox{\parbox[c][4.3cm][c]{0.93\linewidth}{\centering
        Placeholder for log--log verification of
        \(K_2-Pe^{-2}\sim(1-\lambda)^6\)}}}
        \caption{}
        \label{fig:narrowverify_a}
    \end{subfigure}
    \begin{subfigure}{0.49\linewidth}
        \centering
        \IfFileExists{eps/narrow_gap_compensated.pdf}
        {\includegraphics[width=\linewidth]{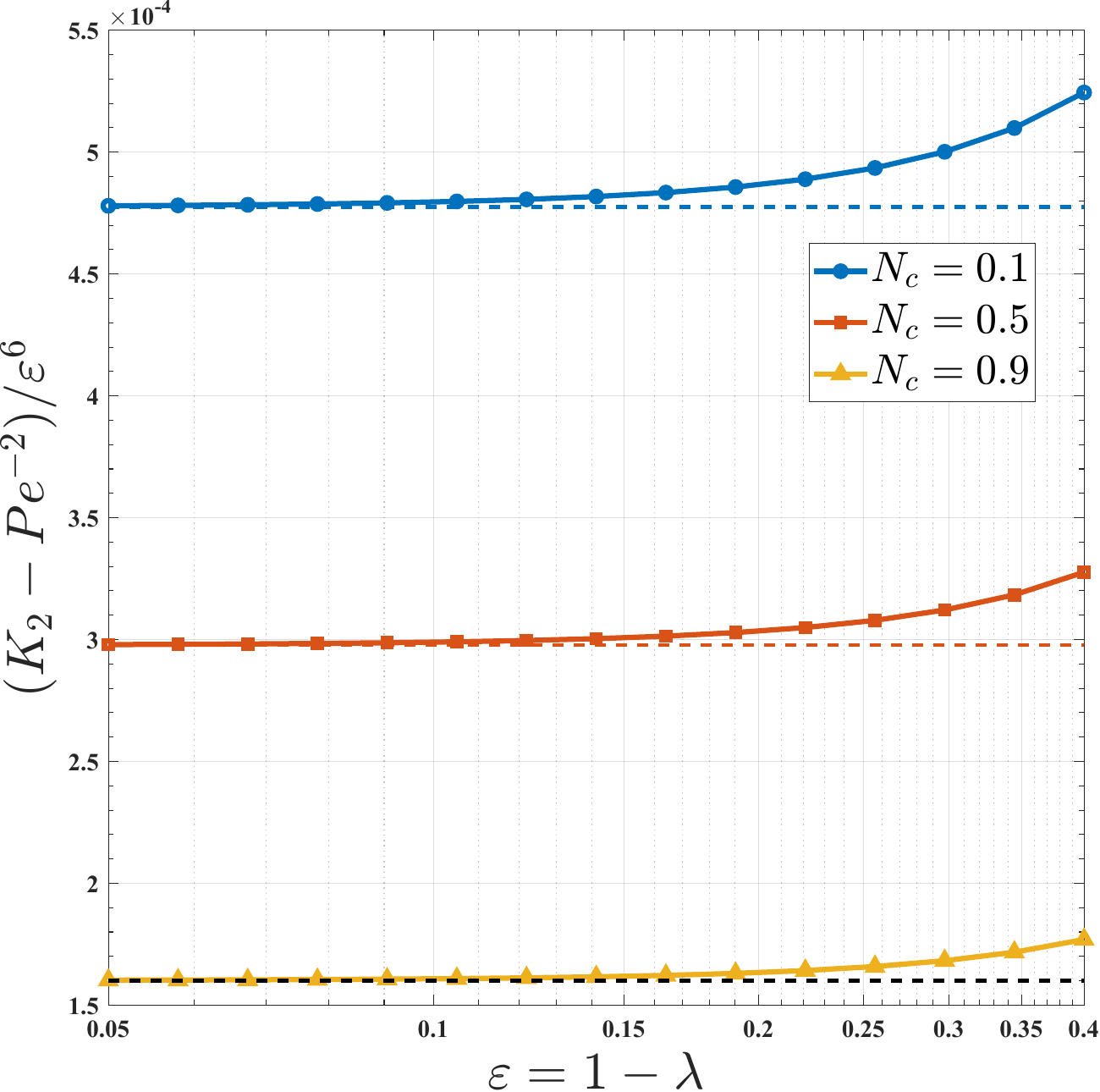}}
        {\fbox{\parbox[c][4.3cm][c]{0.93\linewidth}{\centering
        Placeholder for compensated narrow-gap limit}}}
        \caption{}
        \label{fig:narrowverify_b}
    \end{subfigure}
    \caption{Numerical verification of the narrow-gap law at fixed pressure gradient for $N_c=0.1,\,0.5,\,0.9$ and $m_p=1$, using the steady cell problem for a chemically passive wall. (a) $K_2-Pe^{-2}$ versus $\varepsilon=1-\lambda$ on logarithmic axes; the black dashed line denotes the predicted sixth-power scaling. (b) Compensated quantity $(K_2-Pe^{-2})/\varepsilon^6$; the dashed horizontal lines are the analytical limits $(2-N_c)^2/7560$ for the corresponding values of $N_c$.}
    \label{fig:narrowverify}
\end{figure}

Figure~\ref{fig:narrowverify} verifies the two independent parts of Eq.~\eqref{eq:narrowgap6}. In Figure~\ref{fig:narrowverify_a}, least-squares fits over the five smallest gaps (\(0.050\le\varepsilon\le0.091\)) give exponents \(6.0041\), \(6.0042\), and \(6.0045\) for \(N_c=0.1\), \(0.5\), and \(0.9\), respectively, confirming the predicted exponent six. In Figure~\ref{fig:narrowverify_b}, dividing out \(\varepsilon^6\) isolates the prefactor; the compensated curves approach \((2-N_c)^2/7560\), with relative deviations of only about \(0.10\)--\(0.11\%\) at the smallest gap. Thus, both the exponent and its explicit micropolar amplitude are recovered numerically for the chemically passive wall.

The reaction-dependent curves in Figure~\ref{fig:5} reveal an additional distinction between irreversible uptake and reversible exchange. In the dual-reaction case shown here, increasing \(\beta\) raises the computed \(K_2-Pe^{-2}\) at \(t=1\) (\(8.24\times10^{-3}\), \(1.39\times10^{-2}\), and \(2.97\times10^{-2}\) at \(\lambda=0.01\) for \(\beta=0.01\), \(1\), and \(10\)), whereas absorption acting alone has the opposite effect: for \(\theta=Da=0\) at \(\lambda=0.01\), \(K_2-Pe^{-2}\) falls from \(1.26\times10^{-3}\) to \(3.11\times10^{-4}\) over the same range of \(\beta\). The trend is also time-dependent: at \(t=0.3\), before the retained phase has built up, the order is reversed (\(1.66\), \(1.55\), and \(0.53\times10^{-3}\)). At the largest \(\lambda\), the \(\beta=10\) curve itself becomes negative for \(\lambda\gtrsim0.43\), the dispersive counterpart of the sign reversal in \(-K_1\). This contrast demonstrates that the effect of \(\beta\) on normalized mobile-phase dispersion is not universal once reversible storage is present: retention changes the temporal partition of mobile solute and therefore the concentration weighting that enters the generalized-dispersion coefficients. The present trend is therefore interpreted as a coupled retention--absorption effect rather than as a generic consequence of stronger absorption.

Figures~\ref{fig:5b} and \ref{fig:5d} show the corresponding variation with the
Damk\"ohler number. Varying \(Da\) over \(0.1\le Da\le1.5\) changes the transport coefficients substantially at \(t=1\): at \(\lambda=0.01\), \(-K_1\) falls by \(36\%\) and \(K_2-Pe^{-2}\) rises by a factor of \(6.6\). This is consistent with the role of \(Da\) as a kinetic rate parameter: it controls how rapidly the retained phase responds, whereas \(\theta\) controls the local equilibrium partition. Because the observation time \(t=1\) is comparable to the exchange time \(Da^{-1}\), \(Da\) decides how much of the retention has developed. For \(Da=0.1\) the exchange is too slow to matter by \(t=1\): \(-K_1\) stays within \(0.5\%\) of \(\bar v\) at every \(\lambda\), and at \(\lambda=0.01\) the dispersion coefficient exceeds the inert value by only \(13\%\). The sensitivity to \(Da\) grows with \(\lambda\), because the reactive surface per unit fluid volume increases: the ratio of the dispersion coefficients for \(Da=1.5\) and \(0.1\) rises from \(6.6\) at \(\lambda=0.01\) to \(22\) at \(\lambda=0.5\).

\begin{figure}
    \centering
    \begin{subfigure}{0.49\linewidth}
    \centering
    \includegraphics[width=\linewidth]{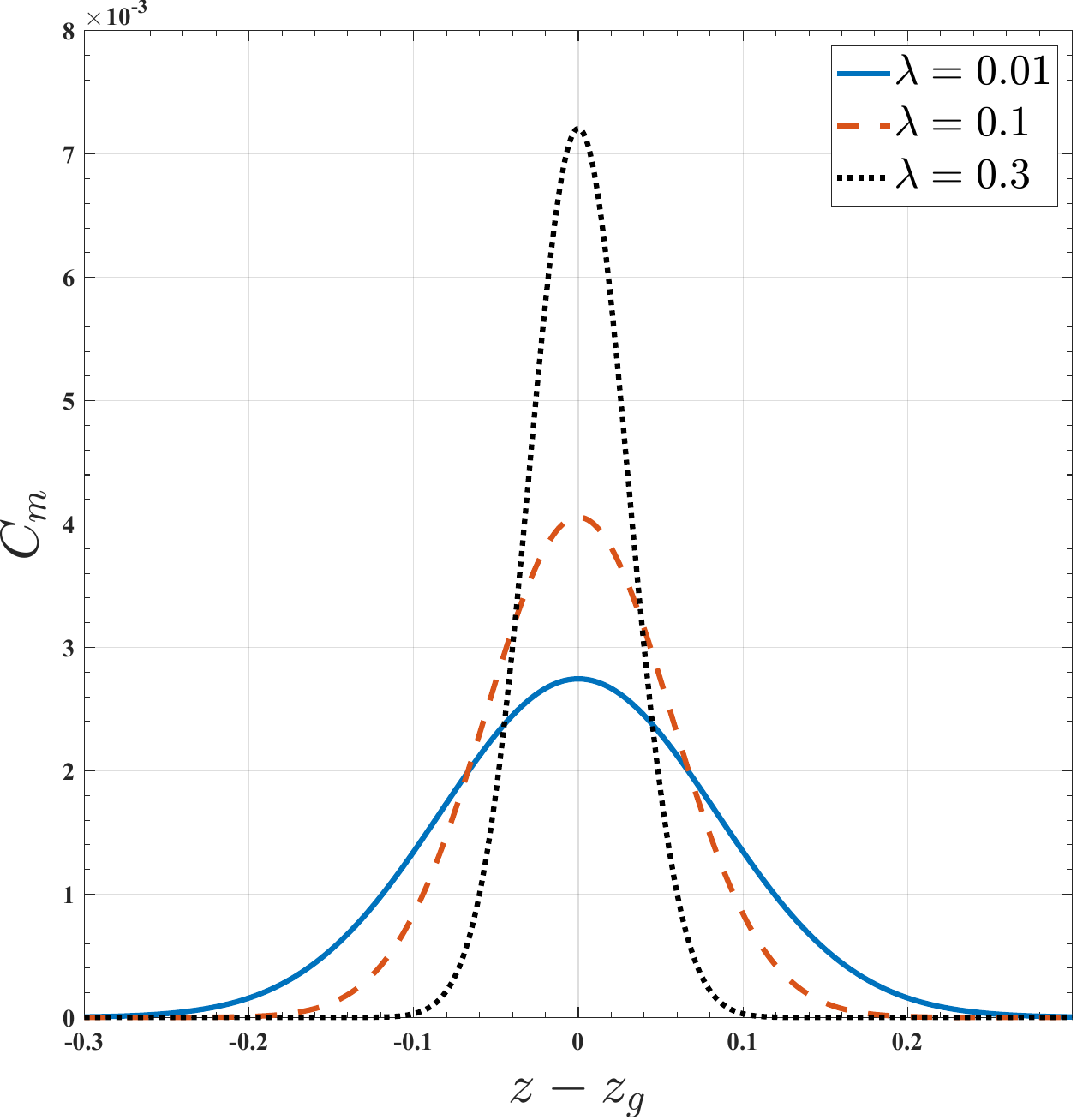}
\end{subfigure}
    \caption{Sectional mean concentration $C_m$ at $t=1$ for different catheter ratios $\lambda$, with $\beta=0.01$, $\theta=0.5$, $Da=1$, $m_p=1$, $N_c=0.5$, and $Pe=1000$.}
    \label{fig:6}
\end{figure}

Figure~\ref{fig:6} shows the corresponding centred mean concentration. Plotting against \(z-z_g\) removes the simultaneous reduction in downstream translation and exposes the changes in mobile mass and accumulated dispersion. As \(\lambda\) increases from \(0.01\) to \(0.3\), the cloud becomes markedly narrower and its peak rises by about a factor of \(2.6\), even though the mobile mass at \(t=1\) is \(5\%\) smaller (\(\Phi_m=0.546\) against \(0.576\)); the accumulated dispersion \(\mathcal D\) falls by a factor of \(7.7\). Hence, catheter enlargement produces a clear geometric trade-off under fixed pressure gradient: it confines the mobile solute much more strongly in the axial direction, but it also slows the centroid transport. The mean-concentration result is therefore consistent with both the pronounced reduction of \(K_2-Pe^{-2}\) and the more moderate reduction of \(-K_1\).

\subsection{Effect of the reaction parameters}
\begin{figure}
    \centering
    \begin{subfigure}{0.49\linewidth}
    \includegraphics[width=\linewidth]{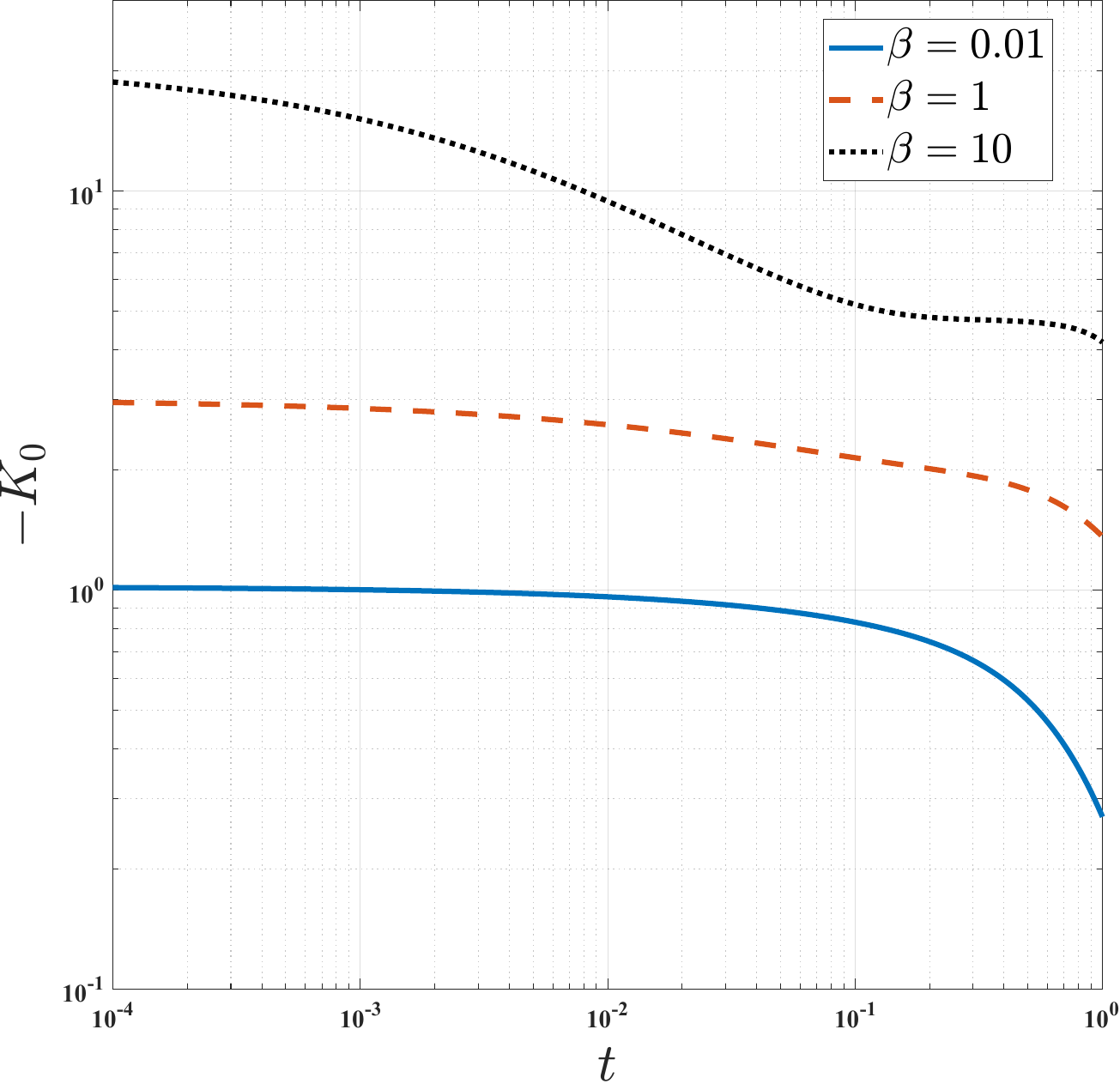}
    \caption{}
    \label{fig:7a}
\end{subfigure}
\begin{subfigure}{0.49\linewidth}
    \centering
    \includegraphics[width=\linewidth]{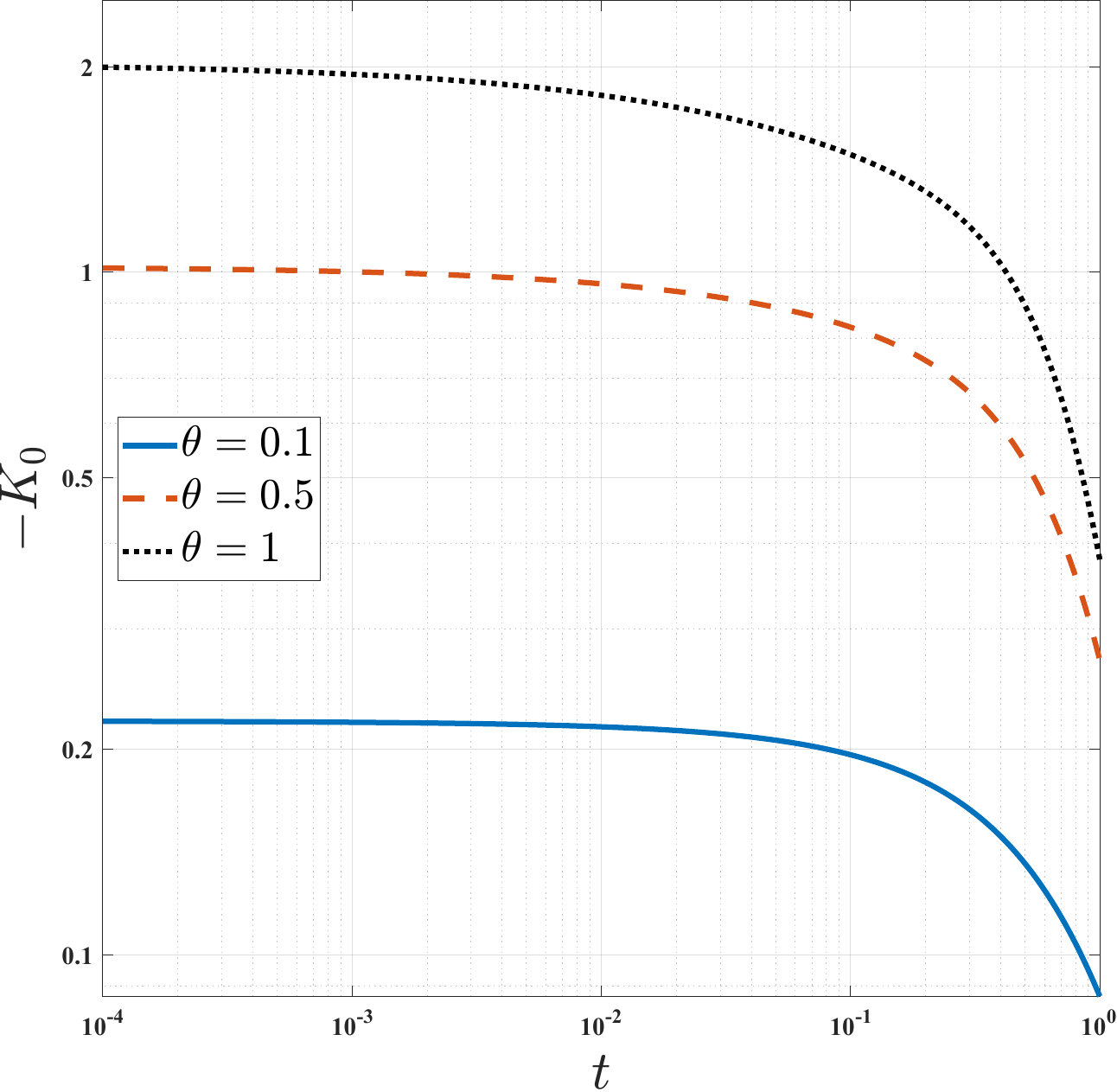}
    \caption{}
    \label{fig:7b}
\end{subfigure}
\begin{subfigure}{0.49\linewidth}
    \centering
    \includegraphics[width=\linewidth]{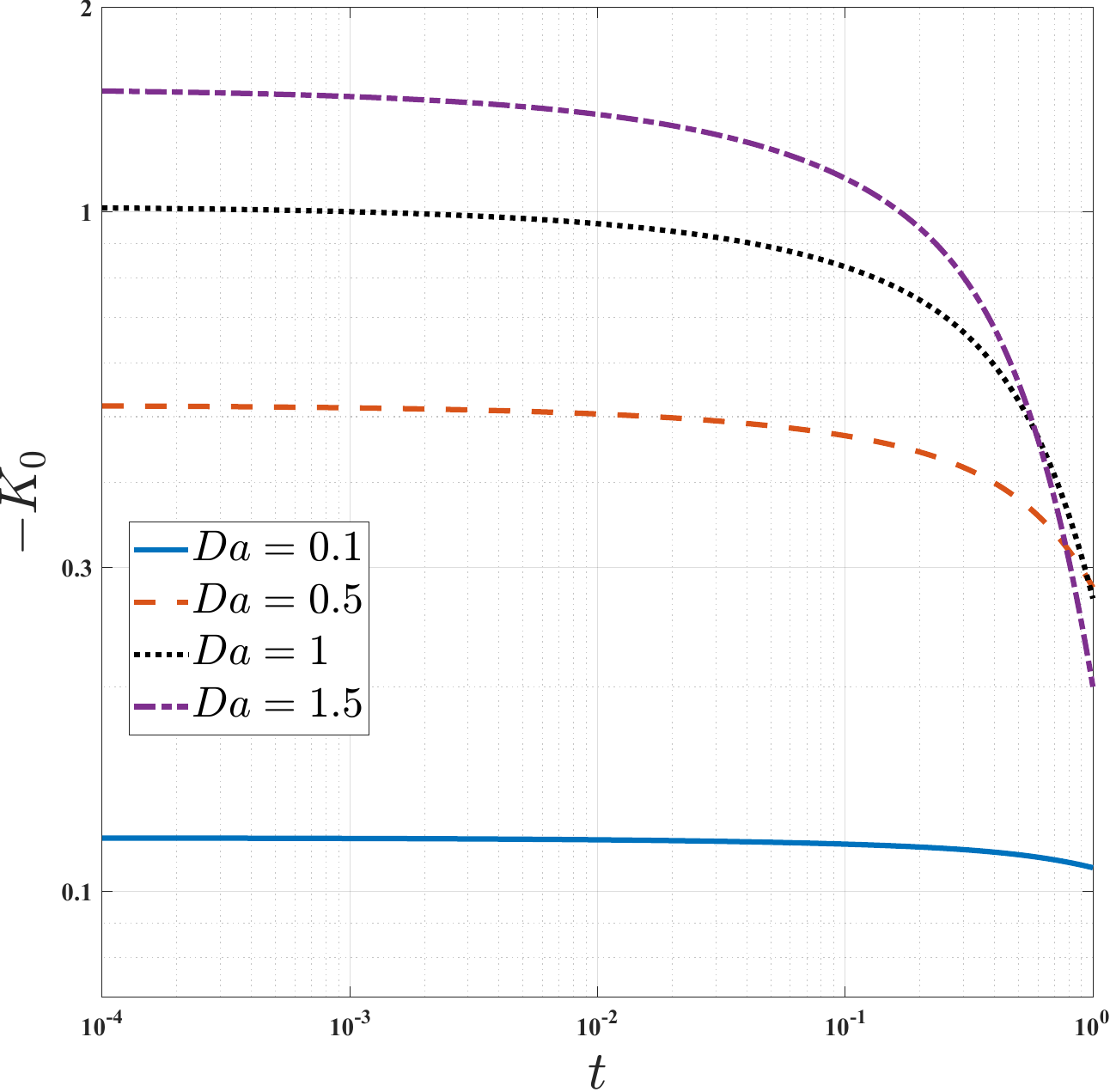}
    \caption{}
    \label{fig:7c}
\end{subfigure}
\caption{Time evolution of the depletion rate $-K_0$ for variations in (a) $\beta$ at fixed $\theta=0.5$, $Da=1$; (b) $\theta$ at fixed $\beta=0.01$, $Da=1$; and (c) $Da$ at fixed $\beta=0.01$, $\theta=0.5$. Other parameters are $m_p=1$, $N_c=0.5$, and $\lambda=0.01$.}
\label{fig:7}
\end{figure}

The three reaction parameters enter the transport problem in physically different ways. The irreversible reaction parameter \(\beta\) removes mobile solute permanently from the fluid system. On the other hand, the retention parameter \(\theta\) fixes the equilibrium tendency of the surface phase relative to the adjacent mobile concentration. Also, \(Da\) sets the timescale on which reversible exchange approaches that local equilibrium. 

Figure~\ref{fig:7} presents the negative exchange coefficient $-K_0$ through time
for $m_p=1$, $N_c=0.5$ and $\lambda=0.01$, while $\beta$, $\theta$ and $Da$ are varied individually. Each curve begins at a nearly constant plateau value, after which it declines. The plateaus are ordered by the strength of the wall kinetics,
taking values of $1.02$, $3.00$ and $21.0$ for $\beta=0.01$, $1$, and
$10$ respectively; $0.220$, $1.020$ and $2.020$ for $\theta=0.1$, $0.5$ and $1$ respectively; and
$0.120$, $0.520$, $1.020$ and $1.520$ for $Da=0.1$, $0.5$, $1$ and $1.5$ respectively. The plateau is shortest for the strongest kinetics: at $\beta=10$ the decline is already underway at the left edge of Figure~\ref{fig:7a}. The subsequent behavior of the curves reveals a considerable difference between the panels. The $\theta$ curves stay
distinct to the end of the window, whereas the $Da$ curves cross: the $Da=1.5$ curve starts highest but has fallen to $0.200$ by $t=1$, below the values $0.280$ and $0.270$ for $Da=0.5$ and $1$.
From the initially uniform release and the empty retained phase,
\begin{equation}
-K_0(0^{+})=\frac{2}{1-\lambda^{2}}\big(\beta+\theta\,Da\big).
\label{eq:K0init}
\end{equation}
Thus, irreversible uptake contributes immediately through \(\beta\), while reversible retention contributes through the forward-transfer product \(\theta Da\). Equation~\eqref{eq:K0init} explains the ordering of the initial plateaus in all three panels and provides an exact short-time check on the coupled bulk--surface implementation. The duration of the plateau follows from the same reasoning. While the depleted layer is thin compared with the gap and the retained phase is still nearly empty, the outer wall acts as a planar absorbing boundary of strength \(\beta+\theta Da\), and
\begin{equation}
-K_0(t)\simeq\frac{2(\beta+\theta Da)}{1-\lambda^{2}}\,
\mathrm{e}^{(\beta+\theta Da)^2t}\,
\mathrm{erfc}\!\left[(\beta+\theta Da)\,t^{1/2}\right].
\label{eq:K0layer}
\end{equation}
The plateau therefore lasts while \((\beta+\theta Da)^2t\ll1\). For \(\beta+\theta Da\le1\), Eq.~\eqref{eq:K0layer} reproduces the computed \(-K_0\) to within \(1\%\) up to \(t=10^{-2}\), whereas for \(\beta=10\) the timescale \((\beta+\theta Da)^{-2}\approx0.009\) is so short that \(-K_0\) has already fallen by \(11\%\) at \(t=10^{-4}\).

The subsequent reduction can be associated with two underlying mechanisms. Solute is consumed rapidly in the region adjacent to the wall, so the near-wall concentration becomes lower than the sectional mean and hence the reaction runs short of reactant. Simultaneously, the wall layer progressively accumulates solute, and once it approaches equilibrium, the backward transfer nearly balances
the forward one. Hence, the reversible pathway becomes inactive. The second
mechanism is what distinguishes the two panels. The Damk\"ohler number controls the rate at which the wall layer becomes filled. A large $Da$ saturates the wall early and then has little left to contribute, so the rapid initial removal ends up giving the steepest decline, and the curves cross. Alternatively, the retention parameter determines how much solute the wall can ultimately retain, a capacity that is independent of time. Consequently, those corresponding curves remain separated throughout the considered interval.

\begin{figure}
    \centering
    \begin{subfigure}{0.49\linewidth}
    \includegraphics[width=\linewidth]{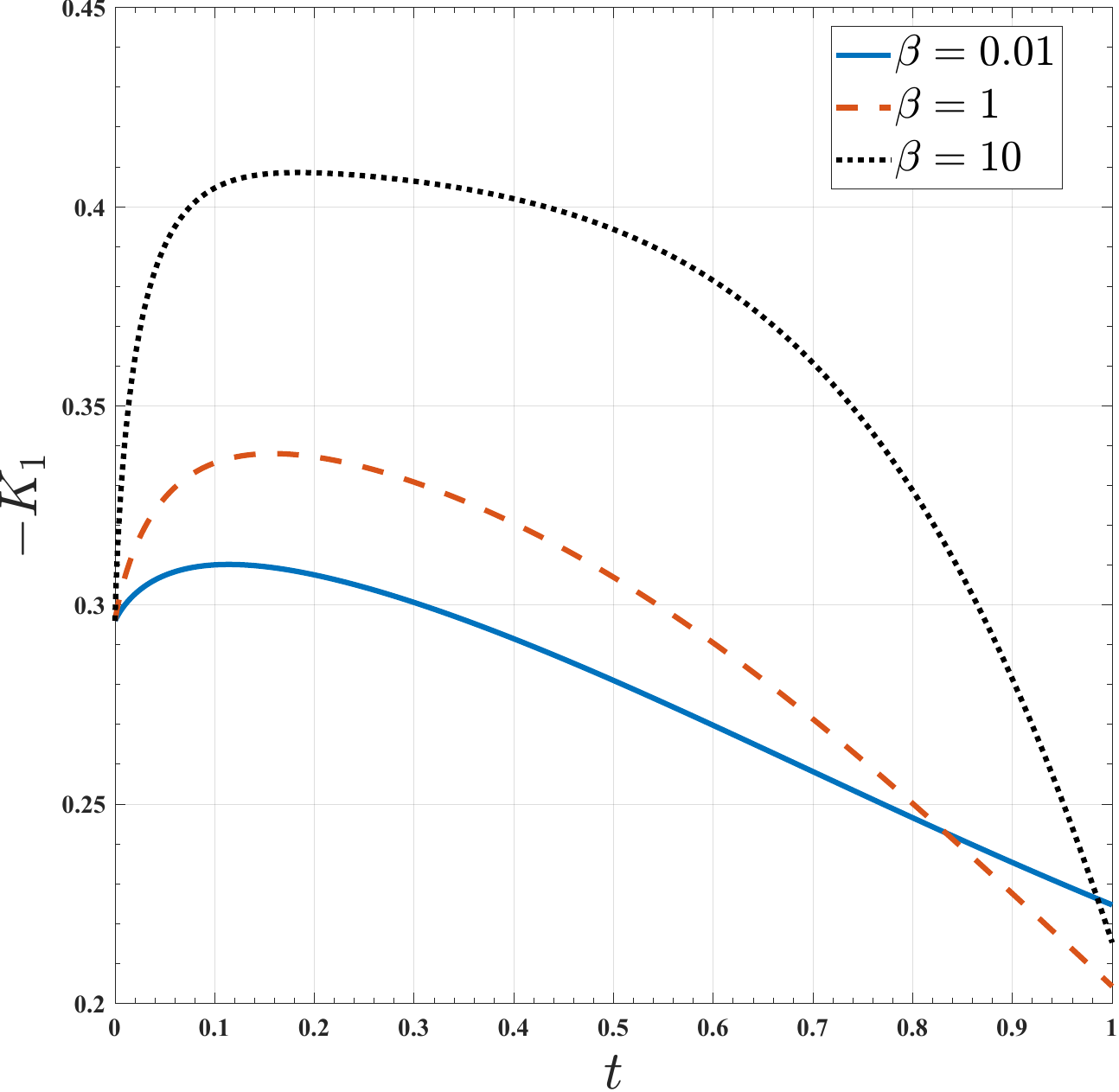}
    \caption{}
    \label{fig:8a}
\end{subfigure}
\begin{subfigure}{0.49\linewidth}
    \centering
    \includegraphics[width=\linewidth]{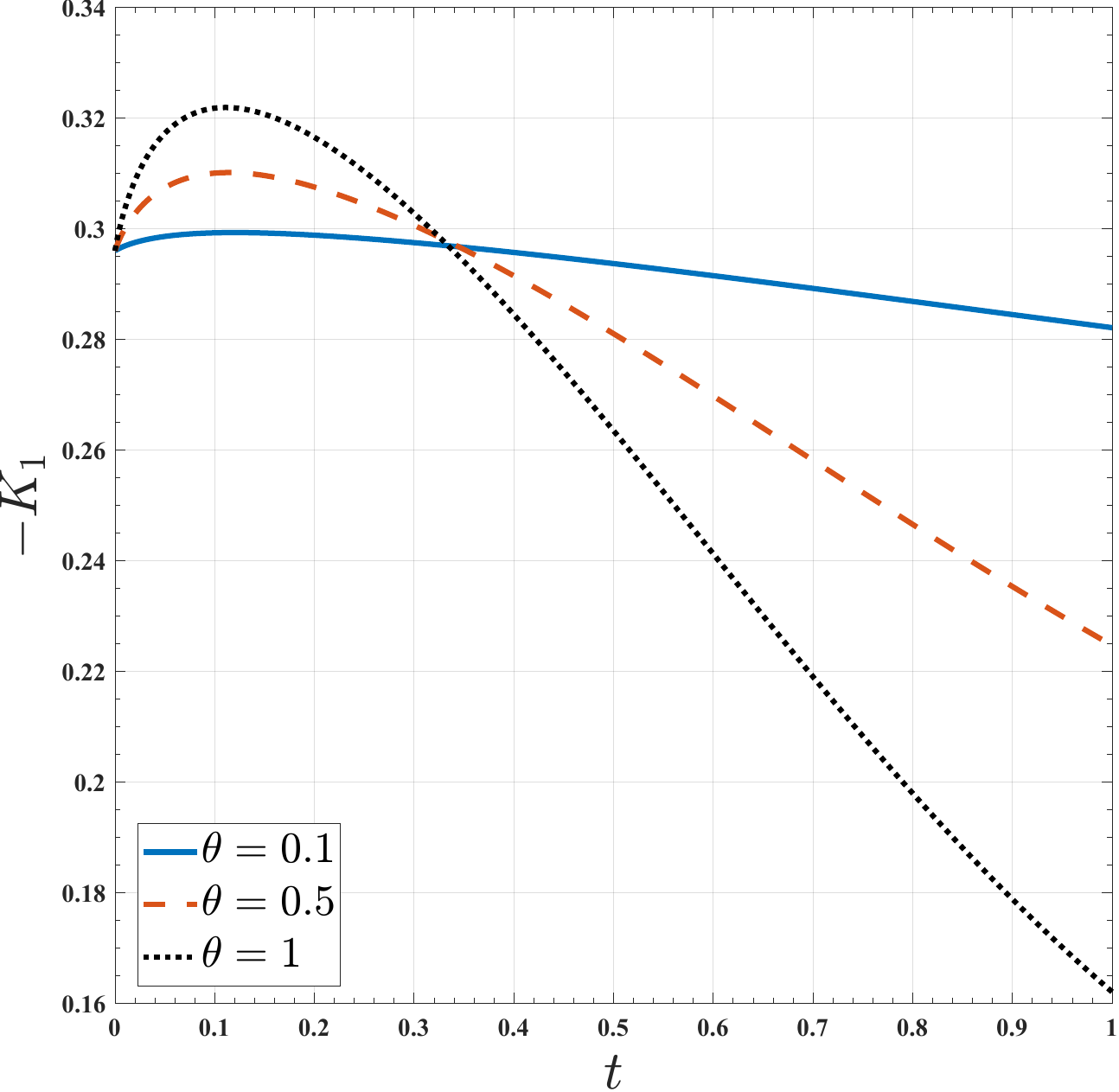}
    \caption{}
    \label{fig:8b}
\end{subfigure}
\begin{subfigure}{0.49\linewidth}
    \includegraphics[width=\linewidth]{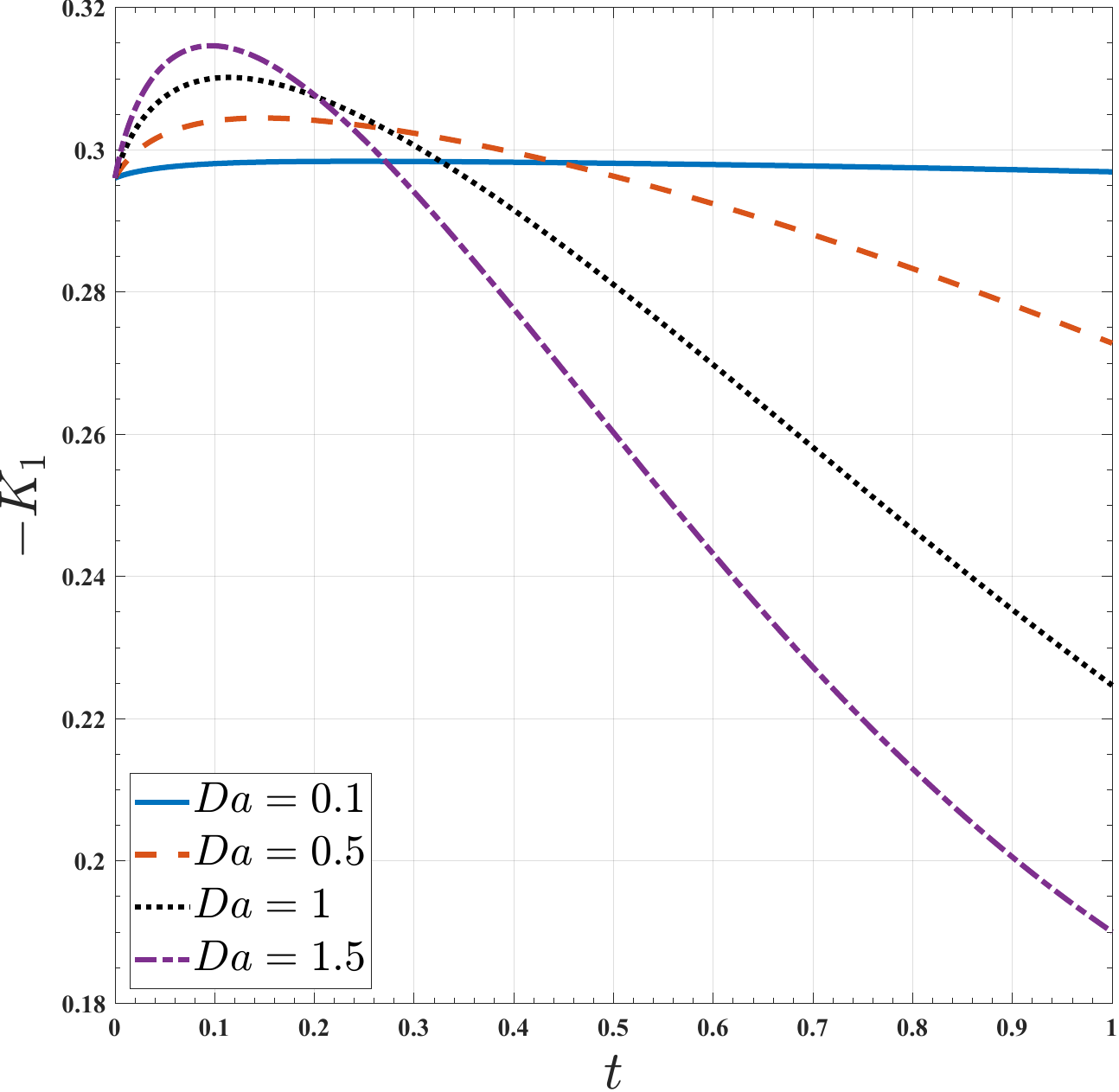}
    \caption{}
    \label{fig:8c}
    \end{subfigure}
\caption{Time evolution of the convection magnitude $-K_1$ for variations in (a) $\beta$ at fixed $\theta=0.5$, $Da=1$; (b) $\theta$ at fixed $\beta=0.01$, $Da=1$; and (c) $Da$ at fixed $\beta=0.01$, $\theta=0.5$. Other parameters are $m_p=1$, $N_c=0.5$, and $\lambda=0.01$.}
\label{fig:8}
\end{figure} 

The behavior of the convection coefficient shown in Figure~\ref{fig:8} is qualitatively different. Every
curve in every panel leaves the same starting point, $-K_1=0.296$, rises to a
maximum between $t\approx0.1$ and $0.24$, and then decreases. The overshoot is
barely visible for weak kinetics, about $1\%$ for $\theta=0.1$ or $Da=0.1$. But its magnitude grows with each of the three parameters, reaching $38\%$ above the initial value ($-K_1=0.409$) at $\beta=10$. The common initial value arises because when the solute is uniformly distributed, it has no choice but to travel at the discharge velocity of the annulus, giving 
\begin{equation}
-K_1(0)=\bar v 
    \label{eq:K1initial}
\end{equation}
identically. Any difference from that value is a measure of how far
the wall reaction has managed to redistribute the solute across the gap or hold it on the wall. The initial rise develops as this redistribution takes place: solute must first be stripped
from the wall region, and the resulting radial gradient must then be carried
across the annulus by molecular diffusion. The time required is the radial relaxation time of the annular cross section, which for $\lambda=0.01$ is $1/14.7\approx0.068$. This is the $(1-\lambda)^{2}$ diffusion time. The maxima are reached after $1.4$ to $3.5$ such times, once the radial structure is fully developed.

The decrease that follows the maximum is produced by the reversible reaction. Solute taken up by the wall does not move with the flow; when it is released, it re-enters the mobile phase behind the advancing cloud and next to the wall, where the fluid moves slowest. Both effects lower the centroid speed of the mobile phase. Once the retained store has grown sufficiently, $-K_1$ falls below $\bar v$ (after $t\approx0.35$ in the baseline case and after $t\approx0.29$ for $Da=1.5$). For weak absorption, as the exchange approaches local equilibrium, the cloud tends to the retarded speed
\[
-K_1\longrightarrow\frac{\bar v}{R},\qquad R=1+\frac{2\theta}{1-\lambda^{2}},
\]
where the retardation factor $R$ is the total (mobile plus retained) capacity per unit mobile capacity. For the baseline case $R\approx2$, and a longer computation gives $-K_1=0.147$ at $t=4$, compared with $\bar v/R=0.148$. The $\beta=10$ case shows the strongest decline rather than a plateau. Rapid absorption depletes the mobile phase so quickly that, by $t=1$, the retained solute exceeds the mobile solute by a factor of $2.8$, and the mobile cloud increasingly consists of solute recently released from the stationary store. Accordingly, $-K_1$ falls from its maximum of $0.409$ to $0.215$ at $t=1$. The $Da$ curves remain well separated at $t=1$, from $-K_1/\bar v=1.00$ at $Da=0.1$ to $0.64$ at $Da=1.5$: faster exchange builds the retained store sooner and therefore produces the retardation earlier.

\begin{figure}
    \centering
    \begin{subfigure}{0.49\linewidth}
    \includegraphics[width=\linewidth]{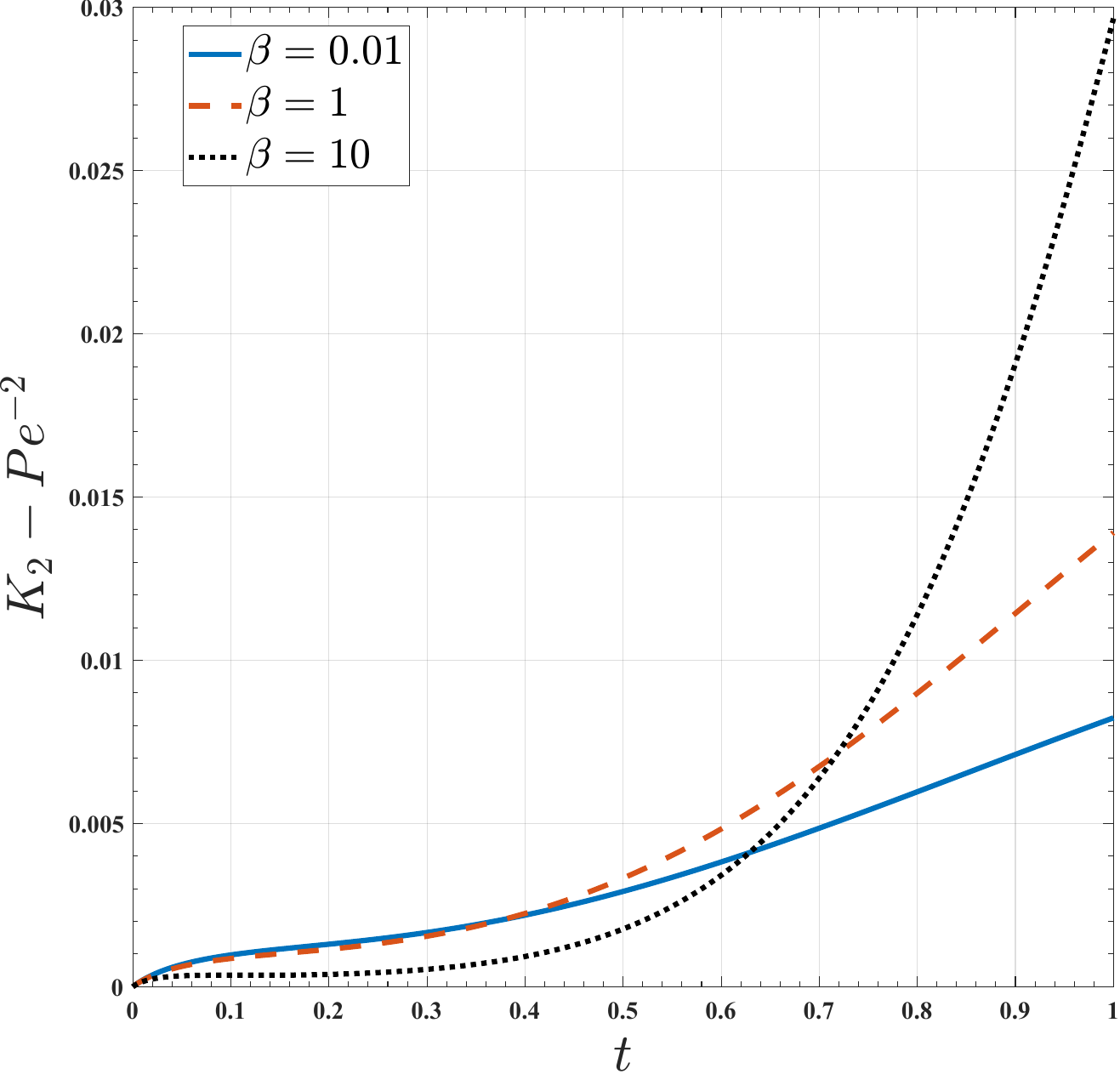}
    \caption{}
    \label{fig:9a}
\end{subfigure}
\begin{subfigure}{0.49\linewidth}
    \centering
    \includegraphics[width=\linewidth]{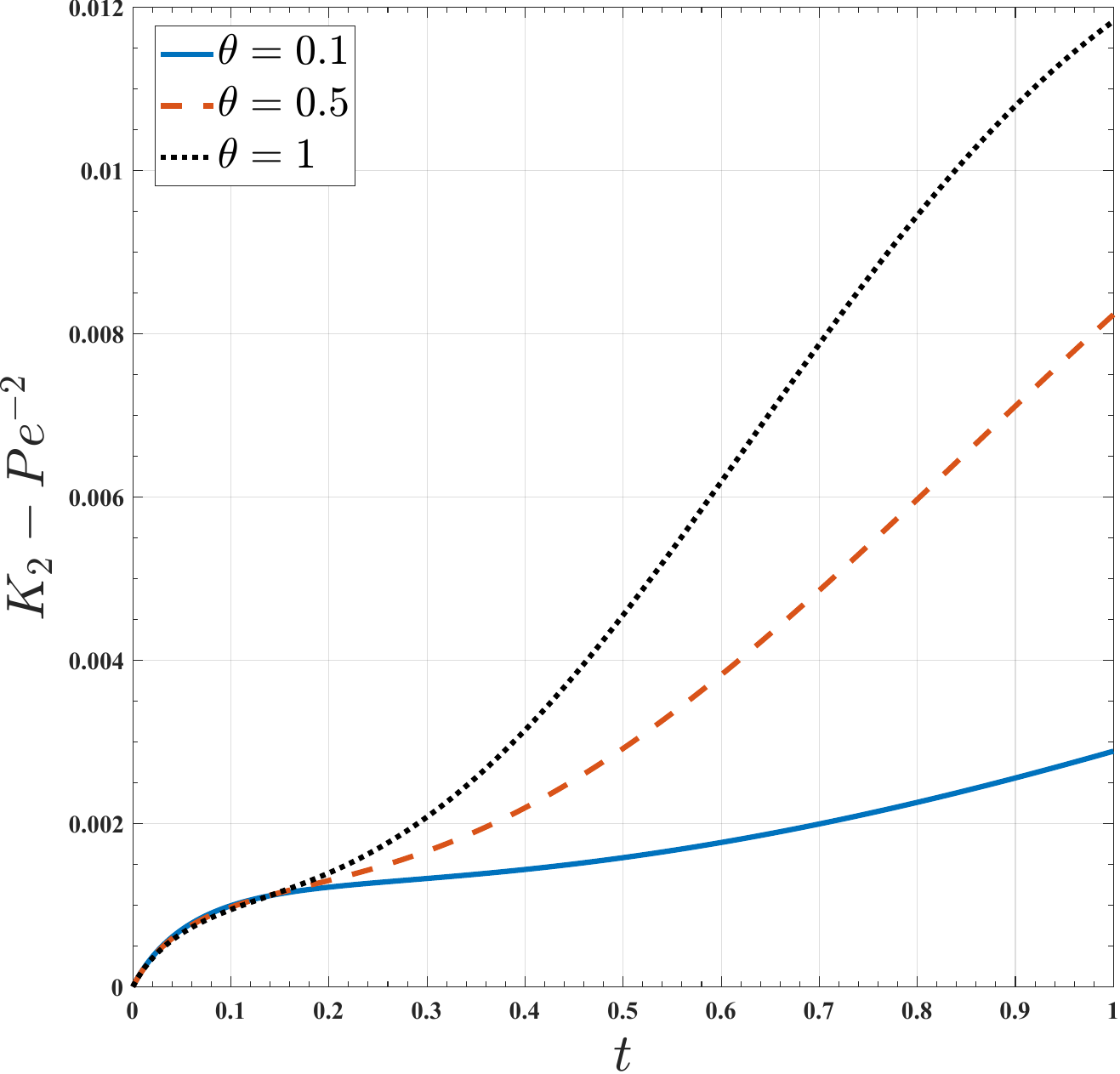}
    \caption{}
    \label{fig:9b}
\end{subfigure}
    \begin{subfigure}{0.49\linewidth}
    \includegraphics[width=\linewidth]{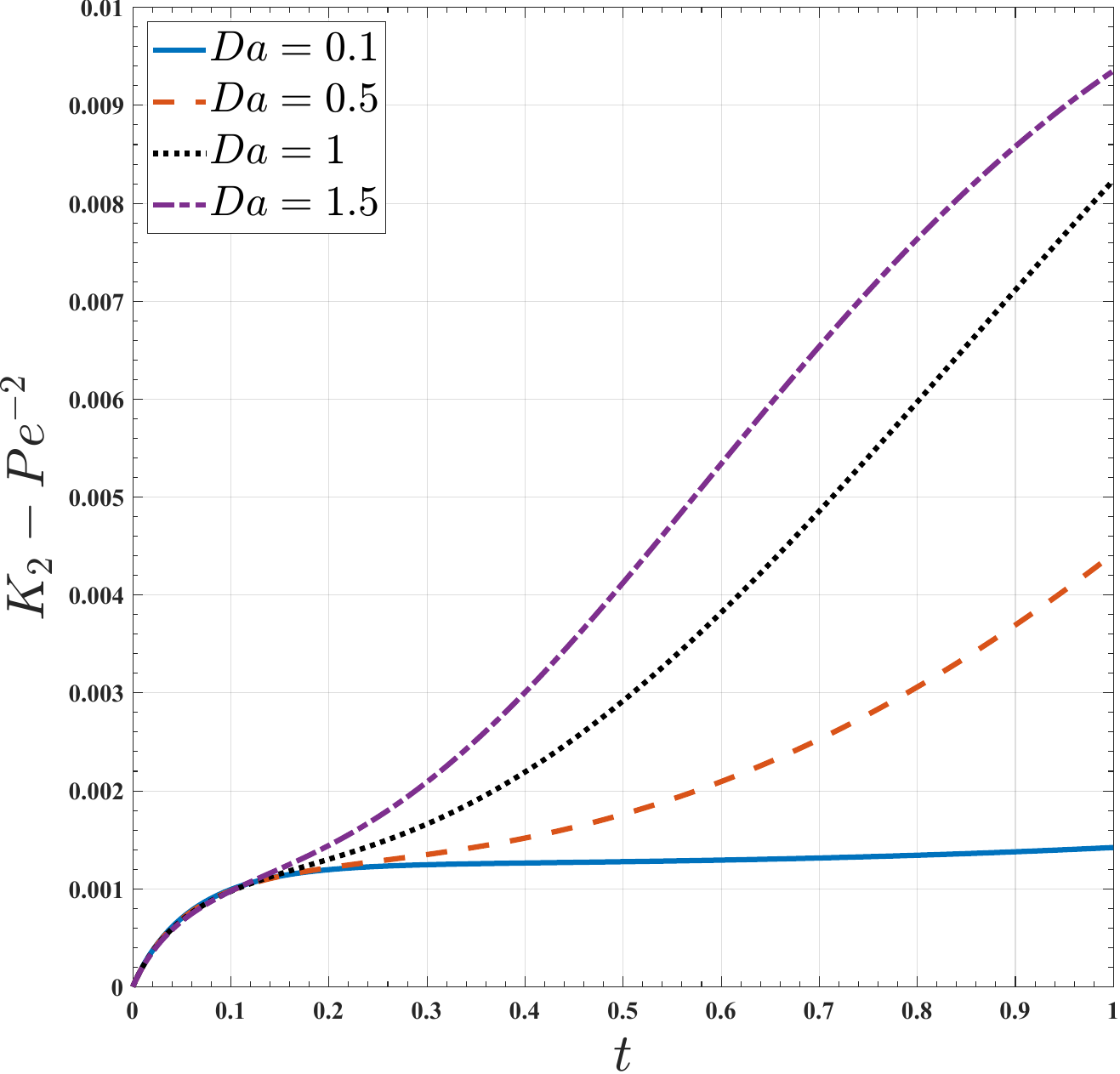}
    \caption{}
    \label{fig:9c}
\end{subfigure}
\caption{Time evolution of the shear-induced dispersion $K_2-Pe^{-2}$ for variations in (a) $\beta$ at fixed $\theta=0.5$, $Da=1$; (b) $\theta$ at fixed $\beta=0.01$, $Da=1$; and (c) $Da$ at fixed $\beta=0.01$, $\theta=0.5$. Other parameters are $m_p=1$, $N_c=0.5$, and $\lambda=0.01$.}
\label{fig:9}
\end{figure}

Figure~\ref{fig:9} isolates the shear-generated part of the dispersion. At release,
\begin{equation}
K_2(0)=Pe^{-2},
\end{equation}
because \(f_1=f_2=0\), so the initial shear-induced contribution is exactly zero even though axial molecular diffusion is already present. Linearizing Eq.~\eqref{eq:fn_hierarchy} about the uniform initial state, with \(f_0(0,r)=1\) and \(K_1(0)=-\bar v\), gives
\[
f_1(t,r)=-(v-\bar v)t+o(t),
\]
and hence
\begin{equation}
K_2-Pe^{-2}
=
\sigma_v^2\,t+o(t),
\qquad
\sigma_v^2
=
\left\langle(v-\bar v)^2\right\rangle
=
\langle v^2\rangle-\bar v^2 .
\label{eq:K2short}
\end{equation}
For the parameters of Figure~\ref{fig:9}, \(\sigma_v^2=2.35\times10^{-2}\). The initial slope is therefore a purely hydrodynamic quantity determined by the cross-sectional velocity variance and is independent of \(\beta\), \(\theta\), and \(Da\). This explains the initial collapse of the reaction-dependent curves, which persists while \((\beta+\theta Da)^2t\ll1\). Importantly, wall exchange is already active at \(t=0^+\), as Eq.~\eqref{eq:K0init} shows; it simply enters the velocity-weighted spreading at higher order than the leading \(O(t)\) shear-dispersion term.

As the transverse concentration structure develops, molecular diffusion progressively counteracts the gradients generated by differential advection and \(K_2-Pe^{-2}\) first approaches the Taylor level set by shear. The chemically passive steady solution gives \(K_2-Pe^{-2}=1.26\times10^{-3}\), providing a useful hydrodynamic reference; at \(t=0.3\) the reactive curves lie between \(0.53\times10^{-3}\) (\(\beta=10\)) and \(2.1\times10^{-3}\) (\(\theta=1\) or \(Da=1.5\)). The reactive curves do not level off at this value, however. Once the retained phase has filled appreciably, exchange with the wall adds a second, kinetically controlled contribution: solute held on the wall is left behind by the moving cloud and released later, which lengthens the mobile cloud. \(K_2-Pe^{-2}\) therefore rises again, to \(8.24\times10^{-3}\) at \(t=1\) in the baseline case, and a longer computation shows that it reaches a plateau of about \(1.5\times10^{-2}\) only near \(t\approx3\), when the exchange has approached local equilibrium. This second stage explains the late-time ordering. Increasing \(\theta\) enlarges the retained store (\(2.89\), \(8.24\), and \(11.8\times10^{-3}\) at \(t=1\) for \(\theta=0.1\), \(0.5\), and \(1\)), and increasing \(Da\) shortens the time needed to build it (\(1.42\), \(4.42\), \(8.24\), and \(9.35\times10^{-3}\) for \(Da=0.1\), \(0.5\), \(1\), and \(1.5\)); for \(Da=0.1\) the second stage has barely begun by \(t=1\). The effect of \(\beta\) reverses in time. At \(t=0.3\) stronger absorption lowers the dispersion (\(1.66\), \(1.55\), and \(0.53\times10^{-3}\) for \(\beta=0.01\), \(1\), and \(10\)), as it would for absorption alone, but by \(t=1\) the order is inverted (\(8.24\times10^{-3}\), \(1.39\times10^{-2}\), and \(2.97\times10^{-2}\)), because faster depletion leaves a larger share of the remaining solute in the retained phase (\(\Phi_s/\Phi_m=0.71\), \(1.08\), and \(2.8\)). The differing late-time responses of \(\beta\), \(\theta\), and \(Da\) therefore reflect how permanent removal, equilibrium retention, and exchange rate control the size and growth rate of the retained store after the universal short-time regime.

\begin{figure}
    \centering
    \begin{subfigure}{0.49\linewidth}
    \includegraphics[width=\linewidth]{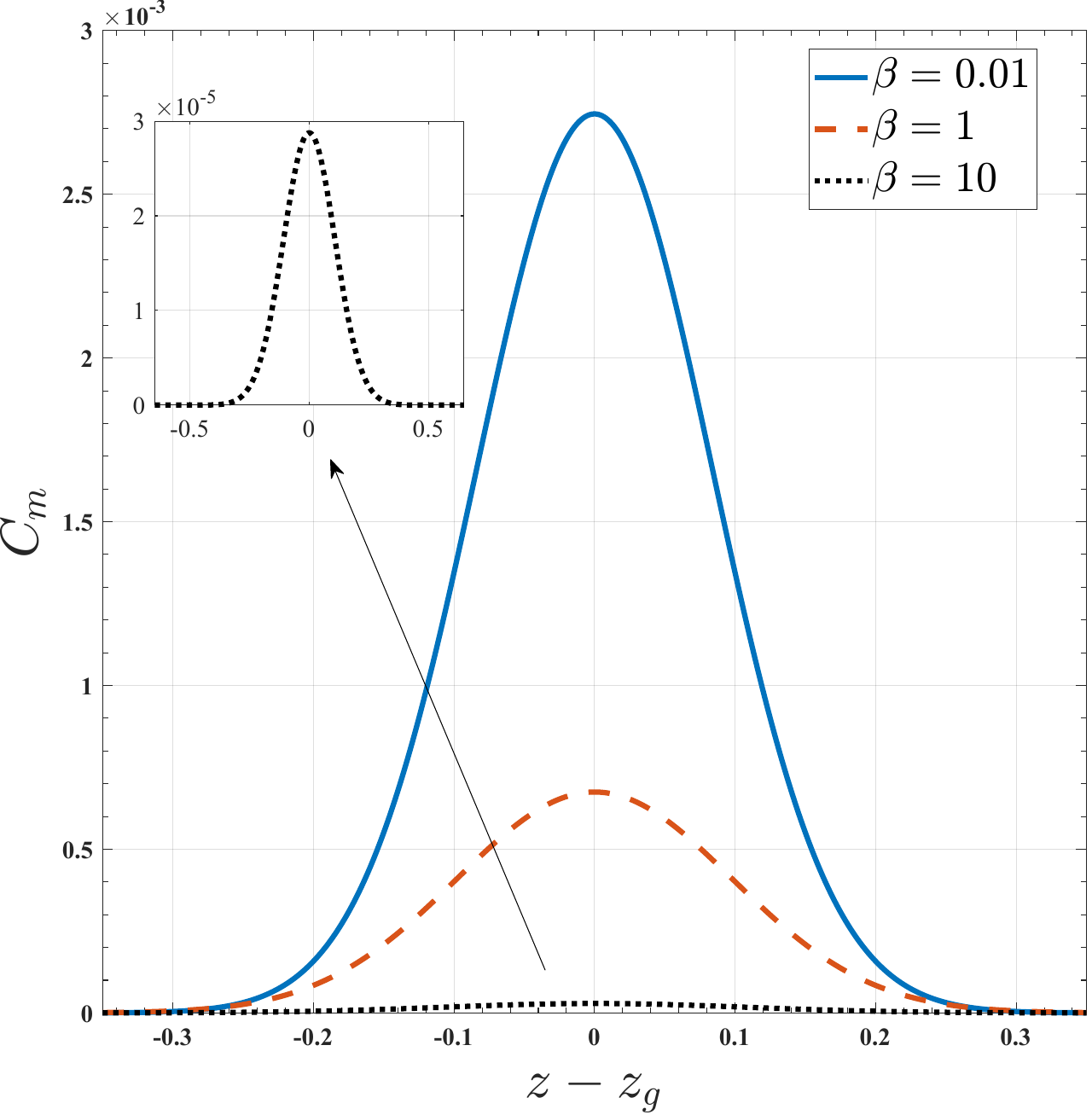}
    \caption{}
    \label{fig:10a}
\end{subfigure}
\begin{subfigure}{0.49\linewidth}
    \centering
    \includegraphics[width=\linewidth]{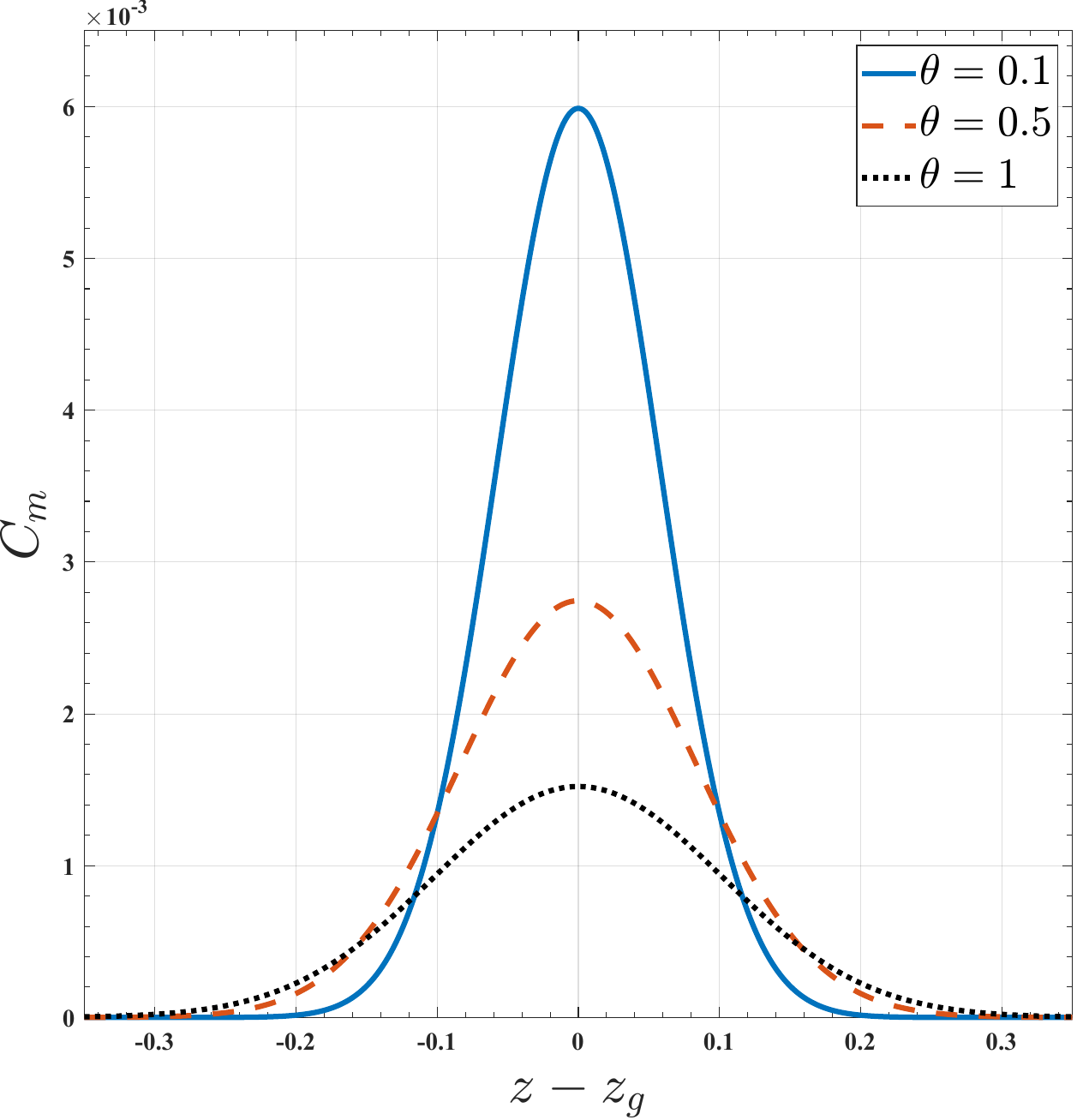}
    \caption{}
    \label{fig:10b}
\end{subfigure}
    \begin{subfigure}{0.49\linewidth}
    \includegraphics[width=\linewidth]{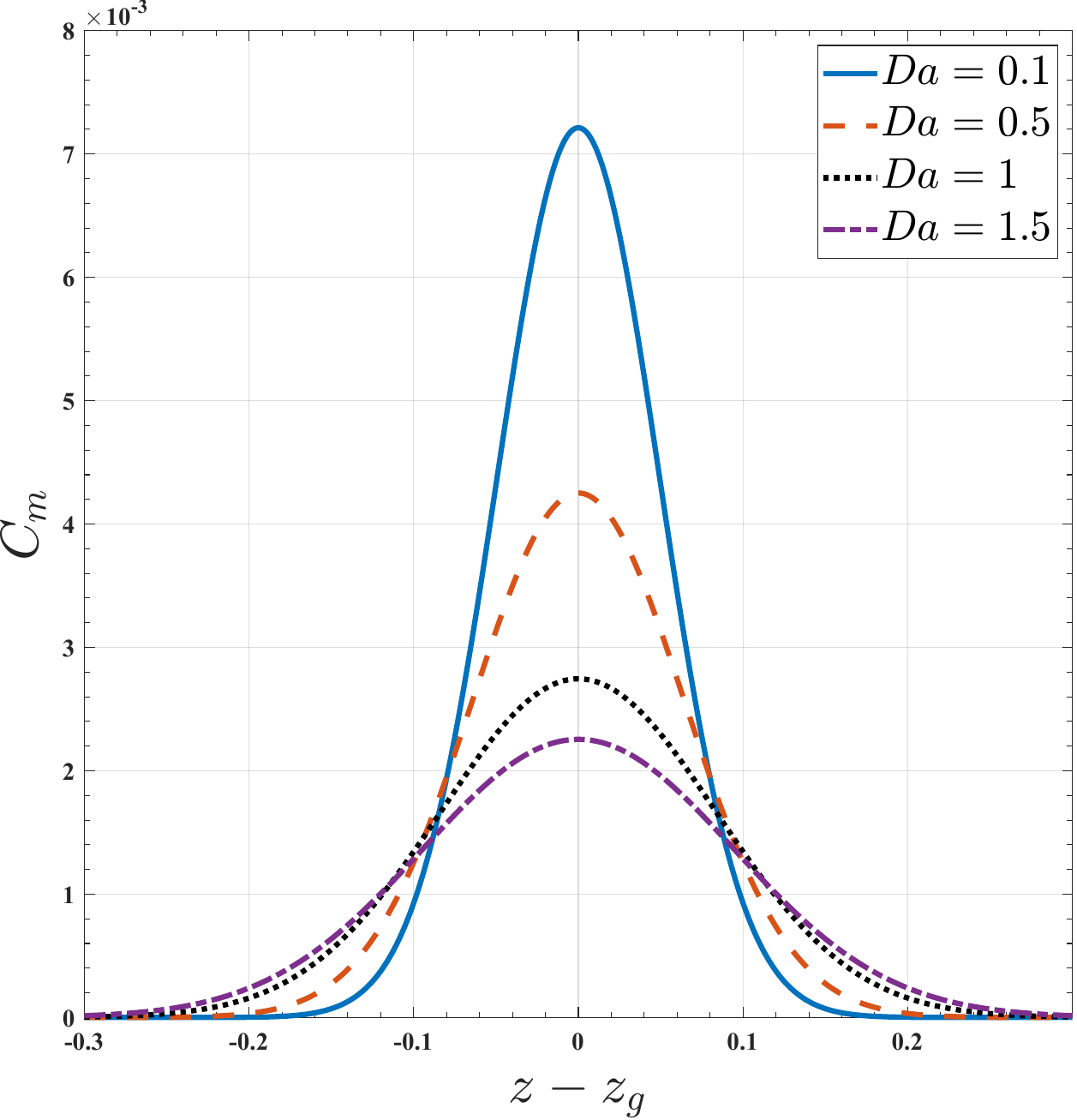}
    \caption{}
    \label{fig:10c}
\end{subfigure}
\caption{Sectional mean concentration $C_m$ at $t=1$ for variations in (a) $\beta$ at fixed $\theta=0.5$, $Da=1$; (b) $\theta$ at fixed $\beta=0.01$, $Da=1$; and (c) $Da$ at fixed $\beta=0.01$, $\theta=0.5$. Other parameters are $m_p=1$, $N_c=0.5$, $\lambda=0.01$, and $Pe=1000$.}
\label{fig:10}
\end{figure}

Figure~\ref{fig:10} shows how the coefficient dynamics combine in the sectional mean cloud at \(t=1\). Because the profiles are centred at \(z_g\), their displayed shape is controlled by the surviving mobile mass \(\mathcal M(t)\) and the accumulated dispersion \(\mathcal D(t)\), rather than by the centroid translation. Increasing any of the three reaction parameters lowers the mobile-phase peak in the calculations shown. The effect of \(\beta\) is strongest: the peak decreases from \(2.75\times10^{-3}\) at \(\beta=0.01\) to \(6.75\times10^{-4}\) at \(\beta=1\), and to approximately \(2.9\times10^{-5}\) at \(\beta=10\). Increasing \(\theta\) from \(0.1\) to \(1\) lowers the peak from \(6.0\times10^{-3}\) to \(1.5\times10^{-3}\), while increasing \(Da\) over the displayed range lowers it from \(7.2\times10^{-3}\) to \(2.25\times10^{-3}\). The underlying mechanisms differ. For \(\beta\), the reduction is almost entirely a loss of mobile mass: \(\Phi_m\) falls from \(0.576\) to \(0.0078\), while the cloud width, proportional to \(\mathcal D^{1/2}\), grows by only \(29\%\). For \(\theta\) and \(Da\), the width changes are comparable to the mass changes: across the displayed ranges the width grows by factors of \(1.77\) and \(1.91\), while \(\Phi_m\) falls by factors of \(2.2\) and \(1.7\). Reversible retention therefore lowers the peak both by storing solute and by spreading the mobile cloud axially, which is the sectional-mean signature of the retention-induced dispersion in Figure~\ref{fig:9}.

Taken together, Figures~\ref{fig:7}--\ref{fig:10} separate three distinct aspects of the transient transport. The coefficient \(-K_0\) measures mobile-phase depletion, \(-K_1\) measures the concentration-weighted centroid speed, and \(K_2-Pe^{-2}\) measures the growth of axial variance generated by shear beyond molecular diffusion. The wall parameters affect these quantities on different timescales. Their influence, therefore, cannot be represented by a single effective reaction rate. The mass partition considered next resolves this distinction further by separating reversible storage from irreversible loss.

\subsection{Partition of mobile, retained and absorbed solute}
\label{subsec:masspartition}

The transport coefficients characterize the mobile cloud, whereas the mass
fractions in Eq.~\eqref{eq:mass_fractions} resolve where the solute resides.
Here $\Phi_m$ is the fraction still available for downstream transport,
$\Phi_s$ is the fraction stored reversibly at the arterial wall, and
$\Phi_a$ is the cumulative irreversible uptake. This decomposition therefore,
separates two wall processes that are combined in the net exchange coefficient
$K_0$.

\begin{figure}
    \centering
    \IfFileExists{eps/mass_partition.pdf}
    {\includegraphics[width=0.72\linewidth]{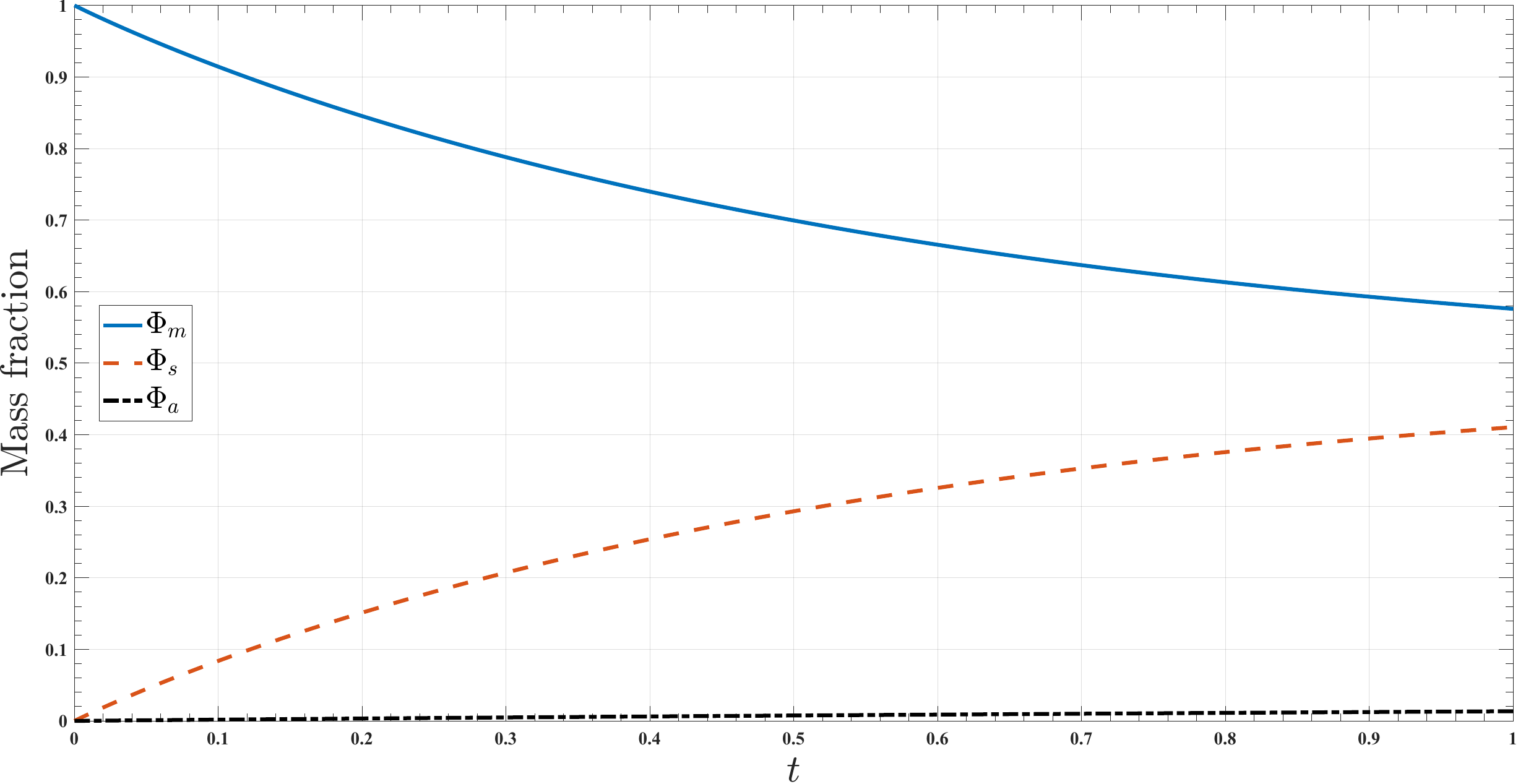}}
    {\fbox{\parbox[c][5.0cm][c]{0.70\linewidth}{\centering
    Placeholder for \(\Phi_m(t)\), \(\Phi_s(t)\), and \(\Phi_a(t)\).\\[0.5em]
    Baseline: \(m_p=1\), \(N_c=0.5\), \(\lambda=0.01\),
    \(\beta=0.01\), \(\theta=0.5\), \(Da=1\).}}}
    \caption{Time evolution of the fractions of the initially injected solute
    remaining mobile, $\Phi_m$, reversibly retained at the arterial wall,
    $\Phi_s$, and irreversibly absorbed, $\Phi_a$, for $m_p=1$, $N_c=0.5$,
    $\lambda=0.01$, $\beta=0.01$, $\theta=0.5$, $Da=1$, and $Pe=1000$.
    The three fractions are evaluated independently and satisfy
    $\Phi_m+\Phi_s+\Phi_a=1$ to numerical accuracy.}
    \label{fig:masspartition}
\end{figure}

Figure~\ref{fig:masspartition} makes the distinction between reversible
storage and permanent removal explicit. The mobile fraction decreases
monotonically as solute is transferred to the wall, while the retained
fraction rises throughout the displayed interval, showing that surface loading
dominates desorption over $0\le t\le1$ for this baseline case. The absorbed
fraction is also monotone, as required by Eq.~\eqref{eq:mass_fractions}, but
remains much smaller because $\beta=0.01$ represents weak irreversible uptake.
At $t=1$, $\Phi_m=0.576$, $\Phi_s=0.411$, and $\Phi_a=0.0135$; thus, reversible
storage accounts for about $41\%$ of the initially injected mass, whereas only
about $1.35\%$ has been permanently absorbed. The size of the retained store
relative to the mobile phase is what controls the retardation and the late-time
dispersion discussed in the previous subsection. The retained solute exceeds the
mobile solute at $t=1$ for $\beta=1$ ($\Phi_s/\Phi_m=1.08$), $\theta=1$ ($1.53$),
and $\beta=10$ ($2.8$), the last being the case in which the mobile cloud is
increasingly supplied by desorption and $-K_1$ declines most strongly. The independently evaluated
fractions satisfy Eq.~\eqref{eq:mass_conservation} with a maximum residual of
$3.4\times10^{-9}$, providing a stringent global check on the coupled
bulk--surface calculation.

\subsection{Spatial structure of the solute cloud}
\label{subsec:spatial}

The transport coefficients compress the cross-sectional dynamics into
one-dimensional quantities, whereas the reconstruction in
Eq.~\eqref{eq:local_reconstruction} retains the local radial structure. To
isolate the second-order contribution generated by differential advection,
we define
\begin{equation}\label{eq:Cdisp_definition}
C_{\mathrm{disp}}(t,z,r)
=
f_1(t,r)C_{m,z}(t,z)
+
f_2(t,r)C_{m,zz}(t,z),
\end{equation}
so that the second-order reconstruction is
\(C\approx f_0C_m+C_{\mathrm{disp}}\). This definition avoids interpreting
\(C_{\mathrm{disp}}\) as a separate species: it is the signed local correction
to the reaction-modified leading field. Because
\(\langle f_1\rangle=\langle f_2\rangle=0\), it also satisfies
\(\langle C_{\mathrm{disp}}\rangle=0\) and therefore redistributes solute
without changing the sectional mean.

Writing \(\xi=z-z_g(t)\), the Gaussian field in
Eq.~\eqref{eq:Cm_green} gives
\begin{equation}\label{eq:Cdisp_normalized}
\frac{C_{\mathrm{disp}}(t,z,r)}{C_m(t,z_g)}
=
\exp\left(-\frac{\xi^2}{4\mathcal D}\right)
\left[
-\frac{\xi}{2\mathcal D}f_1
+
\left(
\frac{\xi^2}{4\mathcal D^2}
-\frac{1}{2\mathcal D}
\right)f_2
\right],
\qquad \xi=z-z_g(t).
\end{equation}
The two contributions in Eq.~\eqref{eq:Cdisp_normalized} have distinct axial symmetry: the \(f_1C_{m,z}\) term is antisymmetric about \(z=z_g\), whereas the \(f_2C_{m,zz}\) term is symmetric. Their superposition therefore explains the paired positive-negative lobes that appear in the reconstructed field. For comparisons between different catheter radii, we use the normalized gap
coordinate
\begin{equation}\label{eq:eta_coordinate}
\eta=\frac{r-\lambda}{1-\lambda},
\qquad 0\leq\eta\leq1,
\end{equation}
so that \(\eta=0\) denotes the catheter surface and \(\eta=1\) the reactive
arterial wall.

\begin{figure}
    \centering
    \IfFileExists{eps/Cdisp_lambda_time.pdf}
    {\includegraphics[width=\linewidth]{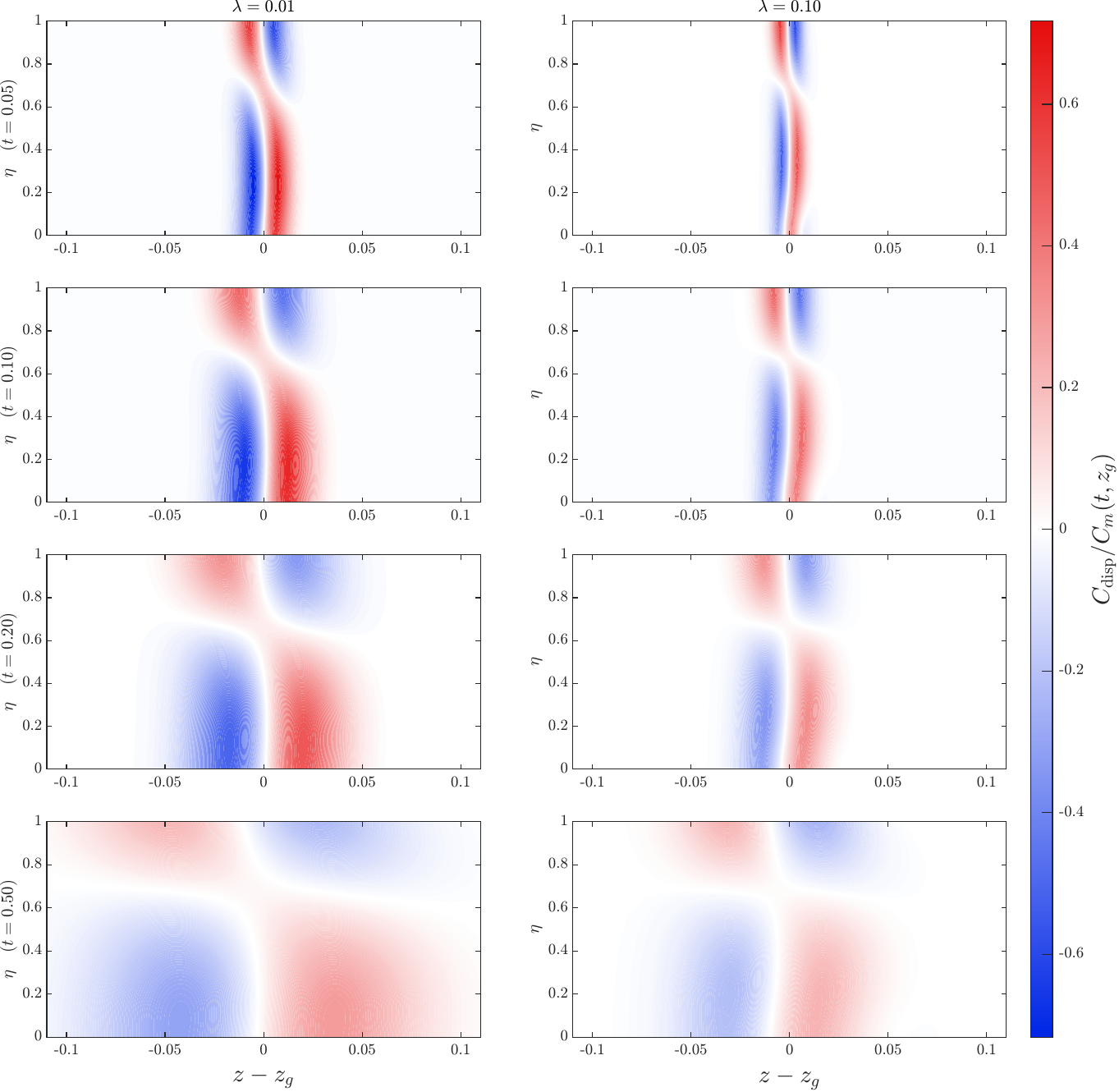}}
    {\fbox{\parbox[c][10.0cm][c]{0.94\linewidth}{\centering
    Placeholder for a $4\times2$ signed heat-map figure of
    $C_{\mathrm{disp}}/C_m(t,z_g)$.\\[0.6em]
    Rows: $t=0.05,\,0.10,\,0.20,\,0.50$.\\
    Columns: $\lambda=0.01$ and $0.10$.\\
    Vertical coordinate: $\eta=(r-\lambda)/(1-\lambda)$.\\
    A common zero-centred colour scale is used for all panels.}}}
    \caption{Signed dispersive correction
    $C_{\mathrm{disp}}/C_m(t,z_g)$ in the centred $(z-z_g,\eta)$-plane. Rows correspond to $t=0.05$, $0.10$,
    $0.20$, and $0.50$, while the columns compare $\lambda=0.01$ and $0.10$. Other parameters are $m_p=1$, $N_c=0.5$,
    $\beta=0.01$, $\theta=0.5$, $Da=1$, and $Pe=1000$.}
    \label{fig:Cdisp}
\end{figure}

Positive and negative values in Figure~\ref{fig:Cdisp} represent local
enrichment and depletion, respectively, relative to the leading
reaction-modified profile $f_0C_m$. Because
$\langle C_{\mathrm{disp}}\rangle=0$, the opposing lobes redistribute solute
within the cross-section without changing the sectional mean. At early times,
the correction is concentrated close to the cloud centroid, reflecting the
initial cross-sectional segregation generated by differential advection. As time
increases, transverse diffusion broadens the signed structure and reduces its
peak magnitude while the axial extent grows. Comparison of the two columns
shows that increasing the catheter ratio from $\lambda=0.01$ to $0.10$
suppresses the correction at each displayed time. This local attenuation is
consistent with the simultaneous reduction of $K_2-Pe^{-2}$ in the global
dispersion measure, and provides a field-level interpretation of how catheter
confinement weakens shear-induced redistribution.

Peak-to-peak transverse-variation diagnostics have been used to quantify departures
from cross-sectional uniformity in multiscale dispersion analyses
\citep{wu2014jfm,poddar2021rsa}. To complement the signed correction in Figure~\ref{fig:Cdisp} with a scalar
measure of the total radial heterogeneity of the reconstructed cloud, we adapt
this diagnostic to the present annular geometry and defines the transverse non-uniformity
\begin{equation}\label{eq:transverse_nonuniformity}
\mathcal N_{\perp}(\xi,t)
=
\frac{
\displaystyle \max_{\lambda\le r\le1} C(t,z_g+\xi,r)
-
\displaystyle \min_{\lambda\le r\le1} C(t,z_g+\xi,r)
}{
C_m(t,z_g)
},
\qquad
\xi=z-z_g(t).
\end{equation}
The normalization by the peak sectional mean is convenient in an annulus,
where there is no fluid centreline that can provide a unique local reference
concentration. Thus \(\mathcal N_{\perp}=0\) corresponds to a radially uniform
reconstructed concentration at that axial location, whereas larger values
measure stronger cross-sectional segregation relative to the peak sectional
mean. Values exceeding unity are admissible because the numerator is a
peak-to-peak radial concentration difference rather than a concentration
fraction. As with Eq.~\eqref{eq:local_reconstruction},
\(\mathcal N_{\perp}\) is interpreted at the order of the retained
second-order generalized-dispersion reconstruction.

\begin{figure}
    \centering
    \begin{subfigure}{0.49\linewidth}
        \centering
        \includegraphics[width=\linewidth]{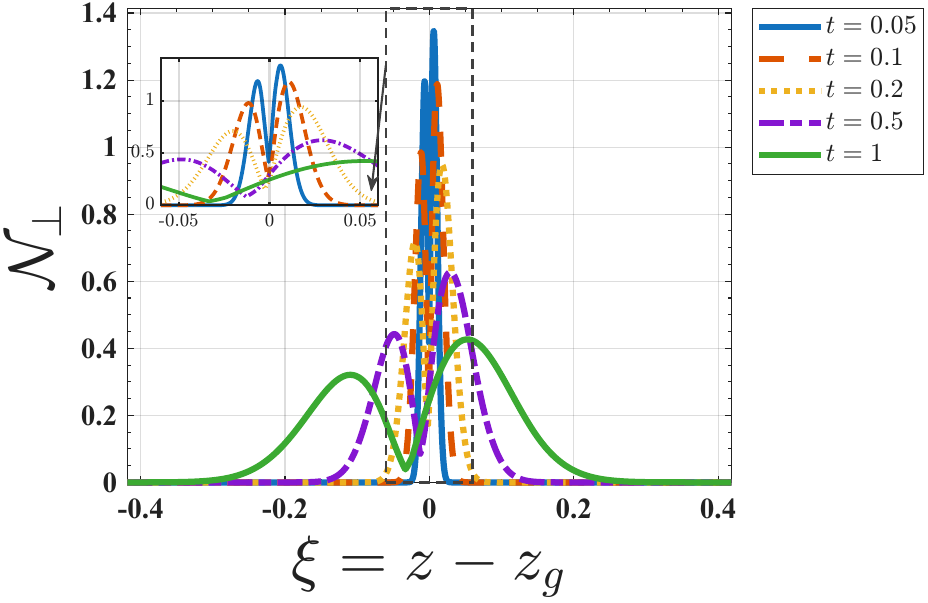}
        \caption{}
        \label{fig:nonuniformity_a}
    \end{subfigure}
    \begin{subfigure}{0.49\linewidth}
        \centering
        \includegraphics[width=\linewidth]{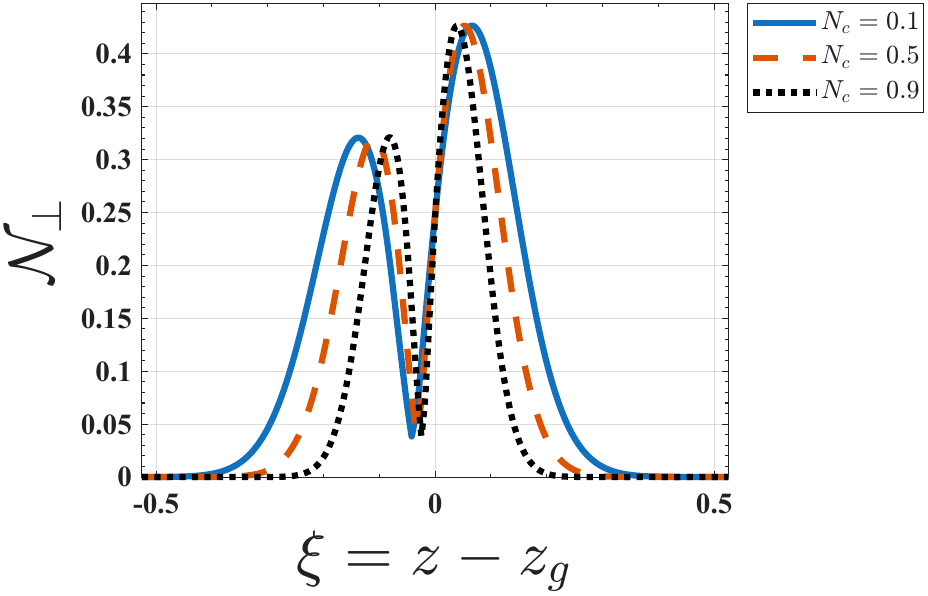}
        \caption{}
        \label{fig:nonuniformity_b}
    \end{subfigure}
    \begin{subfigure}{0.49\linewidth}
        \centering
        \includegraphics[width=\linewidth]{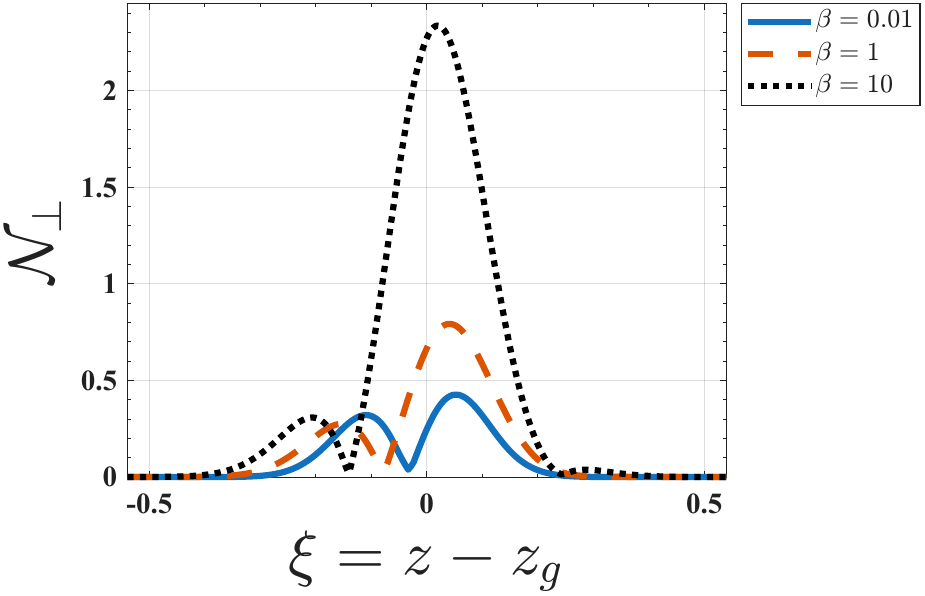}
        \caption{}
        \label{fig:nonuniformity_c}
    \end{subfigure}
    \begin{subfigure}{0.49\linewidth}
        \centering
        \includegraphics[width=\linewidth]{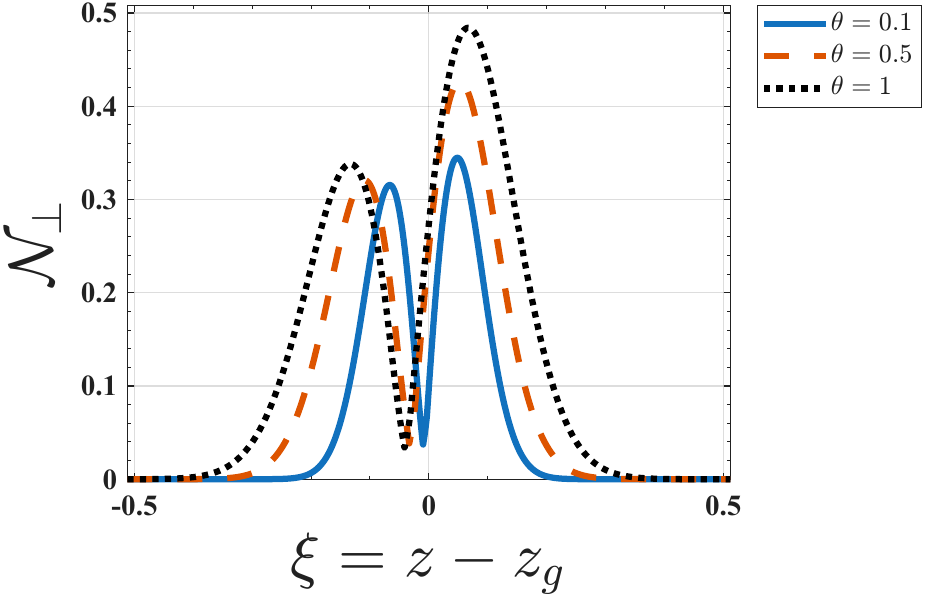}
        \caption{}
        \label{fig:nonuniformity_d}
    \end{subfigure}
    \caption{Transverse non-uniformity \(\mathcal N_{\perp}\) of the
    reconstructed mobile concentration. (a) Temporal evolution for
    \(t=0.05,\,0.10,\,0.20,\,0.50,\) and \(1\) at the baseline parameter
    values; the inset magnifies the near-centroid region \(|\xi|\leq0.06\).
    At \(t=1\), panels (b)--(d) show the effects of (b) the coupling
    number \(N_c\), (c) irreversible uptake \(\beta\), and (d) reversible
    retention \(\theta\). Unless varied in a panel, the parameters are
    \(m_p=1\), \(N_c=0.5\), \(\lambda=0.01\), \(\beta=0.01\),
    \(\theta=0.5\), \(Da=1\), and \(Pe=1000\).}
    \label{fig:nonuniformity}
\end{figure}

Figure~\ref{fig:nonuniformity_a} shows that the strongest transverse
segregation is an early-time effect. The inset resolves the tightly clustered early-time profiles in the immediate
neighbourhood of the moving centroid. The maximum of
\(\mathcal N_{\perp}\) decreases from approximately \(1.35\) at \(t=0.05\)
to \(0.426\) at \(t=1\), while the principal upstream and downstream lobes
move away from the centroid as the cloud spreads. This behavior complements
Figure~\ref{fig:Cdisp}: transverse diffusion progressively weakens the local
radial contrast, but the region over which that contrast is distributed
broadens in the centred axial coordinate.

The flow and reaction parameters affect this heterogeneity in different ways.
In Figure~\ref{fig:nonuniformity_b}, increasing \(N_c\) from \(0.1\) to
\(0.9\) leaves the peak magnitude almost unchanged
(\(\mathcal N_{\perp,\max}\simeq0.426\)) but moves the downstream peak from
\(\xi\simeq0.066\) to \(0.039\) and the upstream peak from
\(\xi\simeq-0.138\) to \(-0.081\). Thus, the principal rheological effect is
to contract the axial extent of the cross-sectional segregation rather than
to alter its maximum strength. The irreversible reaction produces a much
stronger response: Figure~\ref{fig:nonuniformity_c} shows that the maximum
non-uniformity increases from about \(0.426\) at \(\beta=0.01\) to \(0.793\)
at \(\beta=1\), and exceeds \(2\) at \(\beta=10\). Strong uptake therefore,
creates a pronounced relative radial contrast by preferentially depleting
the near-wall mobile phase. Reversible retention has a weaker but systematic
effect. In Figure~\ref{fig:nonuniformity_d},
\(\mathcal N_{\perp,\max}\) increases from approximately \(0.345\) to
\(0.484\) as \(\theta\) increases from \(0.1\) to \(1\), showing that greater
surface storage also enhances the internal cross-sectional structure of the
mobile cloud. These comparisons distinguish changes in local transverse
heterogeneity from the changes in mobile mass and axial spreading measured
by \(K_0\), \(K_2\), and \(\Phi_m\).

\FloatBarrier

\section{Conclusions}
\label{sec:conclusion}

Transient solute transport in a catheterized micropolar annulus has been
analysed with an impermeable inner catheter and an outer arterial wall that
combines irreversible uptake with reversible surface retention. Within the
Gill--Sankarasubramanian generalized-dispersion framework
\citep{gill1970royal,sankara1973royal}, the formulation retains an explicit
surface concentration and consistently couples the mobile and retained phases.
The resulting bulk--surface hierarchy is advanced numerically using
Rannacher-damped Crank--Nicolson time stepping, thereby resolving the transient
exchange, convection, and dispersion coefficients together with the evolving
bulk--surface mass partition.

A structural consequence of the governing hierarchy is that the exchange coefficient \(K_0\) is independent of the micropolar parameters \(N_c\) and \(m_p\), whereas \(K_1\) and \(K_2\) inherit the rheology through the axial velocity field. The initial mobile-phase depletion rate is
\[
-K_0(0^+)=\frac{2}{1-\lambda^2}(\beta+\theta Da),
\]
and it remains close to this value only while \((\beta+\theta Da)^2t\ll1\), after which a depleted wall layer and the filling of the retained phase reduce it.
The shear-induced contribution to dispersion develops according to
\[
K_2-Pe^{-2}\sim\sigma_v^2t
\qquad (t\to0^+).
\]
Thus, wall kinetics act immediately on mobile-phase mass, whereas the leading short-time growth of shear dispersion is governed by the cross-sectional velocity variance.

Micropolarity modifies transport primarily through the velocity amplitude. In the weak-reaction regime, the limiting velocity factor \(A\) gives the approximate scalings \(-K_1\propto A\) and \(K_2-Pe^{-2}\propto A^2\), explaining the greater sensitivity of shear dispersion to microrotational coupling. Catheter confinement produces an even stronger geometric effect under the fixed-pressure-gradient protocol: increasing \(\lambda\) from \(0.01\) to \(0.30\) reduces \(-K_1\) by a factor of about \(2.45\) and \(K_2-Pe^{-2}\) by a factor of about \(6\) in the baseline reactive case; for a chemically passive wall, the shear-induced dispersion falls by a factor of \(24\) over the same interval. In the narrow-gap limit,
\[
\bar v\sim\frac{2-N_c}{6}(1-\lambda)^2,
\qquad
K_2-Pe^{-2}\sim
\frac{(2-N_c)^2}{7560}(1-\lambda)^6 ,
\]
so the classical plane-channel factor \(1/210\) is recovered in the equivalent form \(\bar v^2(1-\lambda)^2/210\), while the fixed-pressure-gradient catheter problem exhibits a sixth-power collapse of shear dispersion. The numerical cell problem independently recovers both the exponent and the prefactor of this asymptotic law.

The two wall processes play distinguishable roles in the transient response. Irreversible uptake depletes near-wall mobile solute permanently, whereas reversible retention governs temporary storage and return to the flow. Retention retards the cloud: after a brief overshoot, \(-K_1\) falls below \(\bar v\) and, for weak absorption, tends to the retarded speed \(\bar v/R\), with \(R=1+2\theta/(1-\lambda^2)\). The dispersion coefficient grows in two stages, first towards the Taylor level set by shear and then, as the retained phase fills, through retention-induced spreading, which at \(t=1\) exceeds the inert Taylor value by a factor of \(6.5\) in the baseline case. Absorption acting alone raises \(-K_1\) and lowers the dispersion coefficient by a factor of four between \(\beta=0.01\) and \(10\); with retention present, the effect of \(\beta\) on dispersion reverses in time, and strong uptake leaves the mobile cloud supplied mainly by desorption, so that for \(\beta=10\) and larger catheter ratios the convection and dispersion coefficients become negative. The exact mass partition \(\Phi_m+\Phi_s+\Phi_a=1\) separates these contributions into mobile, reversibly retained, and irreversibly absorbed fractions and provides an independent conservation check on the surface coupling and the time integration.

The reconstructed signed dispersive field shows directly how catheter confinement weakens local shear-induced redistribution. The resulting picture, therefore, separates the hydrodynamic effects of microrotation and catheter confinement from the kinetic effects of wall retention and absorption, and quantifies how these mechanisms combine to determine transient depletion, advection, and axial spreading.
The transverse non-uniformity measure further shows that this local structure is strongest at early times, is amplified markedly by irreversible uptake and more moderately by reversible retention, while microrotational coupling primarily changes the axial extent of the heterogeneity rather than its peak magnitude.

From an application perspective, the analysis identifies annular clearance and wall kinetics as separate controls on the downstream availability and axial spread of a transported solute. Under a fixed pressure gradient, the sixth-power narrow-gap law implies that reducing the fluid gap can suppress shear-driven spreading far more rapidly than it suppresses mean advection, while \(\Phi_m\), \(\Phi_s\), and \(\Phi_a\) distinguish solute that remains available for downstream transport from that temporarily retained or permanently removed at the arterial wall. These scalings and mass-partition measures therefore provide quantitative benchmarks for interpreting drug or contrast-agent transport in catheterized vessels and for designing reduced-order experiments or models that separate geometric confinement from reactive-wall effects. Because the present formulation assumes a straight, rigid, concentric vessel, steady flow, and first-order wall kinetics, these implications are mechanistic rather than patient-specific; pulsatility, catheter eccentricity, and wall compliance provide natural extensions.

\backsection[Acknowledgements]{
Mr. Sohel Ahmed sincerely acknowledges the Council of Scientific and Industrial Research (CSIR), India, for financial support under grant number 09/1219(16556)/2023-EMR-I.}

\backsection[Funding]{
This work was supported by the Council of Scientific and Industrial Research (CSIR), India (S.A., grant number 09/1219(16556)/2023-EMR-I).}

\backsection[Competing interests]{
The authors declare none.
}

\backsection[Data availability statement]{
The codes used to generate the results in this study are available from the corresponding author upon reasonable request.
}

\bibliographystyle{jfm}
\bibliography{reference}

\end{document}